\documentclass[12pt]{osuthesis}
\usepackage{graphicx}
\pdfoutput=1
\usepackage{subfigure}
\usepackage{caption}
\usepackage{amsmath}
\usepackage{color}
\usepackage{textgreek}
\usepackage{textcomp}
\usepackage{lscape}
\usepackage{multirow}
\usepackage{mathrsfs}
\usepackage{textcomp}
\usepackage{bm}
\usepackage[numbers,sort&compress]{natbib}
\usepackage{cleveref} 				%crossreference multiple figures/equations
\suppressfloats %\nofiles

\title{PARAMETER ESTIMATION OF A NONLINEAR BURGERS MODEL USING NANOINDENTATION AND FINITE ELEMENT-BASED INVERSE ANALYSIS}
\formattedtitle{PARAMETER ESTIMATION OF A NONLINEAR BURGERS MODEL USING NANOINDENTATION AND FINITE ELEMENT-BASED INVERSE ANALYSIS}

\author{SALAH UDDIN AHMED HAMIM}

\degreeone{\ssp Bachelor of Science in Mechanical Engineering \\
        Bangladesh University of Engineering and Technology\\
        Dhaka, Bangladesh\\
        2009}
\degreetwo{\ssp Master of Science in Mechanical and Aerospace Engineering\\
	Oklahoma State University\\
	Stillwater, OK, USA\\
	2011}
\degreesought{DOCTOR OF PHILOSOPHY}
 \degreedate{May, 2016}
\majorfield{Mechanical and Aerospace Engineering}
\begin{document}
\maketitle
\tableofcontents
\listoftables
\listoffigures
%\msp    %========  single space text.
        %========  For the final version, this command should not be used.

%====================== main  body of the dissertation ==========================
\pagenumbering{arabic}\setcounter{page}{1}
 \chapter{INTRODUCTION}\label{chap:intro}

The advent of 21\textsuperscript{st} century has led to the development of several new materials for different engineering applications. Metals or metallic alloys which have been predominantly used for structural applications are being replaced by lighter and stronger composite counterparts. Most often polymers are used as the matrix materials in these composites. Polymers are also being increasingly used in thin film applications. Thin films have already found considerable industrial applications, e.g.~in the production of plane and automobile, as well as in electronic, optical, medical and chemical devices. Unlike metals or ceramics, which exhibit simple elastic--plastic behavior, mechanical behavior of polymeric materials are very complex. Understanding the mechanical behavior of polymers, which exhibit time--dependent responses under applied load, is an important issue in predicting the performance of these materials while in use.

Biomaterials are another important area where knowing the mechanical behavior would greatly improve the quality of understanding of these material systems. The onset of various diseases such as breast cancer~\cite{Samani2007, Paszek2005}, atherosclerosis~\cite{Matsumoto2002,Claridge2009}, fibrosis~\cite{Yeh2002}, and glaucoma~\cite{Last2011} has been found to be related with change in tissue compliance. The application of fast and reliable characterization of biomaterials would not only be beneficial for disease progression, but also in designing improved artificial organs, building virtual surgical simulators and automated robotic surgeon~\cite{Dogan2011,Pal2014}.

Therefore, one of the most important question in today's materials science is understanding mechanical behavior of materials under different loading conditions at different length scales. Traditional mechanical testing methods such as tension, compression, flexural, and bend tests can only provide the macroscale mechanical behavior~\cite{Kunal2010, Hamim2011, Hamim2014, Karumuri2015}. Macroscale test data often differs from the nanoscale test data if the material system is non-homogeneous~\cite{McKee2011}. Moreover, these testing methods require the specimen to be of certain size or shape, which is often difficult to obtain. Complex fixture design and gripping issues are a few more challenges to overcome in case of traditional testing methods.

Nanoindentation is one of the most promising material characterization technique that has the potential to overcome the complexities of conventional testing methods. Nanoindentation involves probing a material with a very small, hard diamond tip of known geometry, while the load and the displacement experienced by the tip is recorded continuously. This load and displacement data is a direct function of material's inherent mechanical properties, and thus makes it theoretically possible to attain mechanical properties from nanoindentation data. The biggest advantage of nanoindentation, which is driving the use of this technique, is that it removes the size or shape restriction placed by the macro or bulk testing techniques. 

However, indenting a material and recording loads and displacements is just the preliminary step in obtaining mechanical properties from the nanoindentation process. As loads and displacements are the only experimentally measurable variables, in order to extract mechanical properties, suitable analytical or numerical methods that relate indentation loads and displacements to material properties are required~\cite{Zhang2010,Engels2014}. This is a challenging task because unlike traditional uniaxial testing methods, nanoindentation load--displacement data comes from complex multi-axial loading, thus making it much more difficult to analyze and subsequently interpret in terms of mechanical properties. 

Past developments in this area has reached to the point where nanoindentation measurements could be related to mechanical properties for materials exhibiting simple elastic or elastoplastic material behavior. However, for materials exhibiting complex material behavior, such as time--dependent material behavior, suitable analysis technique is not available. Understanding mechanical behavior of materials is a root problem, and it carries forward to severely limit applications. For example, accurate mathematical descriptions of the mechanical behavior of soft tissues remain the limiting factor in the advancement of realistic medical simulations and non-invasive diagnostic tools as soft tissues exhibit nonlinear stress-strain behavior at large deformations. 

Developing an analysis technique for nanoindentation of soft materials, such as, polymers, gels, metals at high temperature, and biomaterials, is especially challenging due to the inherent time--dependent mechanical behavior~\cite{Ngan2014}. Time-dependent mechanical behavior, which is known as viscoelasticity or viscoplasticity needs to be taken into consideration in order to accurately predict material behavior under service~\cite{Dean2011}. In case of a viscoelastic or viscoplastic material, the stress state not only depends on the strain, but also the strain rate. 

In chapter 2, a comprehensive review of the existing nanoindentation-based analysis techniques is presented. This includes both analytical and inverse approach for analysis of load--displacement nanoindentation data. Based on the state-of-the-art review few questions are raised, the answers to which if known could significantly improve the applicability of nanoindentation technique for material property characterization. 

In chapter 3, the development of a technique that can be used to characterize nonlinear viscoelastic behavior of soft materials is described. The theories and challenges of the specific techniques is also provided to improve the understanding of the effectiveness of each constituent of the overall technique. 

In chapter 4, an application of the developed technique is presented for an elastoplastic material behavior. This case study is used to understand the overall numerical technique in the context of the nanoindentation experiment. By extending this understanding of the numerical technique, the problem of determining nonlinear viscoelastic constitutive model parameters is solved. Later part of chapter 5 is utilized to draw a conclusion of the study, as well as to report about the possible future works that could improve the robustness and the general applicability of this technique.

 \chapter{LITERATURE SURVEY}\label{chap:chap2}

Nanoindentation, also known as depth-sensing indentation (DSI), is a very popular technique for determination of mechanical properties such as elastic modulus and hardness. It has been extensively used to study the behavior of metallic or ceramic materials in the past couple of decades. Local mechanical properties at the micro- and nanoscale can be effectively characterized by nanoindentation, which is the major advantage of using this technique~\cite{Li2002, Fischer-Cripps2004}. This also makes it ideal to study materials that are otherwise not characterizable by conventional testing methods e.g. thin films, coatings, and localized surface modification of materials~\cite{Karumuri2014, Hamim2014a, Babu2015}. Nanoindentation has also attracted interest for biological material characterization, since it may be used to assess mechanical properties on the cellular scale~\cite{Zhou2012}.

Two different approaches have been primarily used for mechanical characterization of materials by nanoindentation~\cite{Heinrich2009}. The first approach is based on analytical or semi-analytical solutions arising from mathematical contact theories.  The second approach, which is popularly known as `\emph{inverse analysis}', utilizes a combination of finite element methods and numerical optimization algorithms. In inverse analysis the difference between experimental and numerical nanoindentation data, called the objective or error function, is minimized with respect to the material model parameters using numerical optimization. Subsequently, the parameters of the constitutive models are identified as the optimized material properties. Inverse analysis has been found to be applicable in tackling a wide range of problems by the research community~\cite{Buljak2010}. In the next few subsections, a brief review of the nanoindentation based studies is presented for both methods.

\section{Nanoindentation Analysis: Oliver--Pharr Method}

Theoretical studies to characterize the material properties by indentation were first conducted by Hertz. He developed a relationship between the load and indentation depth for spherical elastic bodies. Later Sneddon extended Hertz's work to derive expressions for load, displacement, and contact depth for elastic contacts between a rigid, axisymmetric punch with an arbitrary smooth profile and an elastic half-space~\cite{Sneddon1965}. The first study to use Sneddon's analytical solution and measure the mechanical properties from nanoindentation experiment was conducted by Doerner and Nix~\cite{Doerner1986}. Their study demonstrated that hardness and Young's modulus could be calculated based on the information provided by nanoindentation load--displacement plot. They also pointed out that with the help of suitable analytical procedure plastic properties of a material can also be obtained from nanoindentation. 

In subsequent years, Oliver and Pharr modified the method proposed by Doerner and Nix to find elastic properties of materials~\cite{Oliver1992}. This method has since been cited for more than 13000 times and became more of an unofficial standard for nanoindentation testing. The underlying assumption of this method is that unloading curve of a nanoindentation plot is purely dominated by the elastic properties of the material. Using this method for time-dependent materials would provide inaccurate results since the original assumption does not remain valid. To provide a better understanding of nanoindentation technique a brief overview of this method is followed in next paragraphs.

\begin{figure}[htbp]
   \centering
   \includegraphics[width=0.5\linewidth]{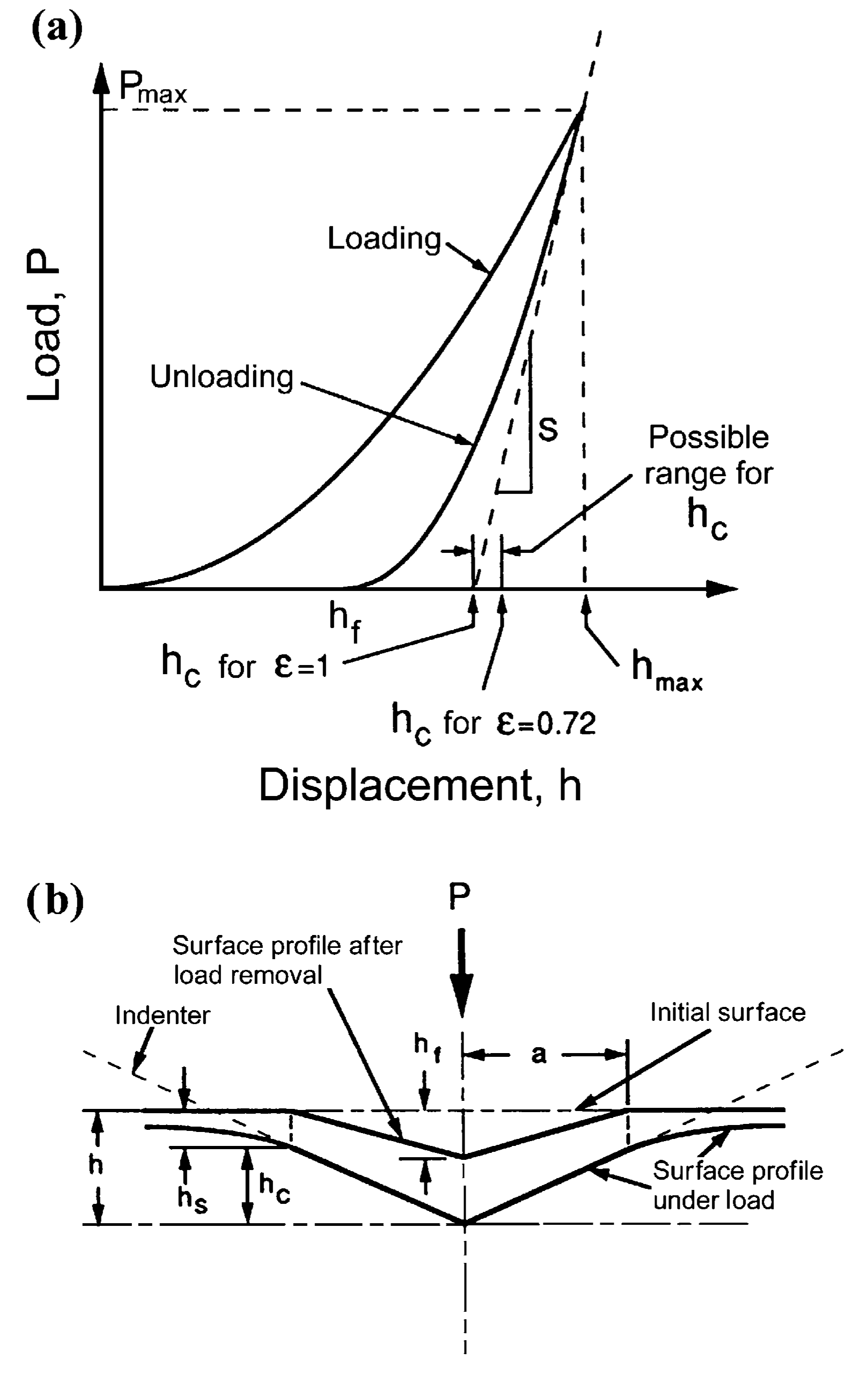} % requires the graphicx package
   \caption{a) Typical nanoindentation load-displacement plot, 
   		b) schematic of the material surface before and after loading~\cite{Oliver1992}}
   \label{fig:indetation_schematic}
\end{figure}

Figure~\ref{fig:indetation_schematic} shows a typical nanoindentation load--displacement plot. In order to extract mechanical properties, such as Young's modulus and hardness, values of contact stiffness, contact depth, and area of contact are required from the nanoindentation plot. The contact stiffness S is the slope of the unloading curve, while the contact depth $h_{c}$ (the depth of actual contact between the indenter and the material) is calculated by Eq.~\ref{eq:h_c}.
\begin{equation}
	h_c=h_{max}-\epsilon~\frac{P_{max}}{S}
	\label{eq:h_c}
\end{equation}
Here, $h_{max}$ is the maximum depth of penetration including elastic deformation of the surface under load, $P_{max}$ is the maximum force, and $\epsilon$ is a geometrical constant associated with the geometry of the indenter~\cite{Oliver1992}. This method of determining the contact depth is commonly referred to as the Oliver--Pharr method; a schematic in Fig.~\ref{fig:indetation_schematic} shows $h_{c}$ and $h_{max}$. 

Once $h_{c}$ is determined, the projected area A of actual contact can be calculated using the cross-sectional shape of the indenter along its length. Determining accurate contact area is found to be crucial for elastic analysis of nanoindentation data~\cite{Poon2008}. This area function could be determined by direct measurement of the imprint geometry under a scanning microscope~\cite{Briscoe1998}, but in practice is normally determined by indenting a reference sample and iteratively fitting the results. The relationship between contact area, A and the contact depth, $h_{c}$ for a Berkovich tip is generally expressed by the following equation--
\begin{equation}
A=24.5h_c^2+C_1 h_c+C_2 h_c^{0.5}+C_3 h_c^{0.25}+\dots
\label{eq:area_function}
\end{equation}
here, the coefficients $C_i$ can be determined by iterative fitting to indentation measurements conducted on reference material such as fused silica. The first term in Eq.~\ref{eq:area_function} represents the area--depth relationship for a perfectly sharp Berkovich indenter, while the other terms account for tip imperfections e.g.~tip roundness. Once the area of contact A is determined, hardness is found using a simple equation-
\begin{equation}
	H=\frac{P_{max}}{A} 
	\label{hardness}
\end{equation}

It is important to note that this hardness is defined using the projected area of contact under load, while macroscopic definition of hardness is force divided by the area of the residual imprint left by the indenter. For most materials the two definitions yields very similar values. However, in case of a material showing little to no plastic flow, the hardness calculated by Eq.~\ref{hardness} tends to be lower than the macroscopic definition.

Once contact area, A and contact stiffness, S is known, Sneddon's solution can be adapted independent of the geometry of the punch, and Young's modulus can be calculated using the following equation for reduced modulus:
\begin{equation}
E_r=\frac{1}{\beta} \frac{\sqrt \pi}{2} \frac{S}{\sqrt A}=\left(\frac{1-\nu_i^2}{E_i}+\frac{1-\nu_s^2}{E_s}\right)^{-1}
\end{equation}

\noindent
here, $\beta$ is a small correction for the non-axisymmetric indenter shape (e.g. $\beta$ = 1.034 for a Berkovich tip). For a perfectly elastic--plastic material with no other form of deformation present, the unloading curve is purely dominated by the elastic recovery of the material. As a result, Young's modulus determination from an unloading curve of a nanoindentation experiment becomes possible. Later Field and Swain developed means of extracting both Young's modulus and yield strength from load--displacement curves of a spherical indentation~\cite{Field1995}.

Approximation of Sneddon's solution is that the indenter is rigid, and therefore, deformation of the indenter is small and insignificant compared to the material being tested. As long as this approximation is valid Sneddon's solution can yield good results for the reduced modulus. However, in case of testing very hard materials, such as diamond-like carbon the deformation experienced by the indenter is substantial, thus violating the approximation Sneddon's solution is based upon. 

Oliver--Pharr method and subsequent developments provided a means for extracting few key material parameters from a nanoindentation plot, namely Young's modulus, hardness, and yield strength. These were groundbreaking developments in terms of characterizing elastic--plastic material behavior. However, these methods are unusable for viscoelastic materials as the underlying assumption of unloading curve purely dominated by elastic recovery no longer holds for viscoelastic materials. In addition to that, hardness, modulus, and yield strength properties are inadequate to represent the full spectrum of behavior for these materials.

\section{Adaptation of Oliver--Pharr Method for Time--Dependent Behavior}

\begin{figure}[htbp]
   \centering
   \includegraphics[width=0.5\linewidth]{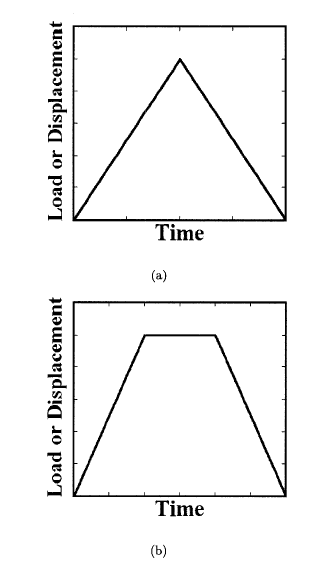} % requires the graphicx package
   \caption{Load-displacement profiles a) triangular, 
   		b) trapezoidal~\cite{Gupta2008}}
   \label{fig:LD_profiles}
\end{figure}

The most widely used indenter load or displacement profiles are the triangular, where the load or displacement is ramped at a certain rate to the maximum value and then unloaded back to zero, as shown in Fig.~\ref{fig:LD_profiles}. For elastoplastic material systems (e.g.~most metals and ceramics) exhibiting little to no time--dependent behavior, the load--displacement nanoindentation curve is insensitive to loading or unloading rates; thus, triangular profiles can be effectively used to characterize these materials. 

However, this is not the case for testing of viscoelastic materials such as polymers and biomaterials due to the fact that viscous behavior of these materials dramatically affect the load--displacement curve. The inherent time--dependency in mechanical response of these materials make the unloading curve of the nanoindentation experiment noticeably different by producing a``nose"~\cite{Briscoe1998,Ngan2002}. 

\begin{figure}[htbp]
   \centering
   \includegraphics[width=0.8\linewidth]{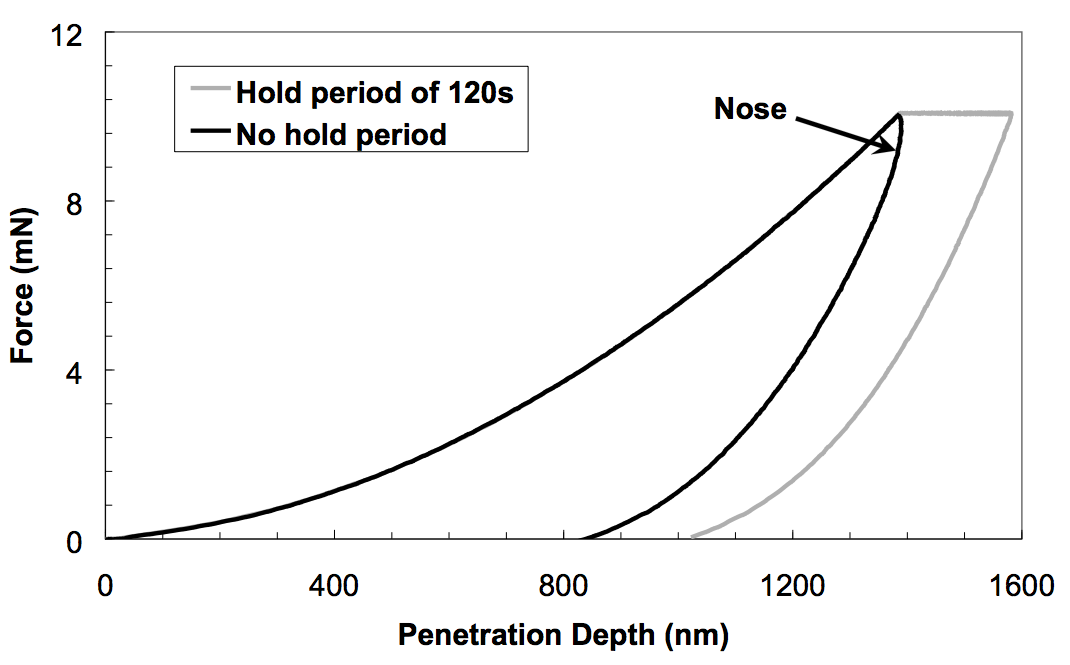} % requires the graphicx package
   \caption{Typical nanoindentation plot for a viscoelastic material~\cite{Conte2006}}
   \label{fig:nose}
\end{figure}

Figure~\ref{fig:nose} shows a typical nanoindentation load--displacement plot for a viscoelastic material. The nose results from excessive creep of a material under the indenter, which dominates over the elastic recovery of the material as the tip retracts from the surface. Applying Oliver--Pharr method on a nanoindentation plot exhibiting nose often provides a negative value for contact stiffness, S, and prevents extracting elastic modulus altogether. Even without the appearance of the nose, the presence of viscoelasticity often leads to overestimation of Young's modulus. 

A trapezoidal load or displacement profile that implements a long enough hold before the unloading has been found to suppress the creep behavior near the initial unloading part~\cite{Ngan2002,VanLandingham2003, Feng2002,Tang2003,Ngan2005}. The holding period ensures complete relaxation of the material, and minimizes the viscoelastic recovery during the unloading. 

An useful modification to Oliver--Pharr method was proposed by Ngan~\emph{et al.} so that true value of Young's modulus for viscoelastic materials can be extracted from nanoindentation experiment~\cite{Ngan2002}. According to their study, for a load-controlled indentation test with hold period prior to unloading, contact stiffness can be corrected using the following equation---
\begin{equation}
\frac{1}{S_e}=\frac{1}{S}+\frac{dh/dt |_{t=tm}}{v_P}
\end{equation}
where dh/dt $|_{t=tm}$ is the indenter displacement rate at the end of the load hold just prior to unloading, S is the contact stiffness found via original Oliver--Pharr method, and $v_P$ = $|dP/dt|$ is the initial unloading rate. 

Although, Ngan~\emph{et al.}'s method provided an useful way to use Oliver--Pharr method for characterizing viscoelastic solids, it simply cannot address the various other important properties of a viscoelastic material~\cite{Cheng2011a}. As a result developing dedicated analysis techniques for viscoelastic material characterization via nanoindentation has been one of the most popular research area of the past decade.

\section{Analytical Approaches for Viscoelastic Materials} 

Analytical solutions capable of characterizing viscoelastic material behavior from nanoindentation load--displacement plot originated from the early works of Radok~\cite{Radok1957}. He was the first to tackle the linear viscoelastic contact problem using the method of functional equations or hereditary integrals, which was later completed by Lee and Radok~\cite{Lee1959}. This method of functional equations solved the viscoelastic problem by replacing the elastic constant with their corresponding viscoelastic operators. This is why this method is also known as \emph{`Correspondence Principal'}. Radok extended the `Laplace transform method' formulated by Lee to eliminate the explicit time dependence of the viscoelastic contact problem and solved it in the Laplace domain~\cite{Lee1955}. Before Lee and Radok, Laplace method was only applicable to problems where displacement and stress boundary conditions are unchanged e.g. flat punch indentation problem. 

The method of functional equations proved to be very successful in formulating analytical solutions for viscoelastic bodies; however, the solutions were only valid as long the penetration depth in a viscoelastic indentation monotonically increased~\cite{Lee1959}. Hence, this method is only valid for the loading portion of the nanoindentation plot. Many researchers attempted to remove this restriction. Hunter was able to remove it for spherical indentation, while Ting's implicit equations were able to remove it altogether for any linear viscoelastic material tested under any axisymmetric indenter shape~\cite{Hunter1960, Ting1966}. However, except for few specific cases applying Ting's formulation is a challenge. As a result, closed form solutions for linear viscoelastic problems are still being formulated using Radok's method of functional equations. 

In 1985, Johnson summarized the correspondence analysis of spherical indentation replacing the elastic constants by the Boltzmann viscoelastic hereditary integral operators. Based on these approaches, in several contributions~\cite{Cheng2000,Cheng2005,Giannakopoulos2006,Huang2006,Jager2007,Lu2003,VanLandingham2005}, the viscoelastic analytical solutions of nanoindentation with different indenter tips were presented. However, the `Correspondence Principal' method is restricted by yielding accurate identification only for specific linear viscoelastic models under fixed experimental processes. 

\subsection{Spring--Dashpot Based Viscoelastic Models}
A key aspect of viscoelasticity is that mechanical behavior of a material can be successfully modeled using a combination of springs and dashpots. Viscoelastic materials demonstrate both elastic and viscous behavior in the same material. If spring represents the elastic behavior and dashpot represents the viscous behavior, a combination of two could be able to model the behavior of a viscoelastic material. The biggest advantage of using spring--dashpot based model is that the viscoelastic models can be tailored to suit specific observations~\cite{Schiessel1995}. 

By putting this idea to use several spring--dashpot model has been proposed in the literature. For a viscoelastic material, stress level is related to both strain level and strain rate in the following general form
\begin{equation}
A_0 + A_1 \frac{d\sigma}{dt} + A_2 \frac{d^2\sigma}{dt^2} + \dots = B_0 + B_1 \frac{d\epsilon}{dt} + B_2 \frac{d^2\epsilon}{dt^2} + \dots
\end{equation}

\noindent
where, $\epsilon$ and $\sigma$ are the strain and stress levels, respectively, and t is the time. $A_i$ and $B_i$ are the coefficients that determine the linear or even non-linear stress--strain behavior.

\begin{figure}[htbp]
   \centering
   \includegraphics[width=0.3\textwidth]{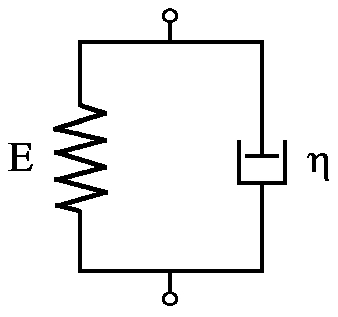} % requires the graphicx package
   \caption{Kelvin-Voigt model}
   \label{fig:KV}
\end{figure}

\begin{figure}[htbp]
   \centering
   \includegraphics[width=0.4\textwidth]{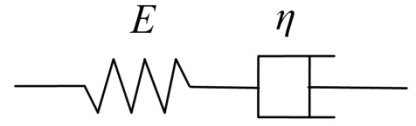} % requires the graphicx package
   \caption{Maxwell model}
   \label{fig:Maxwell}
\end{figure}

In the most simplest of forms, where one spring element and one dashpot element is used to create a model, this technique leads to two well known models, namely Kelvin--Voigt and Maxwell model. These models assume linear stress--strain relationship. \Cref{fig:KV,fig:Maxwell} show Kevin--Voigt and Maxwell model, respectively.


Using the corresponding constitutive equations for spring and dashpot, the following equation can be developed for Kelvin--Voigt model.
\begin{equation}
\sigma=E\epsilon + \eta \dot\epsilon
\end{equation}

\noindent
here, $\sigma$=E$\epsilon$ is the creep part and $\eta\dot\epsilon$ is the recovery part of the model; $\dot\epsilon$=$d\epsilon/dt$ is the strain rate; E is the rigidity modulus, and $\eta$ is the coefficient of viscosity.

Under constant stress conditions, the strain response of the material can then be captured as an exponential decay function
\begin{equation}
\epsilon=\frac{\sigma_0}{E} \left(1-e^{-Et/\eta}\right)
\end{equation}

However, under constant strain rate conditions (stress relaxation part), the Kelvin--Voigt provides unrealistic linear elastic behavior for the viscoelastic material. The Maxwell model, however, povides better approximation for constant stress relaxation.
For Maxwell model, the constitutive equation comes in the following form-
\begin{equation}
\dot\epsilon=\frac{\dot\sigma}{E} + \frac{\sigma}{\eta}
\end{equation}

In case of stress relaxation ($\dot\epsilon$=0), an exponential decay of stress is found,
\begin{equation}
\sigma=\sigma_0 e^{-Et/\eta}
\end{equation}

\noindent
while in a recovery experiment ($\dot\sigma$=0), the model predicts the basic equation of pure Newtonian flow.

\begin{figure}[htbp]
\begin{center}
\includegraphics[width=2in]{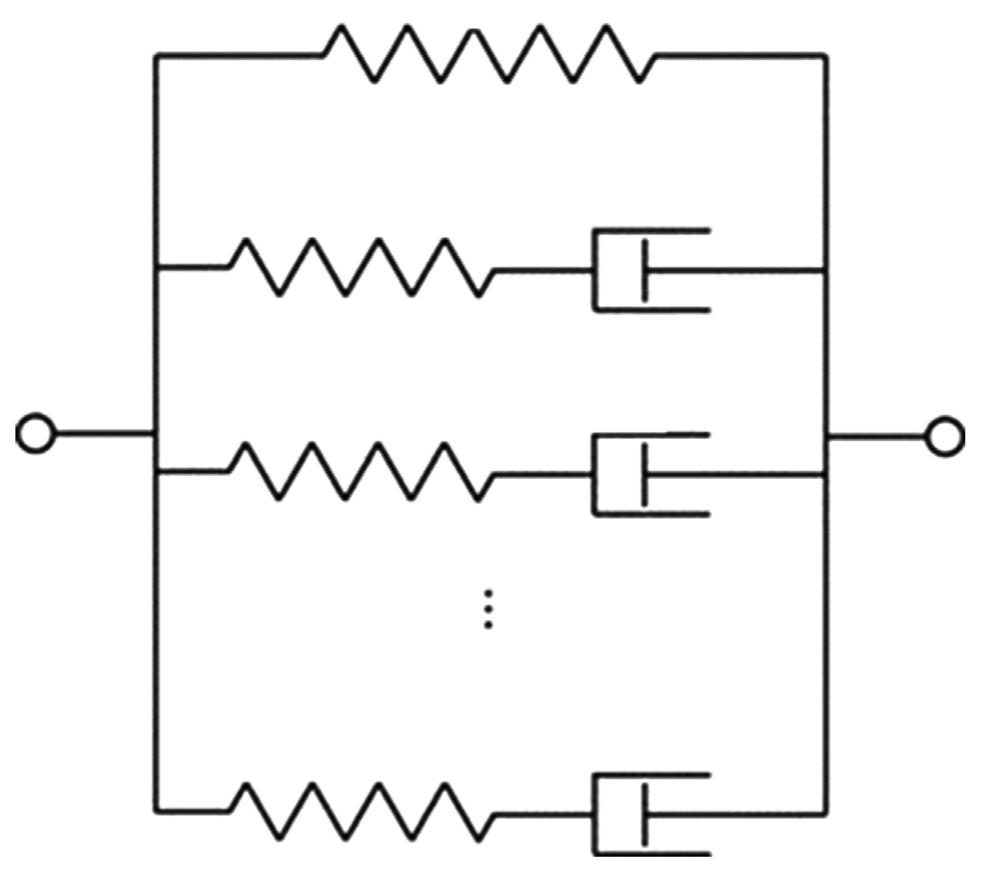}
\caption{Schematic of Generalized Maxwell model}
\label{fig:gen_max_model}
\end{center}
\end{figure}

Real world materials are much more complex in their behavior, and these simplistic models are not sufficient in representing that. Previous studies show that modeling creep and relaxation behavior of complex viscoelastic behavior requires an assembly of multiple spring and dashpot in the model~\cite{Liu2006}. One such model is generalized Maxwell model. Various studies have reported that generalized Maxwell model worked well in terms of modeling the viscoelastic behavior~\cite{Klompen1999,Lu2003,Jo2005,Lin2009,Daphalapurkar2009}. For this model relaxation can be written in the general form
\begin{equation}
\sigma=Y(t)\:\epsilon_0
\end{equation}

\noindent
where, Y(t) is the relaxation function.

The relaxation function can be represented using Prony series having the following expression:
\begin{equation}
Y(t)= E_0 \left( 1 - \sum_{i=1}^n p_i (1 - e^{-t/\tau_i}) \right)
\end{equation}

\noindent
where, $p_i$ is the $i^{th}$ Prony constant, $\tau$ is the Prony retardation constant, $E_0$ is the instantaneous modulus. 
Prony coefficients are usually found by nonlinear regression, which allow adjusting the model with respect to the observed behavior.

\section{Inverse Approach Based Material Behavior Modeling}

Inverse problems are defined as the problems where the output is known and the input or source of output remains to be determined. They are contrary to the direct problems, in which output or response are determined using information from input~\cite{Kubo2006}. In order to analyze inverse problems, experimental data obtained under known boundary conditions are compared with the calculated ones. The combination of nanoindentation and FEA has proved to be a powerful analysis tool for soft polymers such as gels, and coatings, and for soft tissues~\cite{Zhang1997, Ovaert2003,Olberding2006,Lei2007}. 

Inverse analysis requires an optimization algorithm to extract the set of parameters for which the objective function (difference between simulation and experimental load-displacement data) attains the minimum value. The choice of the optimization algorithm for minimizing an objective function is a topic of interest. Whenever possible it is better to employ global optimization techniques. There are many variants of Simulated Annealing or Genetic Algorithm based global optimization scheme, such as evolutionary algorithms, or deterministic algorithms like the Simplex method. These algorithms have proven to be very useful in case of optimization problems where user has no prior information about the location of the solution in the parameter space, thus incapable of making a priori choices about the initial estimates. 

However, in case of finite element analysis, where time required to run one single analysis can range from few minutes to even days, the success of global optimization methods come at a price of astronomical computational cost. In these cases, local optimization algorithms could prove to be useful given that the quality of initial estimates are good. However, these algorithms are gradient-based and involves computationally costly calculations of second order partial derivatives of the objective function. For an objective function \emph{f}, Hessian matrix, \textbf{H} is defined by the Eq.~\ref{eq:hess}. 
\begin{equation}
\textbf{H}_{i,j} = \frac{\delta^2f}{\delta x_i \: \delta x_j}
\label{eq:hess}
\end{equation}
Determining \textbf{H} often requires high computational cost. The cost involving calculation of Hessian matrix can be reduced if it is approximated using the Jacobian matrix, \textbf{J} (Eq.~\ref{eq:jacobian_approx}).
%Jacobian Matrix 
\begin{equation}
\textbf{J}_{i,j} = \frac{\delta f_i}{\delta x_j}
\label{eq:jacobian}
\end{equation}
%Hessian Approximation
\begin{equation}
\textbf{H}_{i,j} \approx 2\textbf{J}^\textsuperscript{T}\textbf{J}
\label{eq:jacobian_approx}
\end{equation}
The biggest disadvantage of gradient-based scheme is that algorithms can sometime get trapped inside a local minima. In addition to that, due to the ill-conditioned nature of inverse analysis, identifying the correct minima from a set of local minima is troublesome. Singularity in approximated Hessian matrix and non-covergence are few other problems that often trouble the local optimization techniques. 

\begin{figure}[htbp]
\begin{center}
\includegraphics[width=0.6\textwidth]{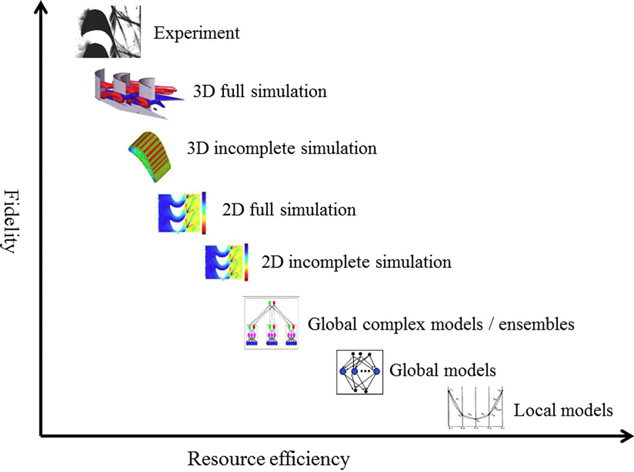}
\caption{An illustration of trade-off between fidelity and computational cost~\cite{Jin2011}}
\label{fig:fidelity}
\end{center}
\end{figure}

\section{Review of Existing Literature}

\subsection{Analytical Approach in Nanoindentation}

A large number of studies have been conducted in an effort to determine viscoelastic behavior using nanoindentation. Cheng \emph{et al.}~derived the analytical solutions for linearly viscoelastic deformation and provided a method to measure viscoelastic properties described by a three-element standard linear solid (SLS) model using a flat-punch indenter~\cite{Cheng2000}. 
Hernandez-Jimenez~\emph{et al}.~studied the viscoelastic behavior of PMMA and PTFE  using Maxwell model~\cite{Hernandez-Jimenez2002}.
Lu~\emph{et al}.~developed methods to measure the creep compliance of PMMA and PC polymers using spherical or Berkovich indenter by deducing closed form analytical solutions using Generalized Maxwell model~\cite{Lu2003}. Prony series parameters for the stress relaxation or creep compliance was found by curve fitting only the loading portion of the nanoindentation plot. Measurement of creep compliance from conventional tension and shear tests were compared with the nanoindentation technique, where reasonable agreement between the values from different techniques was observed. Fisher-Cripps developed creep compliance analytical closed form solutions for three-element Voigt model, four-element Maxwell--Voigt model~\cite{Fischer-Cripps2004}. Gonda~\emph{et al}.~spherical and conical indentations on a thin polymer film on a substrate, where viscoelastic properties found by analytical equations through correspondence principal and the results were verified using finite element modeling~\cite{Gonda2005}. Vanlandingham~\emph{et al}.~investigated linear viscoelastic material analytical solutions for epoxy, PMMA and  PDMS polymers, and compared the values obtained from nanoindentation with values from rheometry measurements~\cite{VanLandingham2005}. Cheng and Cheng derived closed form solutions relating the initial unloading slope, instantaneous relaxation modulus, and contact depth for linearly viscoelastic material under a rigid and arbitrary axisymmetric shape~\cite{Cheng2005b}. In another study, they also derived the relationship between unloading slope, contact depth and instantaneous modulus for conical indentation~\cite{Cheng2005}. Three parameter Standard Linear Solid (SLS) model has also been used to derive equations for spherical nanoindentation of viscoelastic materials~\cite{Cheng2005a}. Cheng~\emph{et al}.~addressed the linear viscoelastic material indentation using three-parameter Maxwell solid~\cite{Cheng2006}. Finite element simulations were conducted using these relationships to verify the solutions. Zhou and Lu developed methods to measure creep compliance and relaxation modulus in both time and frequency domains under constant and ramped loading conditions using a spherical indenter~\cite{Zhou2009}. 
Vandamme~\emph{et al.}~modeled using 3-parameter Maxwell model, the 4-parameter Kelvin-Voigt model and the 5-parameter combined Kelvin-Voigt-Maxwell model and derived the closed form analytical solution for conical indenter~\cite{Vandamme2006}.
Wei~\emph{et al}.~studied the viscous behavior of PMMA and PU materials using a combination of Kelvin--Voigt model and a dashpot~\cite{Wei2008}. Their model accounted for the irreversible delayed plastic (viscoplastic) deformation, irreversible viscous deformation, and reversible delayed elastic (viscoelastic) deformations. Oyen and Cook examined the creep displacements as a function of time for PMMA and a few other polymers using constant loading and unloading rates~\cite{Oyen2006}. They also examined the effect of triangular and trapezoidal loading profiles. For pyramidal indentation tests, a new method for estimation of time-constant was proposed. Liu~\emph{et al}.~developed a model based on Burgers model and applied to understand the viscoelastic behavior of soft polymers like PMMA~\cite{Liu2006}. According to them, Burgers model provided the best agreement with the experimental data in comparison to simple Maxwell or Kelvin-Voigt model. In addition to that, they also indicated that the  nose formation at the beginning of the unloading stemmed from the decrease of the viscosity parameter. 
Jager~\emph{et al}.~characterized viscoelastic properties of bitumen using different spring-dashpot models for real tip geometry of the indenter~\cite{Jager2007}. Linear viscoelastic analysis based on spherical indentation experiment has also been carried out on human tympanic membrane~\cite{Daphalapurkar2009}. 
Lin~\emph{et al}.~studied viscoelastic behavior of PDMS micro pillars using uniaxial, DMA, nanoindentation tests, where generalized Maxwell model was used to describe the viscoelastic behavior of the material~\cite{Lin2009}.

Mencik~\emph{et al}.~analyzed the viscoelastic--viscoplastic behavior of material under indentation for different indenter profiles~\cite{Mencik2010}. They found that materials under sharp indenter undergoes high stresses and exhibits viscoplastic effects. Chen~\emph{et al}.~used dimensional analysis and finite element analysis to understand the effects of residual stress, substrate, and creep behavior on the load-displacement data~\cite{Chen2008}. Through development of some analytical solutions they showed that it is possible to obtain not only Young's modulus and hardness, but also viscoelastic properties and residual stress. 

\subsection{Inverse Approach in Nanoindentation}

The very first instance of applying inverse method for an indentation-based study was probably by Knapp \emph{et al}., where they studied the elastic--plastic behavior of Al under nanoindentation~\cite{Knapp1999}. Their study showed that it was possible to extract modulus, yield strength, and hardening coefficient from the nanoindentation data of thin films using FEA based inverse analysis independent of the effect of substrate. Later, Huber~\emph{et al}.~employed Artificial Neural Network (ANN)-based inverse analysis to extract material parameters from an indentation experiment of metals~\cite{Huber2002}. From that point onwards, inverse FEA-based analysis has been used to extract material properties for different classes of materials, such as, isotropic and anisotropic elastic--plastic materials~\cite{Bocciarelli2005, Nakamura2007,Kopernik2009,Moy2011}, linear viscoelastic materials~\cite{Guessasma2008,Liu2009,Valdez-Jasso2010,Abyaneh2013}, hyper-elastic materials~\cite{Lin2008,Cox2008,Su2010,Rauchs2010}, nonlinear viscoelastic materials~\cite{Kucuk2012,Kucuk2012a}, etc.

After being introduced by Knapp~\emph{et al}., inverse FEA technique has been reported in numerous publications dealing with material property extraction for elasto-plastic materials. On the contrary, the number of studies that tackled viscoelastic nanoindentation using inverse analysis is found to be very low. The probable reasons could include the lack of understanding about the viscoelastic constitutive relationships, the high number of model parameters needing to be optimized, etc. While elasto-plastic behavior in materials has been studied for a long time, viscoelasticity is being studied only recently fueled by the recent interests in understanding polymers and biomaterials.  As we are interested in the viscoelastic materials, this part of the literature will only review the related studies in the area of viscoelasticity.

Ovaert~\emph{et al}.~studied on viscoelastic properties of thin polymer coatings using three- and four-parameter viscoelastic models using indentation and inverse analysis~\cite{Ovaert2003}. 
Kim and Srinivasan~\emph{et al}.~extracted Fung's QLV model parameters for soft tissues using two step parameter optimization process~\cite{Kim2005}. Hartmann \emph{et al.}~used uniaxial test data for viscoplastic parameter identification and validated those using indentation test data~\cite{Hartmann2006}. Samur~\emph{et al}.~studied the viscoelastic behavior of pig liver tissues using inverse analysis~\cite{Samur2007}.
Resapu~\emph{et al.}~extracted Prony series parameters for the relaxation behavior of PVC and PE in indentation tests~\cite{Resapu2008}. Guessasma~\emph{et al}.~determined viscoelastic properties of biopolymer composite materials~\cite{Guessasma2008}. Liu \emph{et al}.~characterized viscoelastic behavior of soft gels using Kelvin--Voigt model~\cite{Liu2009}.
Rauchs identified viscous hyper-elastic and elasto-viscoplastic material parameters from indentation tests~\cite{Rauchs2011,Rauchs2010}. 
 Abyaneh~\emph{et al}.~characterized porcine cornia using linear viscoelastic model~\cite{Abyaneh2013}. Viscoelastic Arruda--Boyce constitutive model has also been studied with AFM indentation and inverse FE analysis for porcine zone pellucida~\cite{Boccaccio2014}. Rayleigh dissipative function has been used by Abetkovskaia~\emph{et al}.~to develop AFM based viscoelastic characterization of soft materials~\cite{Abetkovskaia2010}. Valdez-Jasso~\emph{et al}.~used inverse analysis to characterize viscoelastic behaviors of ovine aorta, where the viscoelastic behavior was modeled using arctangent and sigmoid viscoelastic models~\cite{Valdez-Jasso2010}. Recently, Kucuk~\emph{et al}.~used nonlinear Burgers model to characterize the nonlinear viscoelastic behavior of PMMA and PVAc~\cite{Kucuk2012, Kucuk2012a}.

Inverse analysis of nanoindentation data is challenging due to various reasons. One of this big challenge is to find out unique solution. In case of non-unique solutions, two approaches were found to be effective. In the first approach, additional information from the nanoindentation experiment is gathered, and used in the objective function. These information can include imprint geometry~\cite{Bocciarelli2005, Schmaling2012, Arizzi2014} or pile-up/sink-in~\cite{Moy2011} information. The other approach is to use multiple indenters with different geometry. This method has also been found to beneficial in providing unique solutions from the inverse analysis~\cite{Chollacoop2003,Nakamura2007,Heinrich2009}.

\section{Motivation and Objective of the Study}

Nanoindentation has the potential to become a very effective material characterization tool given that appropriate analysis technique with proper constitutive relationship is used. In the last two or three decades this technique has come a long way in terms of applicability for metallic or ceramic material characterization. 

However, suitable analysis technique for materials such as polymer or soft tissues is still lacking. \Cref{tab:soft_tissue_modulus} summarizes the results found from various studies that used nanoindentation technique for characterizing soft tissues. It can be seen that, for almost all the studied tissues the value of Young's modulus varied by few orders of magnitude. Part of the variability comes from the difference in experimental design and sample preparation, while most of it stemmed from the fact that these materials exhibited time--dependent deformation behavior~\cite{McKee2011}. If consistent strain rate were to be used in the experiment a more consistent Young's modulus could probably be found. 

Even if we consider that the Young's modulus could be extracted reliably independent of viscoelastic influence, it would only serve as a partial knowledge about the material system. Young's modulus only quantifies the intrinsic elastic behavior of a material, which limits its usefulness only to metals or crystalline solids. 
\begin{table}[ht!]
\caption{Young's modulus of soft tissues, measured by indentation~\cite{McKee2011}}
\begin{center}
\begin{tabular}{c c c c}
\hline
Tissue & \multicolumn{1}{p{2cm}}{\centering Range \\ (kPa)} & \multicolumn{1}{p{2cm}}{\centering Average \\ Young's \\ modulus (kPa)} & Reference\\
\hline
Liver and Kidney & 0.6-760 & ~ 190 & ~\cite{Constantinides2008,Barnes2007,Tay2006}\\
Artery and Vein & 6.5-560 & ~ 125 & ~\cite{Ebenstein2004,Oie2009,Engler2004}\\
Skin & 6-222 & ~ 85 & ~\cite{Constantinides2008,Ling2007,Pailler-Mattei2008}\\
Cornea anterior base & 7.5-50 & ~29 & ~\cite{Last2009}\\
Breast tissue & 0.167-29 & ~ 8 & ~\cite{Samani2007,Paszek2005,Samani2003}\\
Muscle & 2-12 & ~ 7 & ~\cite{Engler2004,Chen1996a}\\
Spinal cord and gray matter & 0.2-7 & ~ 3 & ~\cite{Elkin2007, Saxena2009}\\
\hline
\end{tabular}
\end{center}
\label{tab:soft_tissue_modulus}
\end{table}

In the attempts to understand or characterize the time--dependent properties inherent to soft polymers and biomaterials, most researchers simplified the behavior of these materials as linear viscoelastic. In fact most of the studies that used nanoindentation to characterize viscoelastic materials used linear viscoelastic theory developed through `Correspondence Principal'. In addition to the fact that material behavior is simplified as linear viscoelastic, correspondence principal based analytical solutions has further limitations \emph{i.e.}~useful only till contact area increases monotonically (loading portion of the nanoindentation plot). As a result, this method fails to address how viscous behavior affects the unloading curve of the nanoindentation experiment, although substantial amount of information about the material behavior is present in the unloading portion of the curve. 

To best of our knowledge, no analytical or closed-form solutions (in either differential or integral form) exist for indentation of quasi-linear or nonlinear viscoelastic material. However, soft tissues and polymers are generally nonlinearly viscoelastic~\cite{Fung1993, Jo2005}, where the creep compliance or relaxation modulus are a nonlinear function of both time and applied stress or strain. In these cases, an appropriate constitutive law should be used to describe the distinct behaviors of these materials~\cite{Tranchida2007}. Due to the fact that, no closed form solution can be obtained for nonlinear viscoelastic behavior, many researchers tried modeling the behavior of the material using Fung's Quasilinear Viscoelastic (QLV) model~\cite{Fung1993}. Fung's model however considers the material to be nonlinear only with respect to strain~\cite{Darvish2001}, and fails to represent the full spectrum of nonlinearity of the material. 

The closest work that tackled nonlinear viscoelastic behavior of the material was by Kucuk \emph{et al.}~\cite{Kucuk2012,Kucuk2012a}. In these studies, a nonlinear viscoelastic model based on modified Burgers model was used. The unknown model parameters were then obtained using inverse analysis. However, the authors did not provided any information about the inverse analysis procedure that was followed. Without such key information obtaining the values of the model parameter for other material system is difficult. In addition to that, their study utilized quite a high number of parameters in the nonlinear model without providing any information about whether all the parameters were required to capture the behavior or not. 

To understand the full spectrum of mechanical behavior in soft biomaterials and polymers, an study is thus required which would improve on the limitations of previous studies. Because without understanding the mechanical behavior, it would be impossible to predict the behavior of these materials under complex loading scenarios. 

\begin{figure}[htbp]
\begin{center}
\includegraphics[width=0.65\textwidth]{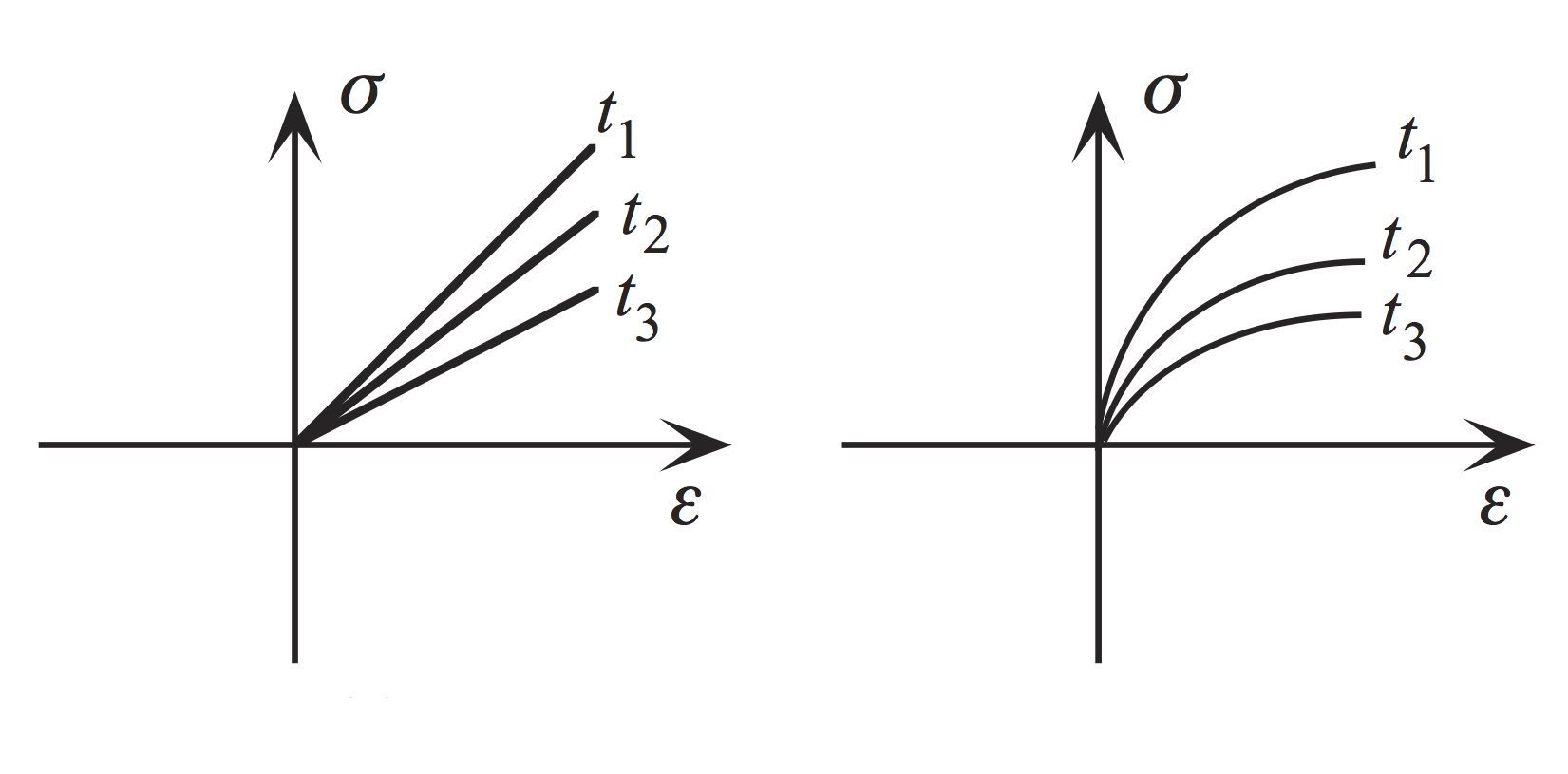}
\caption{Viscoelasticity a) linear and b) nonlinear}
\label{fig:nonlin_visco}
\end{center}
\end{figure}

In order to develop a nonlinear viscoelastic model for soft materials, we propose to implement finite element analysis with inverse analysis. The process is called the inverse analysis because it is the opposite of an ordinary simulation (\emph{i.e.} solving for forces or displacements given material parameters and boundary conditions). The inverse method permit us to treat any material models with nonlinear properties and to include further affecting factors in the numerical model. The rate-dependent properties of materials can be more accurately identified using the inverse method. 

The whole process of developing nonlinear viscoelastic model can be subdivided in few steps. In the first step, a finite element model of the nanoindentation experiment is required which can effectively simulate the experiment. For this work, we have chosen commercially available finite element analysis software--ABAQUS. Confidence was established on the ABAQUS representation of the nanoindentation experiment by comparing the simulation results with the well established analytical solution from contact theory. 

In the next step, an appropriate spring--dashpot system for describing these kind of materials needs to be developed. The associated mathematical model for the spring--dashpot system has to be incorporated in the ABAQUS simulation of the nanoindentation experiment via user-defined subroutine called UMAT. 

In the final step, an optimization based algorithm needs to be established, which will be able to minimize the difference between the simulated and experimentally found load--displacement data. This study will use both the loading and unloading portion of the nanoindentation experiment in the model development process; because unloading curve would provide additional constraints which a successful model must satisfy. In addition, this will provide additional validation of the viscoelastic model parameters extracted from nanoindentation data.

One of the main issue in inverse analysis based model development is the high computation cost associated with the optimization process. In case of a nonlinear model, the number of model parameters that needs to be optimized is usually high. In addition to that, nonlinear FEA analysis requires considerably higher computational cost due to the continuous updating procedure of global stiffness matrix. This updating process is a serious drawback for FEA-based realtime optimization applications~\cite{Dogan2011}. That is why, improving the computational efficiency in the inverse analysis process is another important objective of this study.

 \chapter{DEVELOPMENT OF THE TECHNIQUE}\label{chap:chap3}

\section{Nonlinear Viscoelastic Mathematical Model}\label{subchap:nonlin_model}

There are a few nonlinear viscoelastic models in the literature but to date it appears that none of these models can describe the nonlinearly viscoelastic behavior of a polymer under all loading and environmental conditions~\cite{Findley1989}. Under a given set of loading conditions, however, an appropriate nonlinearly constitutive model can be used to model the viscoelastic response. 

In this study, a spring--dashpot model suggested by Marin and Pao~\cite{Marin1953} was used. In linear case this model is generally called four-parameter Burgers model~\cite{Richter2001} and it is formed by a serial connection of a Maxwell element to a Voigt element as seen in Fig.~\ref{fig:non_burgers}. For an increased relaxation spectrum, the
viscoelastic response can be modeled by increasing the number of Voigt elements. 

The nonlinear characteristic is introduced when the dashpot constants ($m_s$ and $m_t$) take values other than unity. In the three-dimensional model, the total strains are calculated as the summation
of the elastic ($\epsilon$\textsuperscript{e}), transient creep ($\epsilon$\textsuperscript{t}), and steady creep strains ($\epsilon$\textsuperscript{s})~\cite{Shames1997}. In this study, the nonlinear creep deformation is assumed to be incompressible. Under these assumptions, the three-dimensional nonlinearly viscoelastic law can be expressed as:

\begin{figure}[htbp]
\begin{center}
\includegraphics[width=0.6\textwidth]{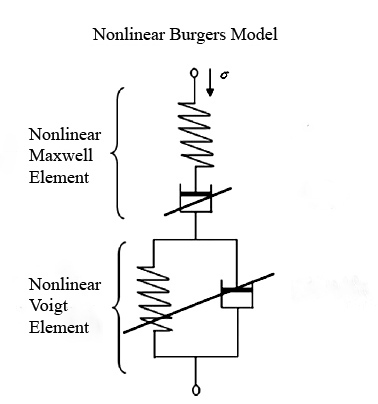}
\caption{Schematic of nonlinear Burger's model}
\label{fig:non_burgers}
\end{center}
\end{figure}

\begin{equation}
	\epsilon_{ij}^e = \frac{1+\nu}{E}\sigma_{ij} - \frac{\nu}{E}\sigma_{kk}\delta_{ij}
	\label{eq:elast}
\end{equation}

\begin{equation}
	\dot \epsilon_{ij}^s = C_s J_2^{m_s} (t)s_{ij} (t)
	\label{eq:steady}
\end{equation}

\begin{equation}
	\dot \epsilon_{ij}^t + \frac{\epsilon_{ij}^t}{t_\epsilon} = \frac{C_t}{t_\epsilon} J_2^{m_t} (t)s_{ij} (t)
	\label{eq:trans}
\end{equation}

\noindent
where $E$, $\nu$ are the Young's modulus and Poisson ratio, respectively; $J_2$ is the second invariant of the deviator stress tensor s; $C_s$, $C_t$, $m_s$, $m_t$, $t_\epsilon$ are the nonlinear material parameters. $\sigma$ is the Cauchy stress tensor; i, j are the indices ranging among 1, 2 and 3. $\delta_{ij}$ is the Kronecker delta which used in the context of summation convention with the well-known property $\delta_{ij}=1$ when i = j and $\delta_{ij}=0$. Small deformations are assumed in the formulation. When more than one Voigt element is included in the model, the total strain components can be given as the sum of elastic, steady creep, and transient creep components for all Voigt elements,

\begin{equation}
	\epsilon_{ij}=\epsilon_{ij}^e+\epsilon_{ij}^s+\epsilon_{ij}^t= \epsilon_{ij}^e+\epsilon_{ij}^s+\sum_{i=1}^n \epsilon_{ij}^{t_i}
	\label{eq:total_strain}
\end{equation}

\noindent
where n is the number of Voigt elements shown in Fig. Equations~\ref{eq:steady} and~\ref{eq:trans} can also be written in integral form:

\begin{equation}
	\epsilon_{ij}^s = C_s \int_0^t J_2^{m_s} (t') s_{ij} (t') dt'
\end{equation}

\begin{equation}
	\epsilon_{ij}^t = \frac{C_t}{t_\epsilon} e^{(-t/t_\epsilon)} \int_0^t J_2^{m_t} (t') s_{ij} (t') e^{(-t'/t_\epsilon)} dt'
\end{equation}

ABAQUS/Standard finite element code is used as the implementation platform. Although ABAQUS has a rich material library for various applications, a nonlinearly viscoelastic model suitable for this work was not available. In this study, a UMAT was developed in order to implement the nonlinear Burgers model. UMAT requires the tangent stiffness matrix of the material model for finite element calculations. For implementation of the nonlinear Burgers viscoelastic model, the UMAT involves mainly temporal discretization. This was done following the procedure implemented by Kucuk \emph{et al.}~\cite{Kucuk2012, Kucuk2012a}. 

A simple, stable integration operator for these equations is the central difference operator:
\begin{equation}
	\dot f_{t+\frac{1}{2} \Delta t} = \frac{\Delta f}{\Delta t}, f_{t+\frac{1}{2} \Delta t} = f_t + \frac{\Delta f}{2}
\end{equation}
\noindent
where $f$ is a function, $f_t$ is its value at the beginning of the increment, $\Delta f$ is the change in the function over the increment, and $\Delta t$ is the time increment. 

Tangent stiffness matrix $\delta \Delta \sigma/\delta \Delta \epsilon$ of the constitutive model, with $\Delta \sigma$ being the stress increments and $\Delta \epsilon$ the strain increments, can be derived by applying central difference operator to the rate-dependent constitutive equations (Eq.~\ref{eq:elast}--\ref{eq:trans}).

Applying the central difference method to the elastic strain component as depicted in Eq.~\ref{eq:elast}, yields
\begin{equation}
	\epsilon_{ij}^e+\frac{1}{2} \Delta \epsilon_{ij}^e = \frac{1+\nu}{E} \left ( \sigma_{ij} +\frac{1}{2} \Delta \sigma_{ij} \right ) - \frac{\nu}{E} \left ( \sigma_{kk} +\frac{1}{2} \Delta \sigma_{kk} \right ) \delta_{ij}
\end{equation}

If the elastic Hooke's law is defined by Eq.~\ref{eq:hooke}, the elastic compliance matrix, C is defined by Eq.~\ref{eq:elast_comp}.
\begin{equation}
	\begin{bmatrix}
	\epsilon_{xx} \\
	\epsilon_{yy} \\
	\epsilon_{zz} \\
	\epsilon_{yz} \\
	\epsilon_{zx} \\
	\epsilon_{xy} \\
	\end{bmatrix}
	 = 
	C \begin{bmatrix}
	\sigma_{xx} \\
	\sigma_{yy} \\
	\sigma_{zz} \\
	\sigma_{yz} \\
	\sigma_{zx} \\
	\sigma_{xy} \\
	\end{bmatrix}
\label{eq:hooke}	
\end{equation}

\begin{equation}
	C=\frac{\delta \Delta \epsilon_{ij}^e}{\delta \Delta \epsilon_{kk}} =
	\begin{bmatrix}
	1/E & -\nu/E & -\nu/E & 0 & 0 & 0 \\
	       & 1/E      & -\nu/E & 0 & 0 & 0 \\
	       &             & 1/E      & 0 & 0 & 0 \\
	& & & \frac{1+\nu}{E} & 0 & 0 \\
	& & & & \frac{1+\nu}{E} & 0 \\
	& & & & & \frac{1+\nu}{E} \\
	\end{bmatrix}_{symmetric}
\label{eq:elast_comp}	
\end{equation}

Similar procedure as applied to Eq.~\ref{eq:steady} for steady creep component gives
\begin{equation}
	\frac{\Delta \epsilon_{ij}^s}{\Delta t} = C_s J_2^{m_s} \left (t + \frac{\Delta t}{2} \right) \left (s_{ij} (t) + \frac{1}{2} \Delta s_{ij} 		\right )
	\label{eq:steady_comp1}
\end{equation}

Assuming $J_2 (t) \approx J_2 ( t+\frac{1}{2} \Delta t )$, we have
\begin{equation}
	\Delta \epsilon^s = \Delta t C_s J_2^{m_s} (t) s_{ij} (t) + \frac{1}{2} \Delta t C_s J_2^{m_s} (t) \Delta s_{ij}
	\label{eq:12.a}
\end{equation}

\begin{equation}
	\frac{\delta \Delta \epsilon_{ij}^s}{\delta \Delta \sigma_{ij}} = \frac{1}{2} \Delta t C_s J_2^{m_s} (t)
\end{equation} 

Since $s_{ij}=\sigma_{ij} - \frac{1}{3} \sigma_{kk} \delta_{ij}$, we have
\begin{equation}
	\frac{\delta \Delta \epsilon_{ij}^s}{\delta \Delta \sigma_{ij}} = \frac{\delta \Delta \epsilon_{ij}^s}{\delta \Delta s_{ij}} \frac{\delta \Delta s_{ij}^s}{\delta \Delta \sigma_{ij}} = 
    \begin{cases}
            \frac{1}{3} \Delta t C_s J_2^{m_s} (t), &         \text{if } i=j,\\
            \frac{1}{2} \Delta t C_s J_2^{m_s} (t), &         \text{if } i\neq j.
    \end{cases}
\end{equation}

The compliance matrix of steady creep then can be written as
 \begin{equation}
	C= \Delta t C_s J_2^{m_s} (t)
	\begin{bmatrix}
	1/3 & 0 & 0 & 0 & 0 & 0 \\
	       & 1/3      & 0 & 0 & 0 & 0 \\
	       &             & 1/3      & 0 & 0 & 0 \\
	& & & 1/2 & 0 & 0 \\
	& & & & 1/2 & 0 \\
	& & & & & 1/2 \\
	\end{bmatrix}_{symmetric}
\label{eq:steady_comp}	
\end{equation}

Finally for the transient creep component as defined in Eq.~\ref{eq:trans}, we have
\begin{equation}
	\frac{\Delta \epsilon_{ij}^t}{\Delta t} + \frac{1}{t_\epsilon} \left (\epsilon_{ij} + \frac{1}{2} \Delta \epsilon_{ij} \right ) = \frac{C_t}{t_\epsilon} J_2^{m_t} \left ( t + \frac{\Delta t}{2} \right )\left ( s_{ij} (t) +\frac{1}{2} \Delta s_{ij} \right)
\end{equation}

\begin{equation}
	\Delta \epsilon^t = \frac{1}{2 t_\epsilon + \Delta t} ( 2 \Delta t C_t J_2^{m_t} (t) s_{ij} (t) - 2 \Delta t \epsilon^t + \Delta t C_t 		J_2^{m_t} (t) \Delta s_{ij} )
	\label{eq:15.b}
\end{equation}

\begin{equation}
	\frac{\delta \Delta \epsilon_{ij}^t}{\delta \Delta \sigma_{ij}} = \frac{\Delta t}{2 t_\epsilon + \Delta t} C_t J_2^{m_t} (t)
\end{equation}

The compliance matrix of transient creep can then be written as
\begin{equation}
	C= \frac{\Delta t}{2 t_\epsilon + \Delta t} C_t J_2^{m_t} (t)
	\begin{bmatrix}
	2/3 & 0 & 0 & 0 & 0 & 0 \\
	       & 2/3      & 0 & 0 & 0 & 0 \\
	       &             & 2/3      & 0 & 0 & 0 \\
	& & & 1 & 0 & 0 \\
	& & & & 1 & 0 \\
	& & & & & 1 \\
	\end{bmatrix}_{symmetric}
\label{eq:trans_comp}	
\end{equation}

From Eq.~\ref{eq:total_strain}, the total compliance is now
\begin{equation}
	\frac{\delta \Delta \epsilon_{ij}}{\delta \Delta \sigma_{kl}} = \frac{\delta \Delta \epsilon_{ij}^e}{\delta \Delta \sigma_{kl}} + \frac{\delta \Delta \epsilon_{ij}^s}{\delta \Delta \sigma_{kl}} + \frac{\delta \Delta \epsilon_{ij}^t}{\delta \Delta \sigma_{kl}}
	\label{eq:total_comp}
\end{equation}

By investigating the total compliance matrix, system tangent stiffness matrix (Jacobian matrix) $\delta \Delta \sigma_{ij}/\delta \Delta \epsilon_{kl}$ can be obtained from Eq.~\ref{eq:total_comp}. It should be noted that the Jacobian matrix in Eq.~\ref{eq:total_comp} accounts only for the elastic deformation and creep deformation caused by load or stress increment. It is seen from Eq.~\ref{eq:12.a} and~\ref{eq:15.b} that the aforementioned creep strain is just a small part of the total steady and transient creep strain. The rest of the creep strain is developed over the time period during the time increment and controlled by the applied stress. An artificial stress increment is introduced to include this creep strain in the system equation. This part of creep strain can be extracted from Eq.~\ref{eq:12.a} and~\ref{eq:15.b} as
\begin{equation}
	\Delta \epsilon' = \Delta t C_t J_2^{m_t} (t) s_{ij} (t) + \frac{1}{2 t_\epsilon + \Delta t} ( 2 \Delta t C_t J_2^{m_t} (t) 	s_{ij} (t) - 2 \Delta t \epsilon^t )
	\label{eq:increment}
\end{equation}

A stress increment $\Delta \sigma' = C \Delta \epsilon'$ is then added into the system equation to account for the creep strain in Eq.~\ref{eq:increment}, with C being the Jacobian stiffness matrix calculated from Eq.~\ref{eq:total_comp}. 

\section{Finite Element Modeling}

The 3D finite element model of nanoindentation experiment was constructed using commercial finite element package ABAQUS (Dassault Syst\'emes, Providence, RI). The indenter in a nanoindentation experiment is made with diamond and possess very high Young's modulus. So, it is possible to model the indenter as analytical rigid body. Finite element solver does not require calculating stress and strains in an analytical rigid body, hence reduces the computational time. 

Berkovich indenter can also be modeled as a 2D axisymmetric conical indenter with an effective cone angle~\cite{Fischer-Cripps2004}. The effective cone angle is calculated in a manner so that it provides the same area to depth relationship as the actual Berkovich indenter. The benefit of using a 2D model is that it requires less computation time compared to a 3D model. Nonetheless, a 3D model was implemented in this study to achieve higher accuracy in simulating the nanoindentation experiment. 

Even after adopting few simplifications, modeling nanoindentation experiment is still very challenging due to the several nonlinearities associated with the experiment (boundary, geometrical and material nonlinearity). Studies showed that, in case of modeling complex geometries, it is beneficial to model rigid elements as discrete rigid body rather than analytical rigid body. So, the Berkovich indenter was modeled as discrete rigid body, while the sample was modeled as deformable body. 

To ensure accuracy of the simulation results, the sample was modeled with finer mesh near the contact area where the stress and strain generated was much higher due to the singularity dominated zone. The contact between the indenter and the sample was defined as surface-to-surface contact, where the indenter was designated as \emph{master surface} and the sample was as designated as \emph{slave surface}. The element types for the sample was chosen from the eight-node brick element family (C3D8). Material behavior of the sample was defined in the model using a subroutine called (UMAT). The mathematical development of a nonlinear viscoelastic constitutive relationship was required to code the UMAT, which development was covered in the previous section. 

\section{Inverse Analysis} 

In order to facilitate the identification of global solution in the parameter space, our study implemented surrogate modeling approach. Surrogate models, also known as metamodels are particularly useful in case of finite element based inverse analysis. 

\begin{figure}[htbp]
\begin{center}
\includegraphics[width=0.8\textwidth]{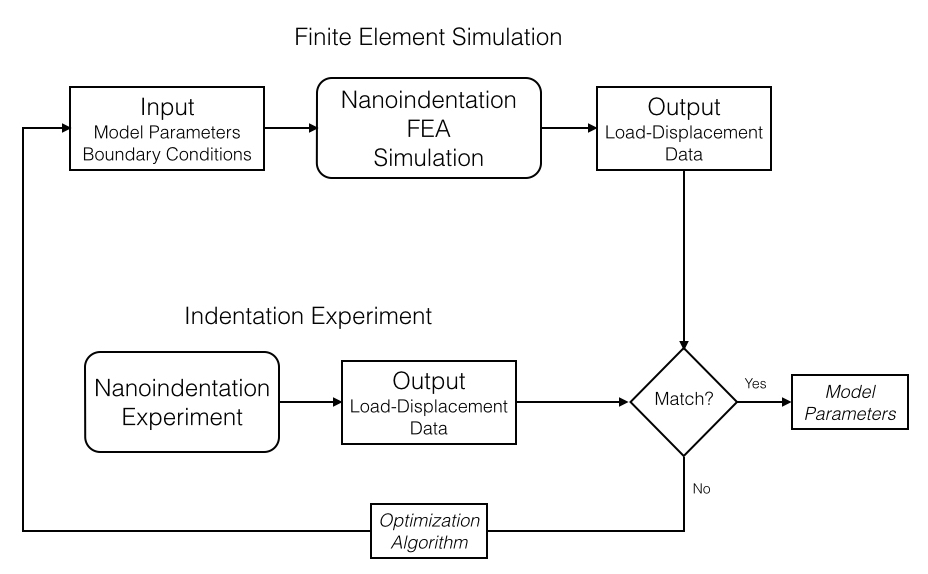}
\caption{Typical inverse method flowchart}
\label{fig:inv_flowchart}
\end{center}
\end{figure}

Figure~\ref{fig:inv_flowchart} shows the typical workflow of an inverse analysis for nanoindentation based model parameter extraction. Due to the fact that in every iteration of the inverse analysis one finite element analysis is required, the high computational cost involving the inverse analysis becomes the limiting factor in determining the correct solution. If finite element analysis can be replaced with a surrogate model, which is a numerical approximation of the input--output relationship, the total computational cost can be dramatically reduced. In a nutshell, use of surrogate model can effectively reduce the computational cost while still keeping the fidelity of the solution adequately high.

In this study, surrogate model is built by utilizing two numerical techniques named as Proper Orthogonal Decomposition (POD) and Radial Basis Function (RBF). Proper Orthogonal Decomposition (POD), also known as Principal Component Analysis (PCA) technique can be used with either experimental or simulated field data to derive a reduced-order set of basis functions capable of being used in a numerical representation of the system~\cite{Aquino2007}. POD reduced-order approximation has been shown to provide accurate numerical representations for complex systems with minimal computational cost~\cite{Han2003a,Holmes1996, Lieu2006}. In addition, POD has been applied to several inverse problem methodologies, such as optimal control ~\cite{Atwell2001,Leibfritz2006,Ly2001,Ravindran2000}, and nondestructive testing and system identification ~\cite{Banks2002,Kopp1997,Larson2008, Ostrowski2005a,Ostrowski2008}. However, work has yet to be shown (to our knowledge) for using POD reduced-order modeling for inverse viscoelastic material characterization from quasi-static indentation testing.

As a means of correlating unclear data using only spatial lines and planes, the concept of proper orthogonal decomposition (POD) was first developed over a century ago as a statistical tool by Pearson~\cite{Pearson1901}. Since then POD has been redeveloped under various names and has been used in numerous different applications from signal processing and control theory, human face recognition, data compression, parameter estimation and many others~\cite{Fic2005}. POD is also known as Karhunen--Loeve Decomposition (KLD), Principal Component Analysis (PCA) or Singular Value Decomposition (SVD)~\cite{Liang2002,Fic2005,Chatterjee2000}. In the recent past, POD has been increasingly used in many engineering applications ranging from computational fluid dynamics (CFD) to modeling of heat transfer problems due to its ability to reduce computational burden while maintaining adequately high accuracy.

For simple understanding of the POD technique, one should imagine a collection of vectors inside a Cartesian coordinate system. If these vectors are parallel to one another it could be assumed that these are correlated. On the other hand, uncorrelated would mean that these vectors are orthogonal (or perpendicular) to one another. 

POD's major objective is to rotate the coordinate system in such a manner so that the least amount of coordinates can be used to define the system. As an example, we know that a vector in cartesian coordinate system requires two projections (x- and y-axis projection) to be effectively defined. However, if the coordinate system is rotated only one projection can define the same vector. In case of complicated data sets, the number of rotated  coordinates would be higher for effective representation of the data. In such cases, POD captures the maximum projection of the vectors in the first rotated coordinate, which is commonly referred to as the first principal component. The second axis in the POD frame, called the second principal component, captures the next orthogonal direction with the largest projection and so on. 

POD is completely data dependent and does not assume any prior knowledge of the process that generates the data. This property is advantageous in situations where a priori knowledge of the underlying process is insufficient. POD does not neglect the nonlinearities of the original vector field. If the original system is nonlinear, then the resulting POD reduced order model will also be nonlinear.

\subsection{Proper Orthogonal Decomposition (POD) Theory}

If a function $\textbf{U}(x)$ is needed to be approximated over some domain of interest, it can be written as the following equation through a linear combination of few basis functions $\boldsymbol\varphi^i (x)$.
\begin{equation}
\textbf{U}(x) \approx \sum_{i=1}^M a_i. \boldsymbol\varphi_i(x)
\end{equation}

Here $a_i$ represents the unknown coefficients. Once basis functions are known, the coefficient values are obtained in a least square means.
\begin{equation}
\text{min} \left|\left| \textbf{U}(x) - \sum_{i=1}^M a_i. \boldsymbol\varphi_i(x) \right|\right|^{L^2}
\end{equation}

For any function $\textbf{U}(x)$, number of choices can be made regarding the selection of basis function. Based on one's expertise and prior knowledge about the system being represented,  one can often opt for a basis constructed from polynomial, trigonometric, or exponential functions. Proper Orthogonal Decomposition (POD) is one such technique that can be used to construct the optimal basis for a function under investigation in a least square sense. 

The derivation of POD presented in subsequent paragraphs refers to arbitrary case of vectors with dimensionality \emph{N}\textgreater 2. The notations presented in this section is congruent with Buljak~\cite{Buljak2012a}. 

POD starts with the idea of snapshots. Snapshots can be defined as an one-to-one relationship between the input and output of a system. In a typical scenario of an inverse finite element nanoindentation simulation, snapshots are the relation of material model parameters and output load--displacement data. In more concrete definition, a snapshot will be a collection of \emph{N} discrete values of a certain state variable resulting from a simulation (which represents a system) collected in vector $\textbf{u}_i$, corresponding to some input parameters (collected in vector $\textbf{p}_i$) on which the solution depends. A system can also be represented by an experiment, where snapshot will store the measurements taken from an experiment.

\begin{figure}[htbp]
\begin{center}
\includegraphics[width=0.8\textwidth]{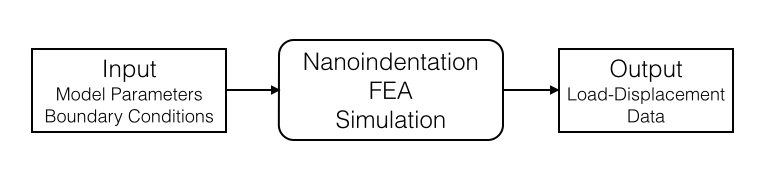}
\caption{Input output relationship in a typical system}
\label{fig:system}
\end{center}
\end{figure}

Further, a set of \emph{M} different snapshots, corresponding to different input parameters, can be collected in a rectangular \emph{N}$\times$\emph{M} matrix \textbf{U}, called the snapshot matrix. 
\[\textbf{U} = \begin{bmatrix}
  u_1^1 & u_1^2 & \cdots & u_1^M \\
  u_2^1 & u_2^2 & \cdots & u_2^M \\
  \vdots  & \vdots  & \ddots & \vdots  \\
  u_N^1 & u_N^2 & \cdots & u_N^M
  \end{bmatrix}\]

Therefore, a snapshot matrix represents a collection of responses of one system, under given conditions, corresponding to different values of parameters on which the solution depends. This snapshot matrix can be interpreted as a set of \emph{M}, \emph{N}-dimensional vectors. Each vector corresponds to one parameter combination. In the context of inverse finite element analysis for model parameter extraction, it can be said that the inputs to the system that are changing from one snapshot to another are some parameters entering into the constitutive model of material, while the boundary conditions and initial conditions are the same for all of the snapshots. So, \textbf{u}$_i$ contains \emph{N} number of individual displacement data for \emph{N} number of corresponding load increments, while the whole snapshot matrix, \textbf{U} contains \emph{M} number of individual finite element simulations. 

It is reasonable to expect that there will be a strong correlation between these snapshot vectors since they represent the outputs of the same system where just some material model parameters are changed. The POD theory can be effectively applied on the snapshot matrix, allowing to construct a new basis in which the dimensionality can be drastically cut-down to \emph{K} $\ll$ \emph{N}. POD finds the most accurate representation in some subspace W with the dimension of \emph{K}$\ll$\emph{N}. If we denote $\boldsymbol\varphi_1$, $\boldsymbol\varphi_2$,\dots, $\boldsymbol\varphi_K$ as the orthonormal basis of the subspace W, then each vector from the original set can be written as
\begin{equation}
\textbf{u}_i \approx \sum_{j=1}^K \bar a_{ij}. \boldsymbol\varphi_j, i=1,\dots, N
\end{equation}
where $\bar a_{ij}$ are amplitudes corresponding to $i^{th}$ vector in new subspace W, and $\bar\Phi$ is matrix that collects all the orthonormal basis of the subspace $\boldsymbol\varphi_j$. In a least square sense the error of approximation then becomes
\begin{equation}
\text{error} = \left|\left| \textbf{u}_i - \sum_{j=1}^K \bar a_{ij}. \boldsymbol\varphi^i(x) \right|\right|^{L^2}
\label{eq:error}
\end{equation}
Eq.~\ref{eq:error} provides the error for only the $i^{th}$ vector. For all the vectors in the snapshot, total error is expressed by the Eq.~\ref{eq:total_error}:
\begin{equation}
E = \sum_{i=1}^N \left|\left| \textbf{u}_i - \sum_{j=1}^K \bar a_{ij}. \boldsymbol\varphi^i(x) \right|\right|^{L^2} = \sum_{i=1}^N ||\textbf{u}_i||^2 - 2 \sum_{i=1}^N \sum_{j=1}^K a_{ij} \textbf{u}_i^T \boldsymbol\varphi_j + \sum_{i=1}^N \sum_{j=1}^K a_{ij}^2
\label{eq:total_error}
\end{equation}

The orthonormal basis has to be chosen in such a manner so that total error is minimized. To do that, first order derivate of total error with respect to all the unknowns (namely $\bar a_{ij}$) are needed. Taking partial derivative of total error:
\begin{equation}
\frac{\delta E}{\delta a_{lm}} = -2 \textbf{u}_l^T \boldsymbol\varphi_m +2a_{lm}
\end{equation} 
\begin{equation}
a_{lm} = \textbf{u}_l^T \boldsymbol\varphi_m 
\end{equation}

By substitution of $a_{lm}$ in Eq.~\ref{eq:total_error}:
\begin{equation}
E = \sum_{i=1}^N ||\textbf{u}_i||^2 - 2 \sum_{i=1}^N \sum_{j=1}^K (\textbf{u}_i^T \boldsymbol\varphi_j) \textbf{u}_i^T \boldsymbol\varphi_j + \sum_{i=1}^N \sum_{j=1}^K {(\textbf{u}_i^T \boldsymbol\varphi_j)}^2
\end{equation}

Few more mathematical manipulation provides:
\begin{equation}
E = \sum_{i=1}^N ||\textbf{u}_i||^2 -  \sum_{i=1}^N \sum_{j=1}^K {(\textbf{u}_i^T \boldsymbol\varphi_j)}^2 = \sum_{i=1}^N ||\textbf{u}_i||^2 - \sum_{j=1}^K \boldsymbol\varphi_j^T \textbf{C} \boldsymbol\varphi_j
\label{eq:total_error_1}
\end{equation}

\noindent
where \textbf{C} is called the covariance matrix defined as $\textbf{C} = \textbf{U}\textbf{U}^T$. The first part of Eq.~\ref{eq:total_error_1} is a scalar constant which depends on the original set of snapshots. So, in order to reduce the error of approximation one has to maximize $\sum_{j=1}^K \boldsymbol\varphi_j^T \textbf{C} \boldsymbol\varphi_j$ under the constraint of orthonormality of the new basis $\boldsymbol\varphi_j^T \boldsymbol\varphi_j = 1, j = 1, \dots, K$. By using Lagrange multipliers method the constrained problem can be converted into
\begin{equation}
\text{max} \sum_{j=1}^K \boldsymbol\varphi_j^T \textbf{C} \boldsymbol\varphi_j - \sum_{j=1}^K \lambda_j (\boldsymbol\varphi_j^T \textbf{C} \boldsymbol\varphi_j - 1)
\label{eq:max}
\end{equation}

In order to maximize Eq.~\ref{eq:max} first order derivatives with respect to $\varphi_j$ is required. By doing that we find
\begin{equation}
\frac{d}{d \boldsymbol\varphi_j} \sum_{j=1}^K \boldsymbol\varphi_j^T \textbf{C} \boldsymbol\varphi_j - \sum_{j=1}^K \lambda_j (\boldsymbol\varphi_j^T \textbf{C} \boldsymbol\varphi_j - 1) = 2\textbf{C}\boldsymbol\varphi_j - 2\lambda_j \boldsymbol\varphi_j = 0
\label{eq:max_1}
\end{equation}

Eq.~\ref{eq:max_1} is only satisfied if $\boldsymbol\varphi_j$ is eigenvector and $\lambda_j$ is the corresponding eigenvalue of matrix \textbf{C}. Now from taking Eq.~\ref{eq:max_1} and Eq.~\ref{eq:total_error_1} into consideration the total error equation can be changed to
\begin{equation}
E = \sum_{i=1}^N ||\textbf{u}_i||^2 - \sum_{j=1}^K \lambda_j
\end{equation}

Recalling that the first term is a constant it results that the error of approximation is minimized if the new basis is constructed of \emph{K} eigenvectors that are corresponding to the first \emph{K} largest eigenvalues of covariance matrix \textbf{C}
\begin{equation}
\bar \Phi = [ \boldsymbol\varphi_j], \qquad j=1,\dots, K
\end{equation}

If the subspace W is constructed with all the eigenvectors of matrix \textbf{C}, there is no error of approximation because in that case all the vectors $u_i$ are just expressed in a different coordinate basis. Approximation in any other subspace that uses smaller number of eigenvectors the error of approximation is found using the following equation
\begin{equation}
E= 1 -\frac{\sum_{i=1}^K \lambda_i}{\sum_{i=1}^M \lambda_i}
\label{eq:truncate}
\end{equation}
\noindent
which is the ratio between the summation of kept eigenvalues and summation of all the eigenvalues.

In this study, POD is used to determine the displacement of the indenter tip inside the material, by finding the correction from results of FE simulations of the nanoindentation experiment with different material model parameter sets. This process is called the method of snapshots~\cite{Buljak2012a}. The snapshot matrix, \textbf{U} then consists of the resulting indenter displacement that are expected to be somewhat correlated. 
 
\subsection{Radial Basis Function Theory} 

Radial Basis Functions (RBF) are very effective in providing an output approximation of a multivariable function for an unknown input point in the parameter space through interpolation of information from the known points~\cite{Buhmann2003}. In this section a very brief description of RBF is provided. The procedure through which RBFs can be combined with the information from POD to solve the inverse problem is also be discussed in the following paragraphs. 

As mentioned earlier RBF is a very effective interpolation technique. To illustrate the idea of RBF, let us assume a function \emph{f}(\textbf{x}) for which we only know \emph{N} number of input--output relations. Let us also assume, \textbf{x} is a point in the parameter space for which we want to approximate the function's value, where \textbf{x} is a \emph{M}-dimensional vector. Classical interpolation methods use only the information around the point \textbf{x} to provide the approximation. The biggest difference that RBF provides in a similar scenario is that it uses all the \emph{N} number of input--output relationship to build one continuous function over the whole domain. Therefore, the actual function \emph{f}(\textbf{x}) is approximated as a linear combination of some function $g_i$
\begin{equation}
f(\textbf{x}) = \sum_{i=1}^N \beta_i  . g_i (\textbf{x})
\end{equation}

\noindent
where $\beta_i$ are coefficients of this combination. This equation is complete when the basis functions $g_i$ and the coefficients $\beta_i$ are known. Various Radial Basis Functions can be chosen as basis function $g_i$. Most notable few are given below---
\begin{equation}
\text{Linear splines}, ||\textbf{x}-\textbf{x}_j||
\end{equation}

\begin{equation}
\text{Cubic splines}, ||\textbf{x}-\textbf{x}_j||^3
\end{equation}

\begin{equation}
\text{Gaussian}, \text{exp} \left( \frac{-||\textbf{x}-\textbf{x}_j||}{c_j^2} \right)
\end{equation}

\begin{equation}
\text{Multiquadratic}, \sqrt{1+\frac{||\textbf{x}-\textbf{x}_j||^2}{c_j^2}}
\end{equation}

\begin{equation}
\text{Inverse Multiquadratic}, \frac{1}{\sqrt{||\textbf{x}-\textbf{x}_j||^2 + c_j^2}}
\end{equation}

For an unknown point \textbf{x} in parameter space, the linear spline RBF will provide the basis $g_i$ using the following manner
\begin{equation}
g_i (\textbf{x}) = g||\textbf{x}-\textbf{x}_i||, \qquad i=1,2,\dots,N
\end{equation}

For determining the coefficients $\beta_i$, known \emph{N} values of the function in the $\textbf{x}_i$ nodes are used in such a manner so that the RBFs approximate exact values of the function at the known points. This is solved using the following equation
\begin{equation}
f(\textbf{x}_j) = y_j = \sum_{i=1}^N \beta_i . g_i (\textbf{x}_j), \qquad j=1,2,\dots, N
\label{eq:approximation}
\end{equation}

\noindent
where $y_j$ are the known values of the function. In compact manner Eq.~\ref{eq:approximation} can be written as
\begin{equation}
\boldsymbol\beta .\textbf{G} = \textbf{Y}
\label{eq:RBF}
\end{equation}

\noindent
where \[\textbf{G} = \begin{bmatrix}
  g_1(\textbf{x}_1) & g_2(\textbf{x}_1) & \cdots & g_N(\textbf{x}_1) \\
  g_1(\textbf{x}_2) & g_2(\textbf{x}_2) & \cdots & g_N(\textbf{x}_2)  \\
  \vdots  & \vdots  & \ddots & \vdots  \\
  g_1(\textbf{x}_N) & g_2(\textbf{x}_N) & \cdots & g_N(\textbf{x}_N) 
  \end{bmatrix}\]
\begin{equation*}
\boldsymbol\beta = [ \beta_1,\beta_2,\cdots,\beta_N]^T
\end{equation*}
\begin{equation*}
\textbf{Y} = [ y_1,y_2,\cdots,y_N]^T
\end{equation*}

Eq.~\ref{eq:RBF} can be solved for unknown interpolation coefficients $\beta_i$, which can then be used to obtain approximated values of the function in any given points in the parameter space. For a particular sampling set \emph{N}, $\beta_i$ is only need to be determined once. In matrix notation Eq.~\ref{eq:RBF} can be written as
\begin{equation}
\textbf{B}.\textbf{G} = \textbf{Y}
\label{eq:RBF_final}
\end{equation}
As RBF takes into account the whole set of input--output relationship of a system, it can provide much more informed approximation compared to the classical local interpolation schemes. Another important advantage of using RBF is that the sampling of \emph{N} in the parameter space need not to follow any particular distribution (in other words, can be scattered). However, particular distribution of sampling points help to keep the error of approximation under control. 

\subsection{Combining POD--RBF for Approximation}

The ability of POD is to create a reduced order model by truncating orthogonal basis or dimensions. In a manner, POD works as a data compression tool where the loss of data is negligible. On the other hand, RBF provides the ability to approximate a function with high fidelity in between the known values in a multivariable parameter space. If both techniques are combined we can get a tool that can essentially provide high quality output approximation without incurring the huge computational cost associated with finite element analysis during an inverse analysis. 

In context of nanoindentation study, let us assume vector \textbf{p} collects the material model parameters and \textbf{u} collects the output of the simulation (load or displacement data). Our goal is to find a function such that \emph{f}(\textbf{p}) = \textbf{u}. This function needs to approximate the output of the simulation over some domain in parameter space. Following the theory related to POD, a reduced dimension model of snapshot matrix, \textbf{U} can be developed where $\bar{\textbf{A}}$ represents the reduced order of amplitudes. In reduced dimension, the aforementioned equation can be written as
\begin{equation}
f_a(\textbf{p}) = \bar{\textbf{a}}
\label{eq:reduced}
\end{equation}

\noindent
where, the relation between the reduced model and full model is given by the following equation
\begin{equation}
f(\textbf{p}) = \bar{\Phi} . f_a(\textbf{p}) = \textbf{u}
\end{equation}

If RBF is applied Eq.~\ref{eq:reduced} can be expressed in following manner
\begin{equation}
f_a(\textbf{p}) = \textbf{B}.\textbf{g(p)}
\end{equation}

Once the basis function is known, interpolation coefficients collected in matrix \textbf{B} is solved in the reduced space using the following equation
\begin{equation}
\textbf{B}.\textbf{G} = \bar{\textbf{A}}
\end{equation}
Then the final equation that will provide the approximation of the system response for any arbitrary set of parameter in the subspace is given by
\begin{equation}
\textbf{u} \approx \bar{\Phi}.\textbf{B}.\textbf{g(p)}
\label{eq:final_pod_rbf}
\end{equation}

Eq.~\ref{eq:final_pod_rbf} involves simple matrix multiplication, and thus can provide much faster turnaround time when compared to finite element simulation. This is particularly useful for inverse analysis where a large part of computational effort is directed towards running simulation inside the optimization loop. It is also a simple approach, where the training of the POD--RBF (obtaining the matrices $\bar{\Phi}$ and \textbf{B}) is done only once. Moreover, once trained this technique can provide high enough computation accuracy, which can even be improved with a larger sampling points. \newpage

\section{Taguchi Design of Experiments for Sensitivity Analysis}

In any process where the output is influenced by multiple number of parameters, there is a need for the information that how individual parameters affect the overall output. In other words, it is useful to know the sensitivity of an output to an input parameter change. This need gave rise to an independent area of research inside statistics called \emph{Design of Experiments (DOE)}. 

Traditionally, researchers used to carry out experiments where only one of the parameter was changed within a certain range while keeping the other parameters constant. Then the same process was to be replicated for the other parameters. This method is called full factorial experimental design, where the number of experiments required to perform the sensitivity analysis is astronomical. On the contrary, Taguchi applied the concept of orthogonal arrays, where all factors are changed simultaneously. For an experiment involving three parameters changing in four levels, the number of experiments required by Taguchi method is only 16, while full factorial design requires 64 independent experiments. 

To perform a systematic sensitivity analysis first an experimental design is required. It is done by choosing an orthogonal array depending on the \emph{degrees of freedom} (Eq.~\ref{eq:doe}):
\begin{equation}
df^{exp} = \sum df^{factor} + \sum df^{interaction}
\label{eq:doe}
\end{equation}

If $k^A$ is the number of levels for factor A, then $df^A$ = $k^A$ -- 1. The experiments are conducted based on the chosen orthogonal array. By employing suitable analysis technique, such as Analysis of Variance (ANOVA) one can determine the contribution of individual parameters contribution towards the output. ANOVA is an useful statistical tool for quantitative determination of influence of any given input parameter and it can be used to interpret experimental data.

\chapter{APPLICATION OF THE PROPOSED TECHNIQUE}\label{chap:chap4}

The main objective of this study was to use an inverse analysis technique to extract material model parameters for a nonlinear viscoelastic model from a nanoindentation experiment. There were three big challenges to this problem ---

\begin{enumerate}
  \item Modeling nanoindentation experiment using finite element analysis
  \item Incorporating nonlinear viscoelastic model in the finite element simulation
  \item Developing the optimization routine to extract the model parameters
\end{enumerate}

\begin{figure}[ht!]
\begin{center}
\includegraphics[width=.75\textwidth]{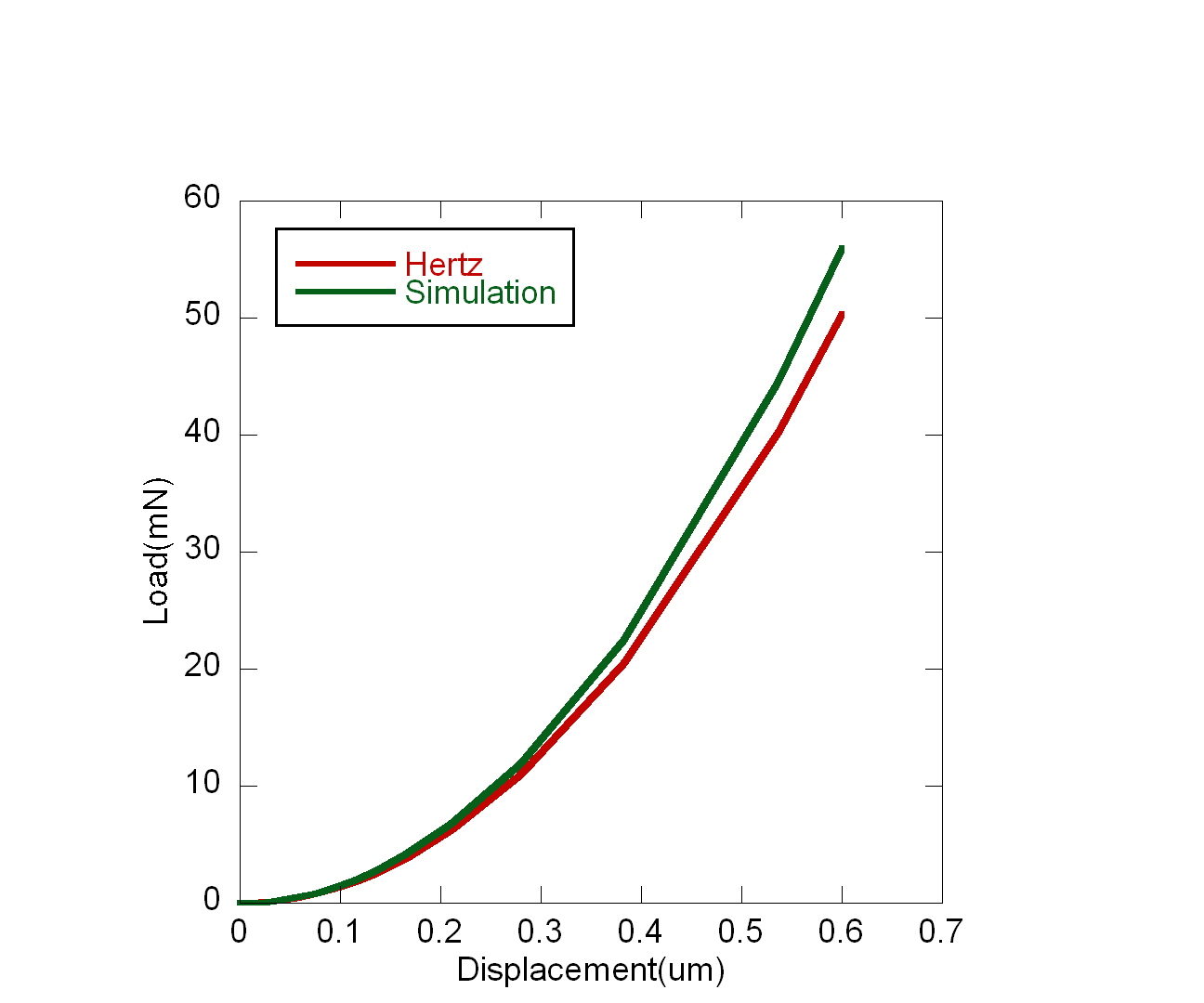}
\caption{Load--displacement behavior of a Berkovich nanoindentation}
\label{fig:verification}
\end{center}
\end{figure}

In last chapter modeling of the nanoindentation experiment for a Berkovich tip using commercial software package ABAQUS was described. To verify that the ABAQUS model was in fact able to simulate the nanoindentation experiment, a simple elastic indentation simulation was performed. From the simulation corresponding load and displacement data were obtained, which was compared against Hertzian analytical solution provided by Sneddon. Figure~\ref{fig:verification} shows the comparison of analytical and simulated load--displacement data. It can be seen that, although it was not a perfect match, simulated data closely followed the analytical data. Attaining perfect match between simulation and analytical data is not very practical as it means going for very fine meshing in the simulation model thus increasing the computational expense exponentially. 

The second challenge which was to incorporate nonlinear viscoelastic model in the finite element simulation has also been solved. This required discretizing the constitutive mathematical model for time step, $\Delta t$. A detailed description of the numerical approach that was involved in developing the equations required for finite element approach has been presented in the previous chapter.

The third challenge, which was the most critical of the three, has been solved by combining two separate technique, \emph{i.e.} Proper Orthogonal Decomposition and Radial Basis Functions. The nonlinear viscoelastic model of our choice has seven unknown parameters of interest that needs to be extracted using the POD--RBF technique. As discussed in the previous chapter, POD--RBF technique needs \emph{snapshots} of the system to become trained in approximating the system. 

The process of training a surrogate model is often referred as sampling. A simple way of sampling the parameter space can be the grid system, where the distance between the sampling points for a parameter is kept constant over the domain of interest. If every unknown parameter is sampled \emph{m} times over its domain, for \emph{n} number of unknown parameters a total of $m^n$ simulations will be required. This would be a large number of finite element simulations to handle. Hence, in order to verify the POD--RBF technique's ability in solving an inverse problem, first a rather simple problem where the number of unknown parameters are less were chosen. This was done in order to investigate the key properties of a POD--RBF based surrogate model, which could later be utilized to increase the confidence in POD--RBF technique for the ultimate application \emph{i.e.} parameter identification of nonlinear Burger's model.

The performance or the ability of the POD--RBF based surrogate model to precisely approximate the FE simulation depends on couple of parameters, namely number of training points used and the choice of RBF. Higher number of training points relating input parameters to system output improves the quality of surrogate model's approximation at the expense of higher number of FE simulations. Although being offline or outside optimization loop, optimizing the number of training points is crucial since it directly relates to the computational cost of the overall inverse analysis. 

There are only handful of articles in the literature that have tackled the inverse problem of nanoindentation-based material model calibration using POD--RBF based surrogate models~ \cite{Bolzon2004,Bocciarelli2005,Bocciarelli2007a,Bocciarelli2009,Bolzon2011,Bolzon2012,Bolzon2013,Bolzon2014}. Furthermore, to our best of our knowledge, none of the previous studies reported if the performance of the surrogate model could be optimized with respect of number of training points. In addition, the choice of an RBF, which affects the performance of the surrogate model, has also not been investigated at depth. Prior studies have typically employed only one kind of RBF in an \emph{ad hoc} manner without providing much analysis into comparative benefits of using different types of RBFs to solve a given problem.

Since a well-trained surrogate model is at the root of solving the nanoindentation-based inverse problem, this study was designed to facilitate the understanding of a POD--RBF based surrogate model's performance with respect to the number of training points and the choice of an RBF. It was expected that the findings of this study would provide a general framework for solving nanoindentation-based material modeling inverse problem using POD--RBF technique.

\section{Case Study}

In this study, nanoindentation was conducted on a standard metallic material. The nanoindentation experiment was then modeled with a finite element analysis software, where a custom elastic--plastic material behavior was incorporated. A range was selected for each parameter within which the values of the parameter would be altered. A Taguchi orthogonal array-based experimental design was formulated by varying each parameter within the range at four equidistant levels. The analysis of variance (ANOVA) technique was employed to recognize the influence of the parameters over the output. The number of levels for the unknown parameters within the specified range were optimized based on the ANOVA results. Training data were generated in a full factorial basis by varying each parameter of the custom material model for the initial and optimized model. A random noise of 1\% and 5\% was appended to the training data to investigate the stability of each surrogate model.

\subsection{Nanoindentation}

The nanoindentation experiment was conducted using an MTS Nanoindenter XP (Agilent Technologies, Santa Clara, CA, USA) using a load-controlled scheme with a Berkovich tip. The maximum load was set to be 4.9 mN for the experiments. A triangular loading profile was chosen with a 15 s duration for both the loading and unloading segments. Before conducting the actual experiments the Berkovich tip was calibrated using a fused silica reference material. Also, the acceptable thermal drift rate was chosen to be 0.15 nm/s. 

The nanoindentation experiment was conducted on a reference material \emph{i.e.}~single crystal aluminum. This sample is commonly used to check the performance of a nanoindenter. The single crystal aluminum sample has Young's modulus of 70.4 GPa and Possion's ratio of 0.345 as provided by the supplier (Agilent Technologies, Santa Clara, CA, USA). 

\subsection{Finite Element Simulation}

A commercial finite element software  (ABAQUS, Dassault Syst\'{e}mes, Providence, RI, USA) was utilized in this study, both for modeling the nanoindentation experiment and for solving the finite element problem. The Berkovich tip was modeled as a 3D discrete rigid body while the sample was modeled as a 3D deformable body. A finer mesh was provided to the sample near the contact region to ensure good convergence and also to improve the quality of the finite element solution. 

\begin{figure}
\begin{center}
\subfigure[Stress--strain relationship of bilinear plasticity model]{
\resizebox*{9cm}{!}{\includegraphics{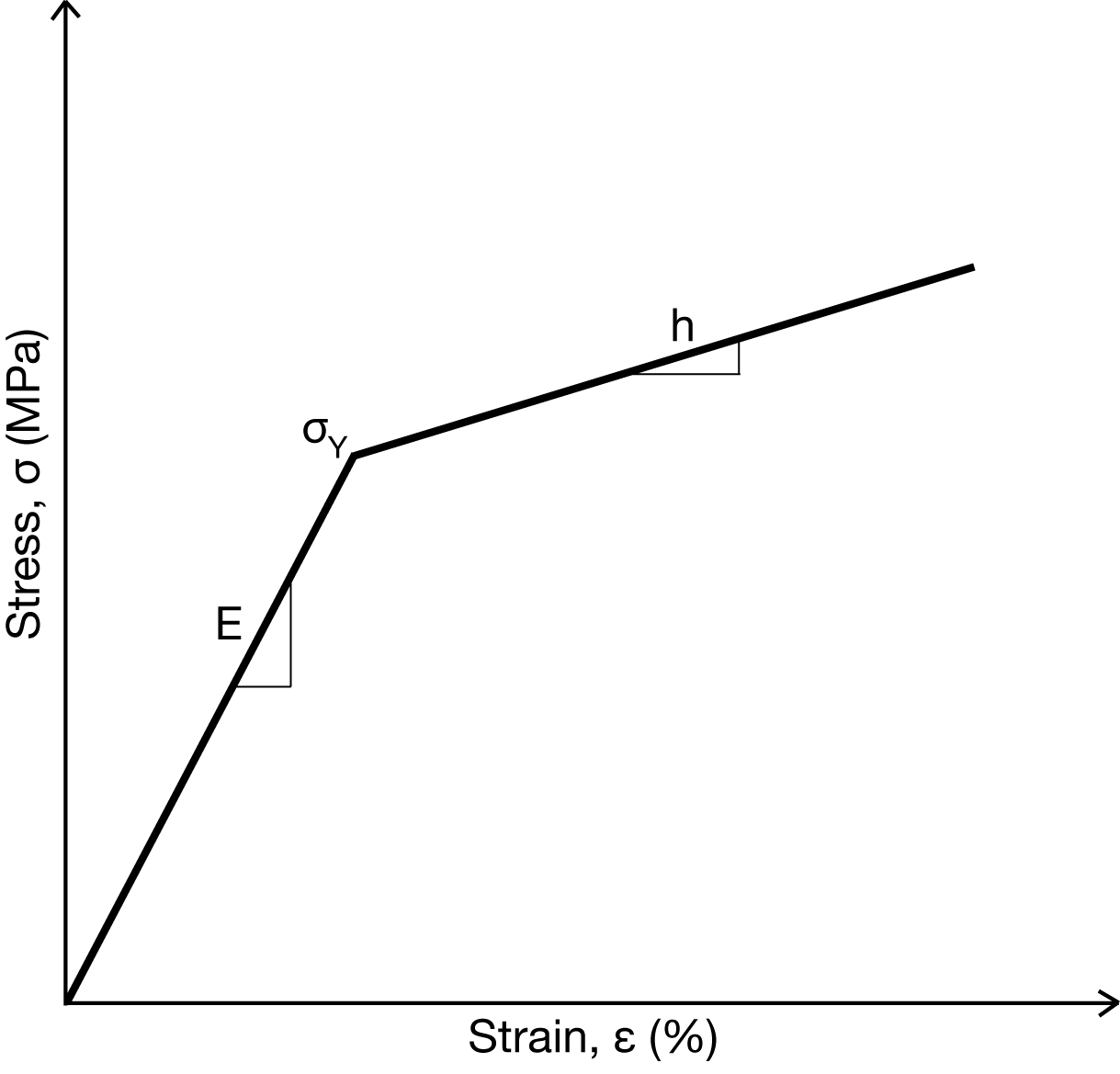}}}\hspace{5pt}
\subfigure[Schematic of ABAQUS finite element model for Berkovich indentation]{
\resizebox*{12cm}{!}{\includegraphics{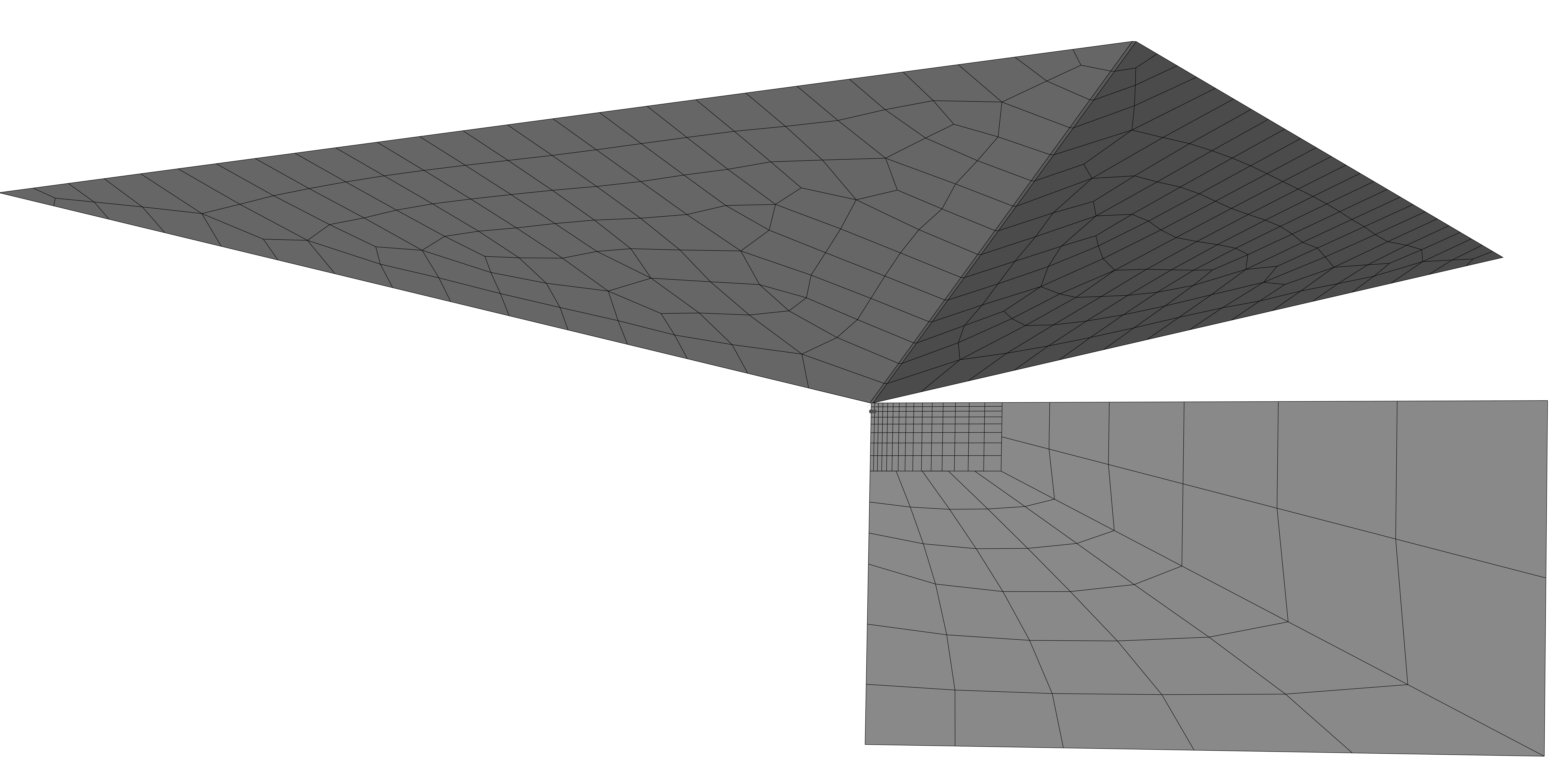}}}
\caption{ABAQUS finite element modeling details}
\label{fig:abq_details}
\end{center}
\end{figure}

\begin{figure}
\begin{center}
\subfigure[Von Mises equivalent stress]{
\resizebox*{7cm}{!}{\includegraphics{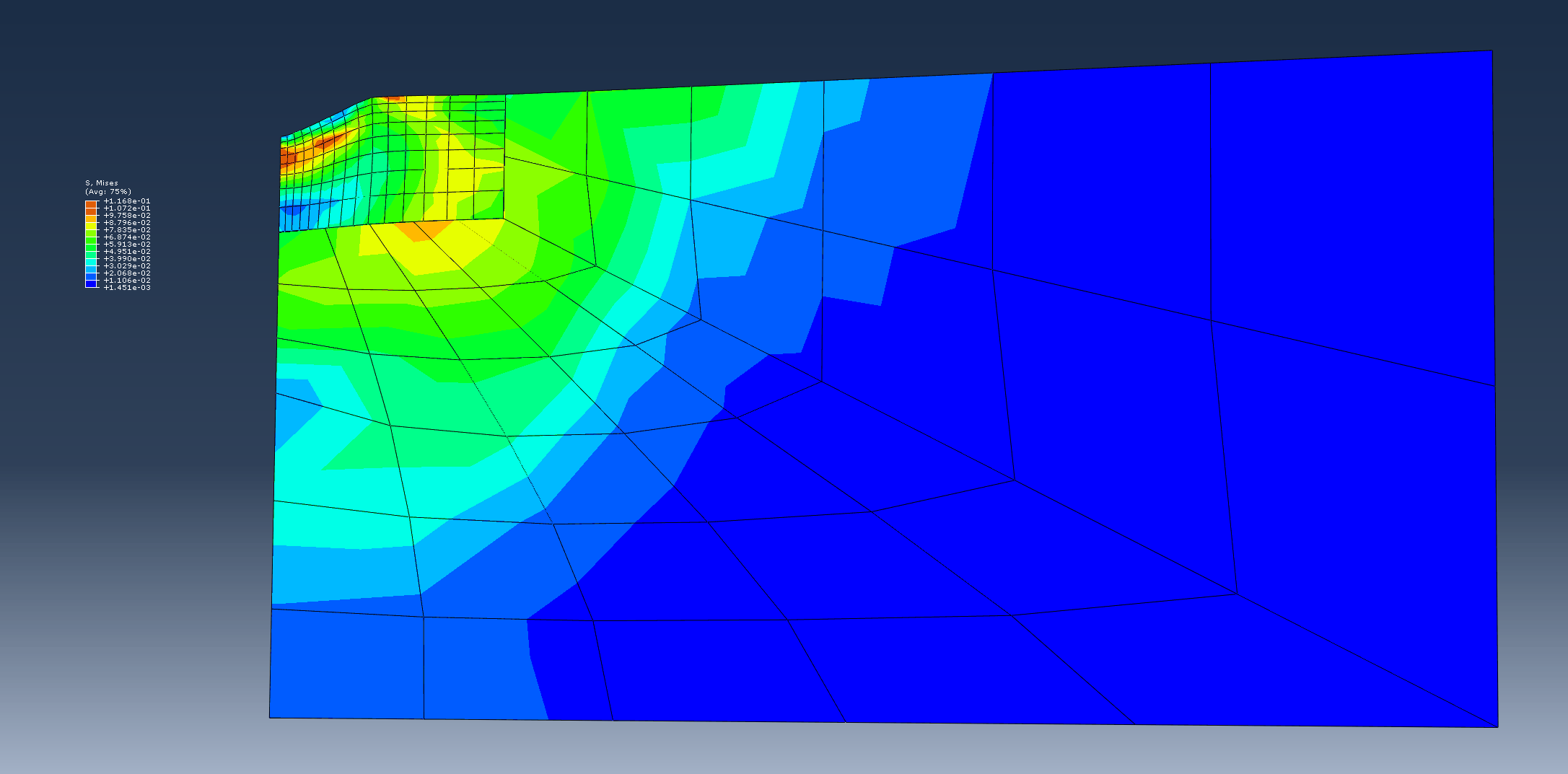}}}\hspace{5pt}
\subfigure[X-axis stress component, $\sigma_{11}$]{
\resizebox*{7cm}{!}{\includegraphics{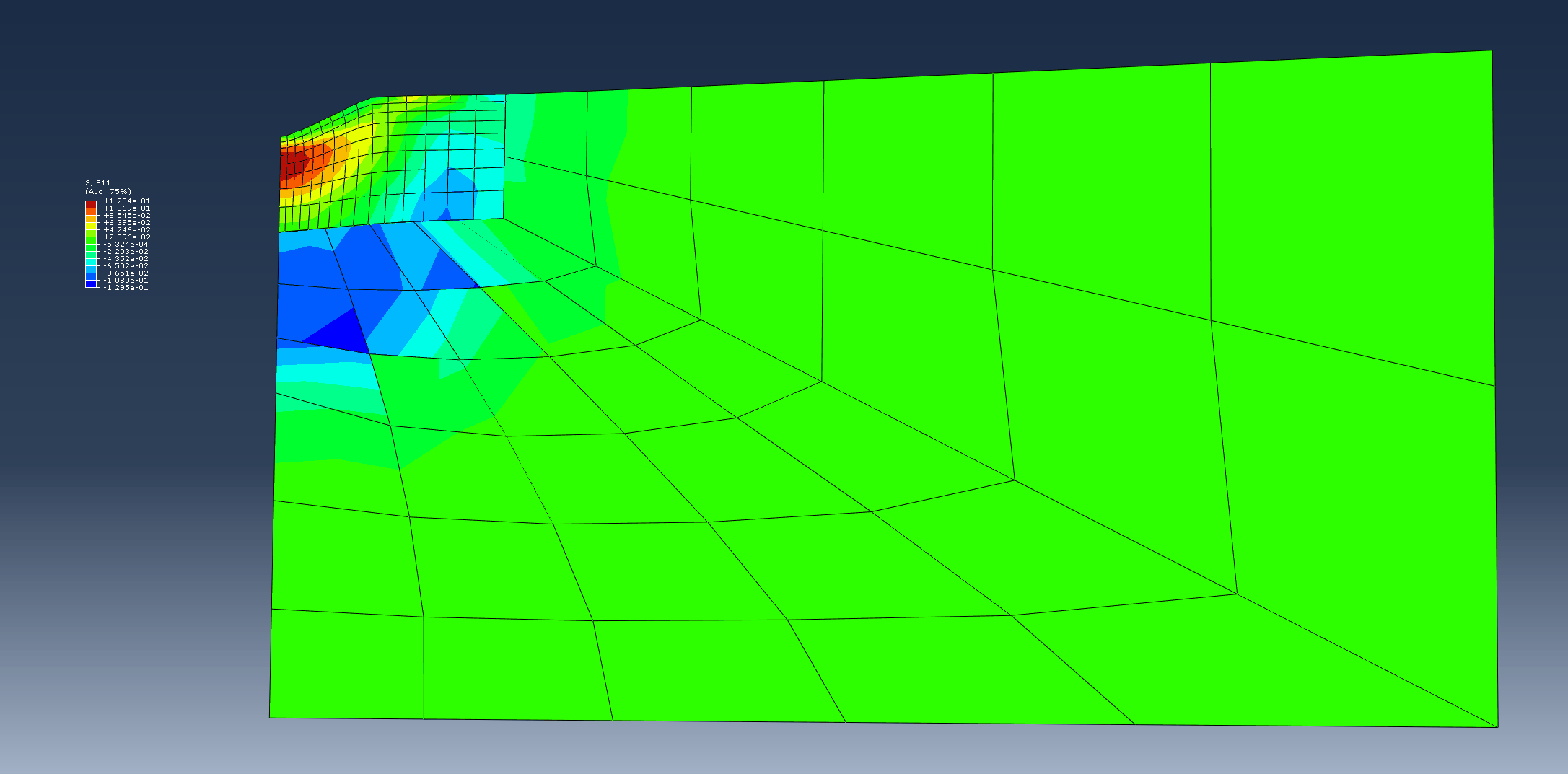}}}
\subfigure[Y-axis stress component, $\sigma_{22}$]{
\resizebox*{7cm}{!}{\includegraphics{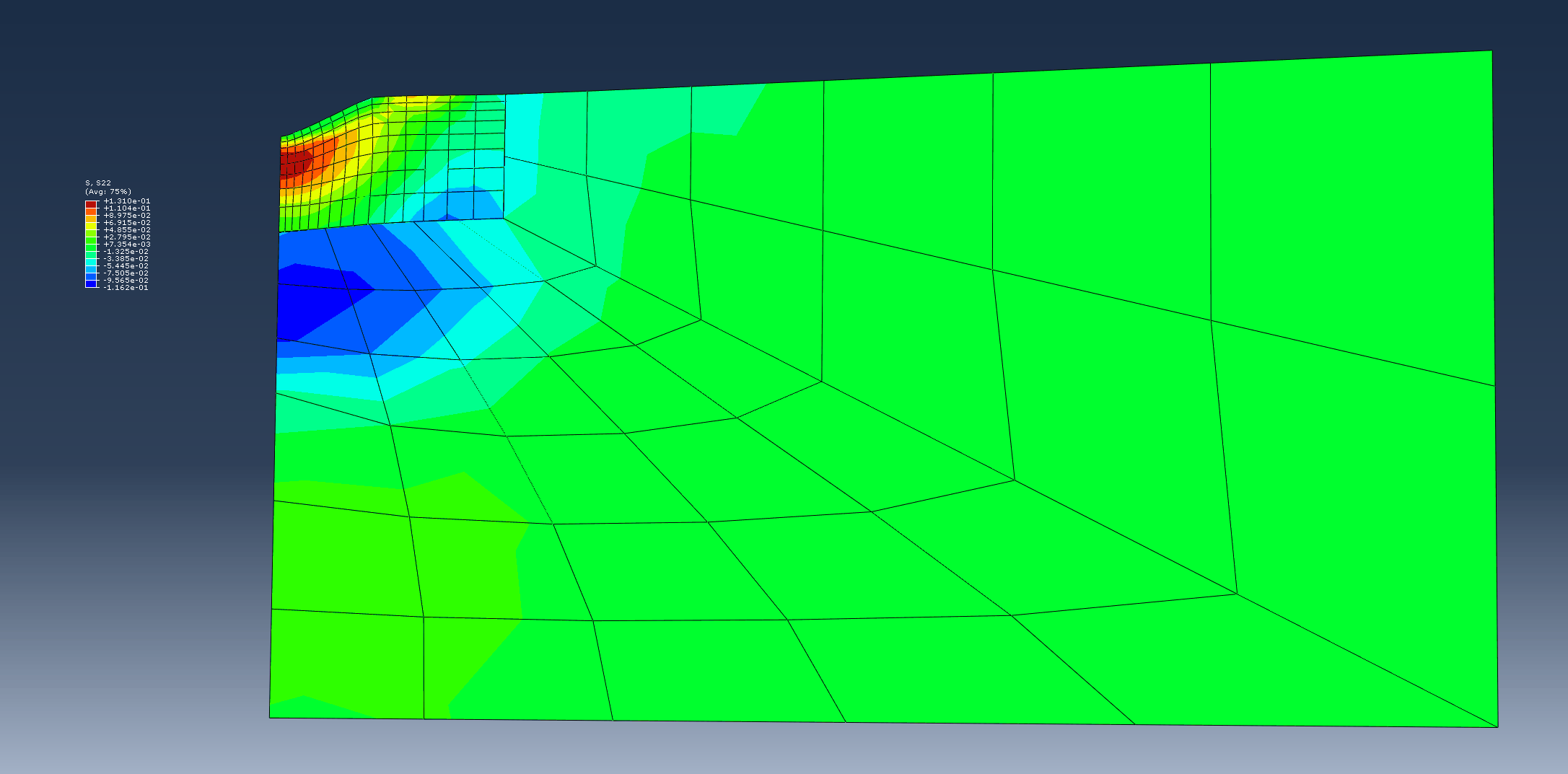}}}\hspace{5pt}
\subfigure[Z-axis stress component, $\sigma_{33}$]{
\resizebox*{7cm}{!}{\includegraphics{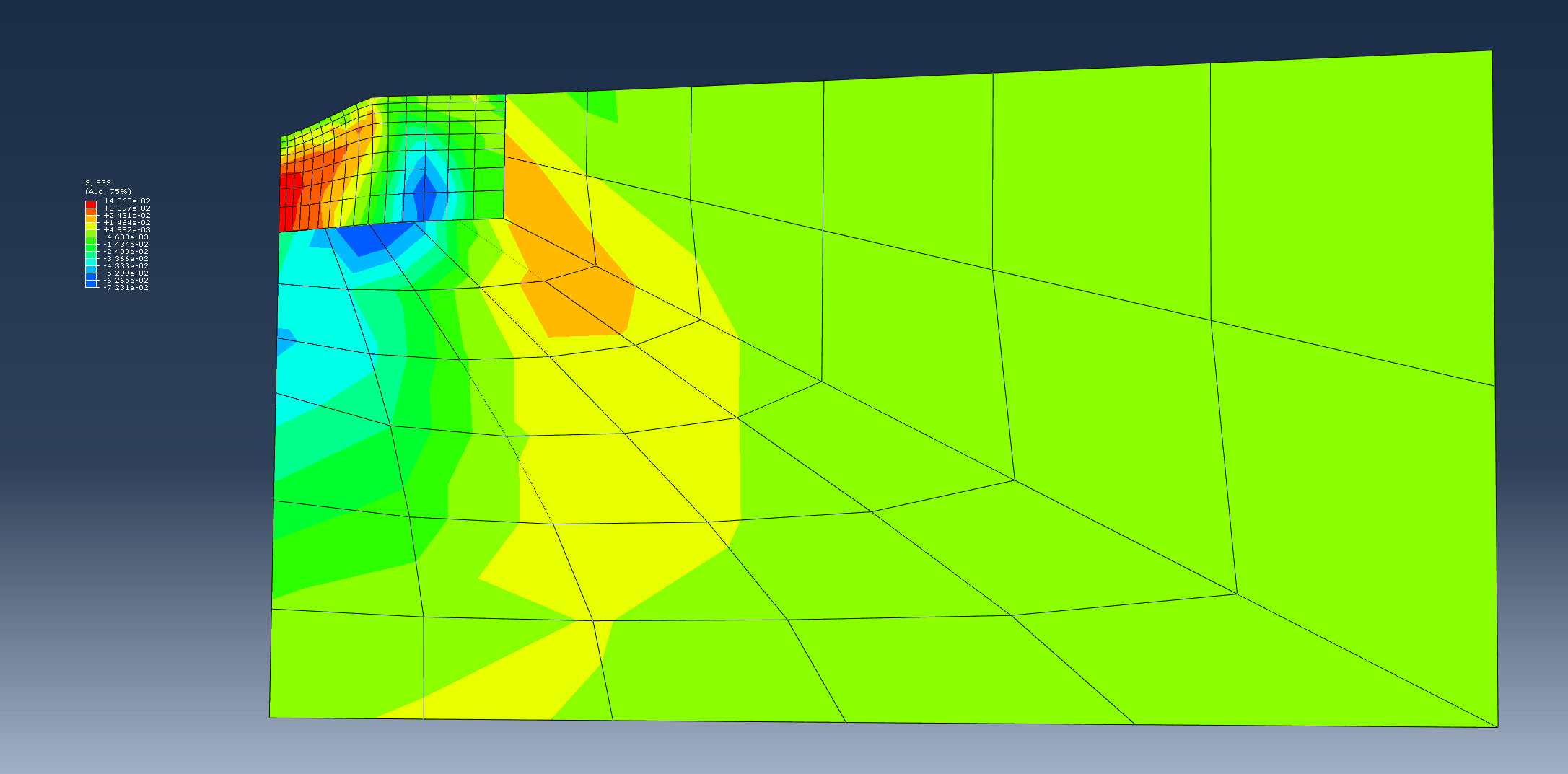}}}
\caption{Stress contours ABAQUS output after unloading for elastic--plastic Berkovich indentation simulation}
\label{fig:stress_contours}
\end{center}
\end{figure}

The contact between the indenter and the sample was assumed sliding contact with a friction coefficient of 0.25 and was defined as \emph{surface-to-surface} contact. The indenter and the sample were assigned as the \emph{master} and the \emph{slave} surface, respectively. In ABAQUS surface-to-surface contact, master surface nodes can penetrate the slave surface (\emph{i.e.} causing deformation to the slave surface), while the slave surface nodes cannot penetrate master surface. In indentation modeling using FE, it is generally assumed that the indenter is much stiffer than the sample surface; hence, deformation of the indenter by the sample surface is neglected. The element type was chosen from the eight-node brick element family (C3D8). The finite element problem consisted a total of 1323 elements and 1817 nodes. Figure~\ref{fig:abq_details} shows a schematic of the FE model generated in ABAQUS and Fig.~\ref{fig:stress_contours} shows the typical ABAQUS simulation's stress contour outputs at the end of a Berkovich indentation.

The elastic--plastic material behavior of the sample was incorporated in the FE software using a UMAT subroutine. The material behavior was chosen as isotropic elastic--plastic with linear hardening as shown in Fig.~\ref{fig:abq_details} (a). This model defines both the elastic and the plastic part of the stress--strain relationship as linear~\cite{Qian1997,Nakamura2000}.  Only four parameters are required to describe this particular material model, which are elastic modulus (\emph{E}), Poisson's ratio ($\nu$), yield strength ($\sigma_Y$), and hardening coefficient (\emph{h}). For the numerical study, Poission's ratio was kept constant at the known value of 0.345. According to prior FE-based studies, Poisson's ratio does not affect the FE simulation of indentation experiment as much as the other model parameters~\cite{Magnenet2008} and hence, kept constant in most of the indentation modeling studies~\cite{
Ma2013, Clement2013}. Table~\ref{tab:range_of_parameters} lists the range of values used in this study for the three parameters.

\begin{table}[!htbp]
\caption{Range of values for the model parameters}
\begin{center}
\begin{tabular}{lcc}
\hline
  Model parameter & Lowest level & Highest level\\
\hline  
  Young's modulus, E & 60 & 75 \\
  Yield strength, $\sigma_{Y}$ & 0.05 & 0.20 \\
  Hardening coefficient, h & 0.4 & 0.7 \\
\hline  
\end{tabular}
\end{center}
\label{tab:range_of_parameters}
\end{table}

\subsection{Taguchi Design of Experiments and Sensitivity Analysis}

When a system's output is governed by two or more independent variables, information about each variable's influence over the output may provide deeper insight into the optimization problem. In other words, it is important to know how the system output is affected by each given input parameter.  By doing so the performance of the overall optimization routine could be greatly improved since this information could subsequently be utilized in reducing the computation expense of the meta-model development. 

In this study, a Taguchi-based design of experiments methodology with ANOVA was adopted to quantify each input parameters contribution towards the overall output or the error function. Employing Taguchi orthogonal arrays instead of a full factorial experimental design help in reducing the number of finite element simulations required in assessing the sensitivity of model parameters.

The first step of performing a sensitivity analysis is to define an experimental design, which involved choosing an appropriate orthogonal array. This was achieved by first calculating the `degrees of freedom' of the experiment, as
\begin{equation}
\text{dof}^{\text{exp}} = \sum \text{dof}^{i} + \sum \text{dof}^{\text{int}}
\end{equation}
where, $\text{dof}^{i} = k^{i} - 1$, $k^{i}$ is designated as the number of levels for the input parameter $i$, and $\text{dof}^{\text{int}}$ are from the interaction between model parameters. In this study four levels were considered for each of the three input parameters, as listed in Table~\ref{tab:levels}. As a result, the degrees of freedom for each factor equaled 3. No interaction was considered among the model parameters. Hence, the total number of degrees of freedom for the experiment was found to be 9. The Taguchi orthogonal array which can successfully accommodate this number of degrees of freedom is modified $L_{16}$. The experimental design for this study according to the modified $L_{16}$ orthogonal array is listed in Table~\ref{tab:L_16}.

\begin{table}[htb]
\caption{Levels of material model parameters}
\begin{center}
\begin{tabular}[l]{@{}cccc}
\hline
  Levels &  \multicolumn{1}{p{2cm}}{\centering Elastic \\ Modulus, \\ E (GPa)} & \multicolumn{1}{p{2cm}}{\centering Yield \\ Strength, \\ $\sigma{Y}$ (GPa)} & \multicolumn{1}{p{2cm}}{\centering Hardening \\ Coefficient, \\ h}\\
\hline  
  Level 1 & 60 & 0.05 & 0.40 \\
  Level 2 & 65 & 0.10 & 0.50 \\
  Level 3 & 70 & 0.15 & 0.60 \\
  Level 4 & 75 & 0.20 & 0.70 \\
\hline 
\end{tabular}
\end{center}
\label{tab:levels}
\end{table}

\begin{table}
\caption{Experimental design based on the modified L$_{16}$ orthogonal array}
\begin{center}
\begin{tabular}[l]{@{}cccc}
\hline
  Experiment & Elastic modulus & Yield strength & Hardening coefficient\\
\hline  
  1 & 1 & 1 & 1 \\
  2 & 1 & 2 & 2 \\
  3 & 1 & 3 & 3 \\
  4 & 1 & 4 & 4 \\
  5 & 2 & 1 & 2 \\
  6 & 2 & 2 & 1 \\
  7 & 2 & 3 & 4 \\
  8 & 2 & 4 & 3 \\
  9 & 3 & 1 & 3 \\
  10 & 3 & 2 & 4 \\
  11 & 3 & 3 & 1 \\
  12 & 3 & 4 & 2 \\
  13 & 4 & 1 & 4 \\
  14 & 4 & 2 & 3 \\
  15 & 4 & 3 & 2 \\
  16 & 4 & 4 & 1 \\
\hline
\end{tabular}
\end{center}
\label{tab:L_16}
\end{table}

For each of these experiments, finite element simulation yielded results in terms of indenter displacement as a function of indentation load. The load increments for the simulation was chosen in such a manner so that it matched with the experimental loading data. The error function, \textdelta~for this study was defined by the following equation.
\begin{equation}
\delta = \sqrt {\frac{1}{n} \sum_{i=1}^n \left [ \frac{(h_i^{sim} - h_i^{exp})}{h_i^{exp}} \right ]^2}
\end{equation}
where, $n$ is the number of data points in the load--displacement plot. By following the Taguchi orthogonal array experimental design the relationship of three model parameters with the system output or the error function was formulated, which was then analyzed using analysis of variance (ANOVA).

\subsection{POD--RBF Based Surrogate Model}

The proper orthogonal decomposition (POD) theory, also known as principal component analysis (PCA), was developed to approximate a function over some domain of interest based on the known relationships between the input and the output~\cite{Chatterjee2000,Liang2002,Ly2001}. This study followed the POD--RBF procedure outlined by Rogers~\emph{et al.}~\cite{Rogers2012}. 

As per POD terminology, the relationship between the input and the output of a particular system is called a snapshot. In the context of this study, snapshots or training points were relation of material model parameters and the output tip displacement data. If $M$ number of simulations were carried out where in each of them at least one input variable was changed, then the snapshot matrix $\bm{U}$ was formulated by combining $M$ number of displacement vectors. Moreover, if the output of the simulation (displacement vector) had $N$ data points, then snapshot matrix $\bm{U}$ can be defined as, 
\begin{equation}
\bm{U} = 
	\begin{bmatrix}
		u_1^1 & u_1^2 & \cdots & u_1^M \\
		u_2^1 & u_2^2 & \cdots & u_2^M \\
		\vdots  & \vdots  & \ddots & \vdots  \\
		u_N^1 & u_N^2 & \cdots & u_N^M
	\end{bmatrix}
\end{equation}

Input material model parameters were collected in the input matrix, $\bm{P}$. The first step towards creating a reduced order model using POD was to generate snapshots of the system for the range of input parameters and subsequently combining all these appropriately in the $\bm{U}$ and $\bm{P}$ matrix. A brief outline of surrogate model training using POD--RBF technique is provided here without detailed mathematical derivations, which can be found in the literature~\cite{Hotelling1933, Sirovich1991}. 

\begin{description}
\item[Step 1:] Develop the covariance matrix $\bm{C}$ for the snapshot matrix $\bm{U}$, where $\bm{C}$ = $\bm{U}^{\bm{T}}$.$\bm{U}$ .

\item[Step 2:] Find the POD orthonormal basis vectors $\bm{\Phi}^j$ (for $j=1, 2, 3,\dots, M$) which would optimally represent $\bm{U}$ . Here, POD basis matrix $\bm{\Phi}$ = $\bm{U}$. $\bm{V}$, and $\bm{V}$ represents the eigenvectors of $\bm{C}$. $\bm{V}$ can be found by solving the eigenvalue problem noted as $\bm{C}.\bm{V} = \bm{\Lambda} . \bm{V}$.

\item[Step 3:] Truncate the POD basis based on the energy of the POD modes and determine $\hat{\bm{\Phi}}$. The subsequent POD model would retain majority of the information about the system, while reducing the dimension of the problem considerably. The truncated POD basis, $\hat{\bm{\Phi}}$ = $\bm{U}$. $\hat{\bm{V}}$ .

\item[Step 4:] Once truncated POD basis matrix is known, the amplitude matrix $\bm{A}$ can be computed as, $\bm{A} = \hat{\bm{\Phi}}^T. \bm{U}$.  $\bm{A}$ is defined as a nonlinear function of $\bm{P}$ matrix. At the time $\bm{A}$ is known, POD reduced order model of the system is ready, and data can be interpolated to find out the surrogate approximation for unknown input parameters.

\item[Step 5:] Compute the coefficient matrix $\bm{B}$ as, $\bm{B} = \bm{A}. \bm{F}^{-1}$, where, $\bm{F}$ is the matrix of interpolation functions or RBFs in the context of this study. $\bm{F}$ is defined as--
\begin{equation}
\bm{F} = 
	\begin{bmatrix}
		f_1(|p^1 - p^1|)  & \cdots & f_1(|p^j - p^1|) & \cdots & f_1(|p^M - p^1|) \\
		f_2(|p^1 - p^2|)  & \cdots & f_2(|p^j - p^2|) & \cdots & f_2(|p^M - p^2|) \\
 		\vdots   & \vdots & \vdots & \ddots & \vdots  \\
		f_i(|p^1 - p^i|)  & \cdots & f_i(|p^j - p^i|) & \cdots & f_i(|p^M - p^i|) \\
		\vdots   & \vdots & \vdots & \ddots & \vdots  \\
		f_M(|p^1 - p^M|)  & \cdots & f_M(|p^j - p^M|) & \cdots & f_1(|p^M - p^M|)
	\end{bmatrix}
\end{equation}
where, $\bm{p}^i$ and $\bm{p}^j$ are input parameter vectors used to generate the i-th and j-th snapshots, respectively.

\item[Step 6:] At an unknown point, $\bm{p}$ in the parameter space, the system output can be computed using the relationship, $\bm{u}(\bm{p}) \approx \hat{\bm \Phi}.\bm{B}.\bm{f}(\bm{p})$, where $\bm{f}(\bm{p})$ is defined as $\bm{f}(\bm{p}) = f_i (|\bm{p}-\bm{p}_i|)$. $\bm{f}(\bm{p})$ is essentially an M-dimensional column vector of RBF values of unknown point $\bm{p}$ with respect to the known input points. 

\end{description}

Table~\ref{tab:RBFs} shows the radial basis functions that were used in this study. It is important to note that, LS and CS are piece-wise smooth functions, while MQ, GS and IMQ are continuously smooth functions. The biggest difference between piece-wise and continuously smooth functions is that the latter creates a continuously smooth function through all the known data points while the former is only smooth in between the data points. Also, the continuously smooth RBFs used in this study utilized a shape parameter denoted by $c_j$. The primary role of shape parameter is to remove ill-conditioning during numerical manipulations. In keeping with literature, the value of the shape parameter $c_j$ was kept constant at 0.5~\cite{Bolzon2011}. 

\begin{table}
\caption{Experimental design based on the modified L$_{16}$ orthogonal array}
\begin{center}
\begin{tabular}{lc}
\hline
  Radial Basis Functions (RBFs) & Equation\\
  \hline
  Linear splines (LS) & $||\textbf{x}-\textbf{x}_j||$ \\
  Cubic splines (CS) & $||\textbf{x}-\textbf{x}_j||^3$ \\
  Multiquadratic (MQ) & $\sqrt{1+\frac{||\textbf{x}-\textbf{x}_j||^2}{c_j^2}}$ \\
  Gaussian (GS) & exp$\left( \frac{-||\textbf{x}-\textbf{x}_j||}{c_j^2} \right)$ \\
  Inverse Multiquadratic (IMQ) & $\frac{1}{\sqrt{||\textbf{x}-\textbf{x}_j||^2 + c_j^2}}$ \\
  \hline
\end{tabular}
\end{center}
\label{tab:RBFs}
\end{table}

\section{Results and Discussions}

\subsection{Sensitivity Analysis}
For each of the three unknown parameters (Young's modulus, yield strength, and hardening coefficient) initially four levels were selected within the range specified in Table~\ref{tab:levels}. This experimental design required the modified L$_{16}$ Taguchi orthogonal array, as listed in Table~\ref{tab:L_16}. The results of ANOVA for computer experiments that were conducted following the modified L$_{16}$ orthogonal array are listed in Table~\ref{tab:anova_results}.

\begin{table}
\caption{Experimental design based on the modified L$_{16}$ orthogonal array}
\begin{center}
\begin{tabular}[l]{@{}lcccccc}\hline
  Source & DF & Adj SS & Adj MS & F-Value & P-Value & \% Contribution\\
  \hline
 Modulus & 3 & 5272 & 1757 & 0.62 & 0.629 & 1.70\\
Yield & 3 & 268843 & 89614 & 31.45 & 0.000 & 87.14\\
Hardening & 3 & 17288 & 5763 & 2.02 & 0.212 & 5.60\\ 
Error & 6 & 17094 & 2849 &  & \\
\hline
Total & 15 & 308498 &  &  & \\
  \hline
\end{tabular}
\end{center}
\label{tab:anova_results}
\end{table}

\begin{figure}
\begin{center}
\subfigure[\% Contribution of individual model parameters]{
\resizebox*{9cm}{!}{\includegraphics{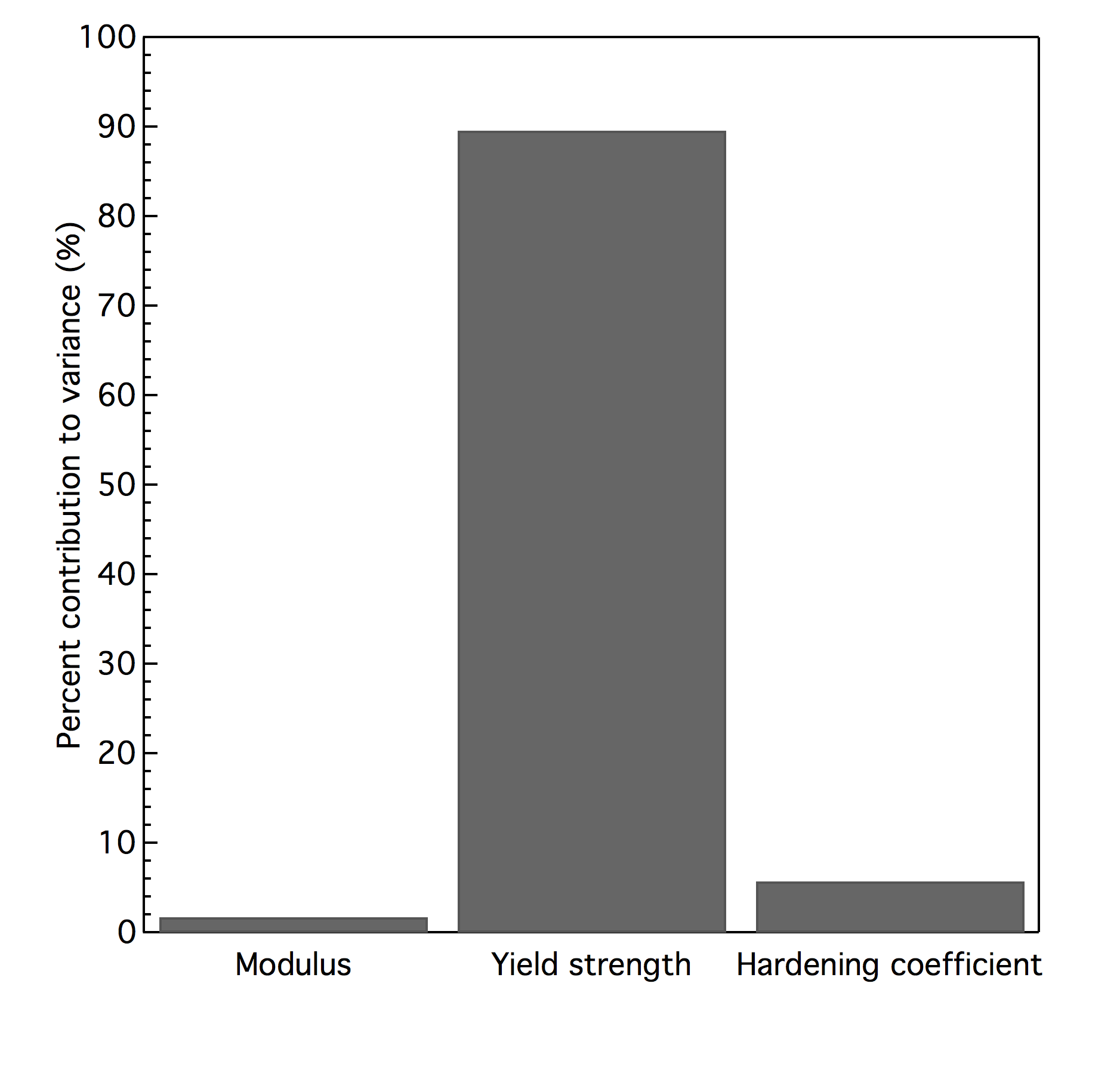}}}\hspace{5pt}
\subfigure[Main effects plot for means of individual model parameter]{
\resizebox*{11cm}{!}{\includegraphics{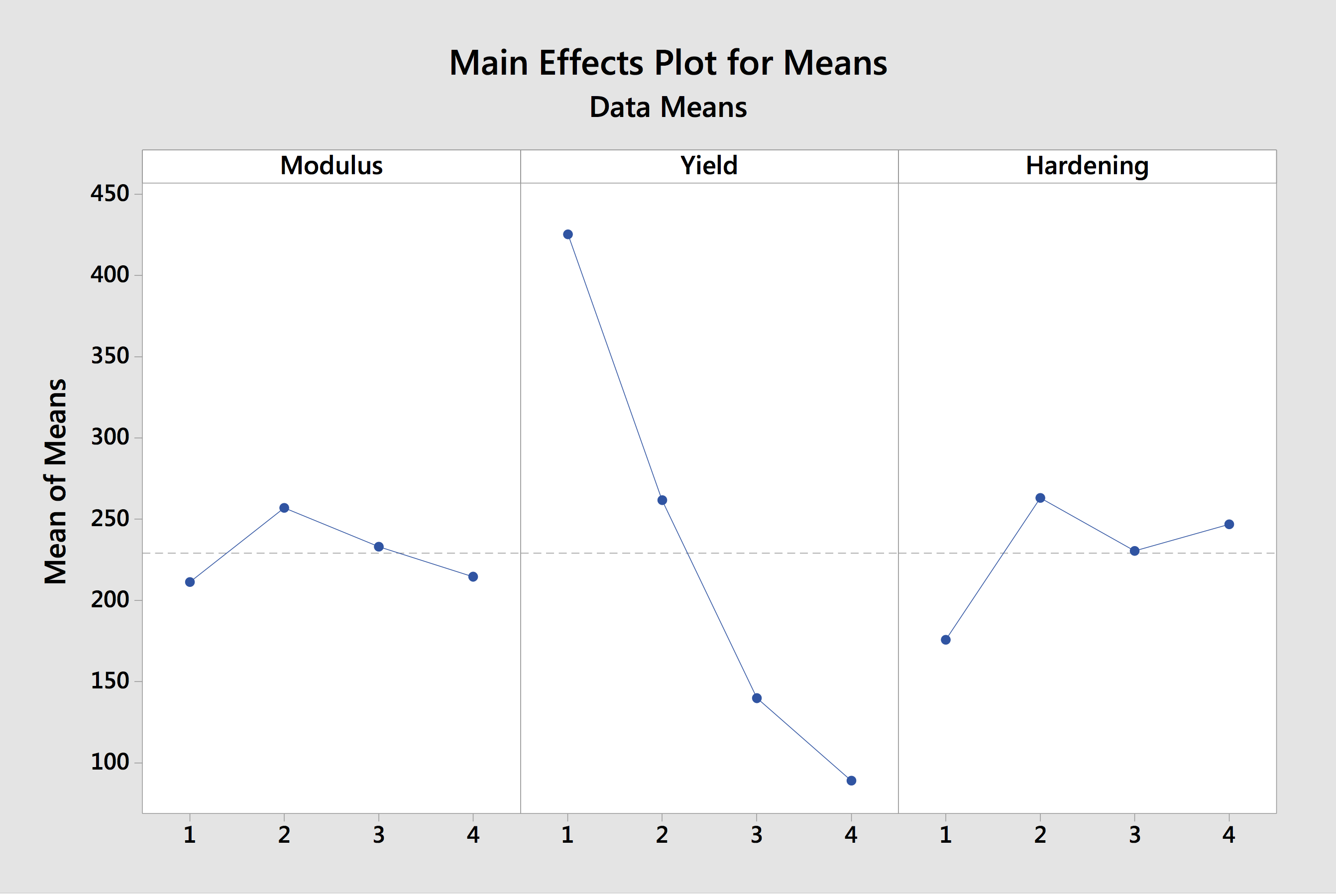}}}
\caption{Individual model parameter's sensitivity towards the output obtained by ANOVA.}
\label{fig:anova_visual}
\end{center}
\end{figure}

From the $P$-values of the ANOVA results it was found that the yield strength parameter significantly affected the output (P-value $\approx$ 0.00 $\textless$ 0.05) at 0.05 level of significance. The other two parameters, Young's modulus and hardening coefficient, however, did not had a significant effect. In ANOVA the sum of squares represents the variance contributed by each parameter. Accordingly, the `\% Contribution' column in Table~\ref{tab:anova_results} shows the percentage contribution of each parameter towards the total sum of squares. Figure~\ref{fig:anova_visual} shows a visual representation of how the output changes within the range of each individual parameters, and the \% Contribution for each parameters found from ANOVA. It can be seen that majority of the variance  originated from the yield strength parameter. Meanwhile, the contributions of the other two parameters were significantly less as compared to the overall variance.  Hence, the most influential input parameter for the bilinear plasticity model was determined to be the yield strength.

This behavior can be explained in terms of the mechanics of the indentation process. For a sharp tip indenter, such as the Berkovich tip, plastic deformation starts dominating the behavior of the nanoindentation response very early in the loading process. Thus, even a small change in yield strength of the material would have a significant effect on the nanoindentation behavior, especially under a sharp tip. 

\subsection{Effect of Training Points and Choice of RBF}
Initially, a total of $4\times4\times4 = 64$ finite element (FE) solutions were used to train the surrogate model. These snapshots or training points had at least one input parameter different in each of the 64 input set. From the simulation a total of 30 data points were extracted to represent the load--displacement data. The final snapshot matrix was then developed with the dimension of 30 row and 64 columns, where each column represented one single simulation. 

POD was then used to reduce the model by identifying the correlation between the 64 different snapshots. The resultant eigenvalues for the snapshot matrix is given below.
\[\lambda = \begin{bmatrix}
  5.31\times 10^{08}  \\
  4.35\times 10^{05}  \\
  1.21\times 10^{04}    \\
  5.66\times 10^{03} \\ 
  3.22\times 10^{03} \\
  2.12\times 10^{03}
  \end{bmatrix}\]
  
It can be seen from the eigenvalues that high degree of correlation existed between the snapshots. This is a normal behavior since the system, boundary conditions and the measurements were the same for all the snapshots, while only few model parameters were changed. Then the dimension of the problem was reduced using the equation provided in the POD theory (Eq.~\ref{eq:truncate}). The reduced model was able to retain 99.91\% data variability by keeping only 1 dimension out of 6. 

Now, the FE nanoindentation output was not as sensitive to the Young's modulus or the hardening coefficient as to the yield strength. Therefore, the number of levels for the Young's modulus and hardening coefficient parameters was reduced to three from the initial number of four. The number of levels for yield strength was kept constant at four. Table~\ref{tab:levels_2} shows the optimized levels for the Young's modulus and hardening coefficient parameters. By employing the same full factorial design to generate set of input parameters, a total of $3\times4\times3 = 36$ different input sets (snapshots or training points) were now used to train the optimized surrogate model. 

\begin{table}
\caption{Experimental design based on the modified L$_{16}$ orthogonal array}
\begin{center}
\begin{tabular}[l]{@{}ccc}\hline
  Level & Elastic modulus (GPa) & Hardening coefficient\\
  \hline
  1 & 60.0 & 0.40 \\
  2 & 67.5 & 0.55 \\
  3 & 75.0 & 0.70 \\
  \hline
\end{tabular}
\end{center}
\label{tab:levels_2}
\end{table}

After the reduced model is established and its corresponding amplitude matrix is determined, RBF was used for the interpolation. As mentioned previously, the choice of RBF is an important step towards developing the inverse technique. In this study, five different RBFs were used initially to find out the best performing RBF for approximating nanoindentation response. It was assumed that the error of approximation would be the maximum between two sampling points. Hence, the validation points or unknown points referred in the subsequent paragraphs were the points that were halfway between two known or training points.

\begin{figure}
\begin{center}
\subfigure[Known data (training) points]{
\resizebox*{9cm}{!}{\includegraphics{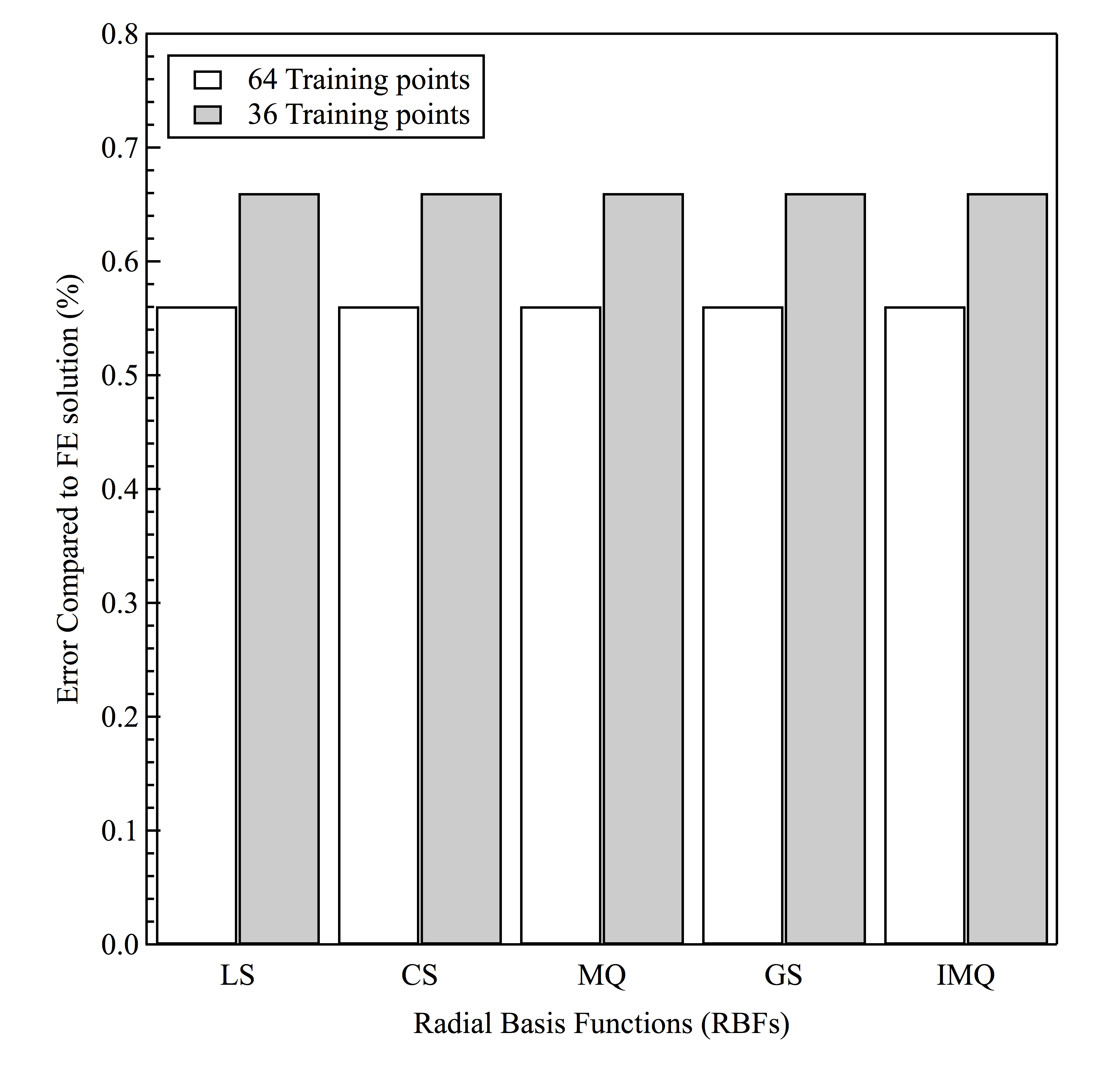}}}\hspace{5pt}
\subfigure[Unknown data (validation) points]{
\resizebox*{9cm}{!}{\includegraphics{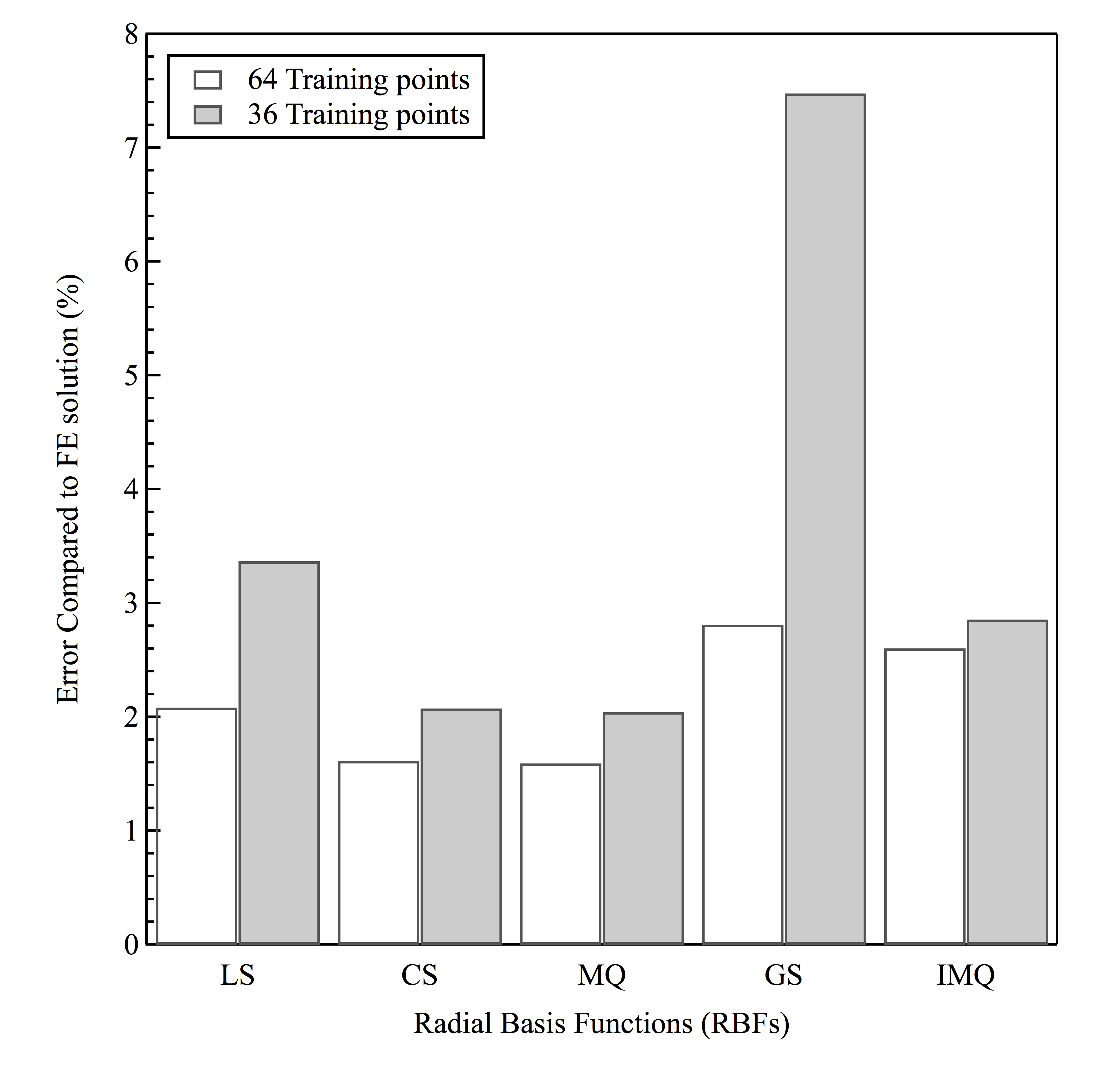}}}
\caption{Comparison of the initial 64-training point and subsequently optimized 36-training point surrogate model performances calculated for training and validation points for five different radial basis functions (RBFs), namely, linear splines (LS), cubic splines (CS), multiquadratic (MQ), Gaussian (GS), and inverse multiquadratic (IMQ).}
\label{fig:performance}
\end{center}
\end{figure}

Figure~\ref{fig:performance} shows the performances of initial 64-training point and subsequently optimized 36-training point surrogate models for training and validation points for various radial basis functions with the five different RBFs. A POD--RBF surrogate model trains itself by combining the input--output relationships of the known points. As a result very small deviations or errors are expected for the approximations of the known training points. Accordingly it was found that both the 64-point and 36-point surrogate models provided good approximation for the training points (Fig.~\ref{fig:performance}a). The $\sim$0.8\% error in approximation could be attributed to the POD model reduction and rounding off errors. The 36-point surrogate model showed a small increase in error for the known points. Nonetheless, this increase in error was relatively small when compared to the reduction in number of training points used for the optimized surrogate model. 

The performance of the surrogate models for validation (unknown) data points is shown in Fig.~\ref{fig:performance}b. Once again, we report on the performances of both 64-point and 36-point surrogate models with five different RBFs. While the different RBFs did not show any differences in the performance of the surrogate models for the known points, significant variations were observed for the validation (unknown) data points. 

For the initial surrogate model trained with 64 training points, the lowest error was found for the multiquadratic (MQ) RBF, while the Gaussian (GS) RBF showed the highest error. The cubic spline (CS) RBF was also found to be very close in performance to the MQ RBF. The difference between the best and worst performing RBF was approximately $\sim$1\%. These errors were magnified for the 36-point model. In this case the difference between the best and worst performing RBF was approximately $\sim$5.5\%. Once again, the MQ RBF provided the lowest and the GS RBF the highest approximation error, respectively. Also, there were only minor differences in performance between the CS and MQ RBFs. In a relative sense CS, MQ and IMQ showed a small increase in approximation error for optimized surrogate when compared to the initial surrogate. 

The variation in performances of these RBFs in approximating the nanoindentation data can be examined in terms of the mechanics of the loading process that is being modeled. A nanoindentation experiment typically yields a nonlinear response in the load--displacement data. In the context of this study, this means that a linear change in any unknown model parameter would lead to a nonlinear change in the measured tip displacement data. The LS and CS RBFs represent piece-wise smooth functions and thus can be expected to have issues when the modeled behavior is non-linear. Accordingly, both the 64-point and 36-point LS RBF-based surrogate models exhibited poor performance. The end effect of nonlinearity on the model's performance was much more pronounced for LS RBF-based optimized surrogate model (with reduced number of points), since the spatial distances between the training points were greater. While the CS RBF is also a piece-wise smooth function, it offers a better performance because of its ability to conform into a nonlinear shape. Thus, the  CS RBF-based models provided better approximations for both 64-point and 36-point models.

The MQ, GS, and IMQ RBFs represent continuously smooth functions and thus should be very capable for modeling nonlinear load--displacement data from a nanoindentation loading process. Hence, all these RBFs should provide comparable approximation error while used in POD--RBF based surrogate models. However, except for MQ, the performance of these RBFs were poor. In fact, GS RBF-based surrogate models exhibited the worse performance with significantly higher approximation error compared to all the other RBFs. Understanding this requires an investigation into the effects of the shape parameter that plays a role in the approximations by continuously smooth functions.

\subsection{Effect of Shape Parameter} 
To understand variations in the performance of continuously smooth RBFs, especially the GS function, a parametric study was conducted to observe the effect of shape parameter. As stated earlier, the primary role of shape parameter is to remove ill-conditioning during numerical manipulations. Although it is desirable to have a high value of shape parameter, beyond certain point RBF approximation becomes unstable due to near singular interpolation matrix~\cite{Buhmann2003}. 

\begin{figure}
\begin{center}
\resizebox*{10cm}{!}{\includegraphics{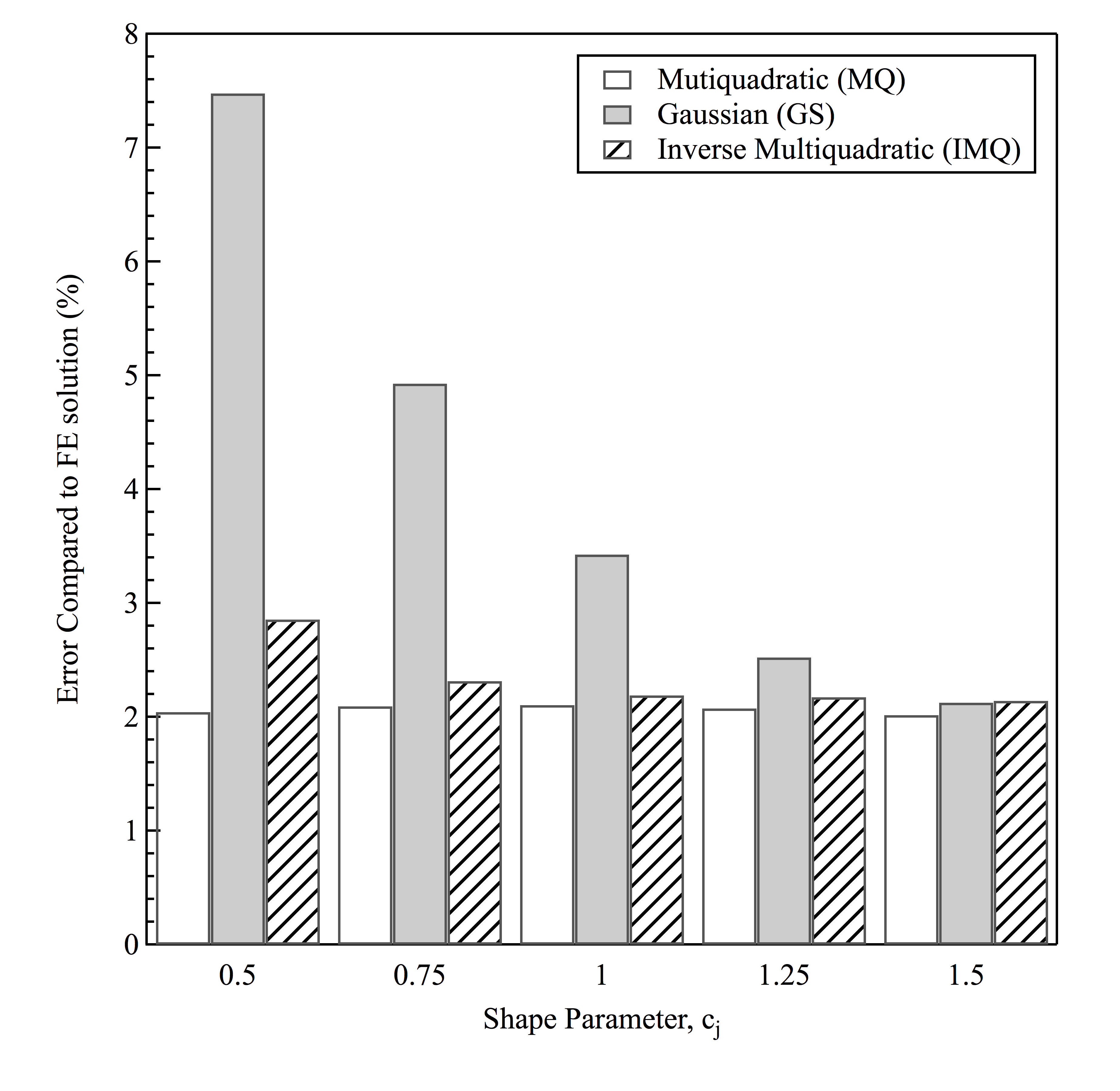}}
\caption{Effect of shape parameter (c$_j$) on the performance 36-point surrogate model for various continuously smooth radial basis functions (RBF's), multiquadratic (MQ), Gaussian (GS), and inverse multiquadratic (IMQ).}
\label{fig:shape}
\end{center}
\end{figure}

For this investigation the shape parameter was varied from 0.5 to 1.5 and the resulting effect on fitting studied for the three continuously smooth RBFs of interest, namely MQ, GS, and IMQ. Figure~\ref{fig:shape} shows the variations in approximation error for these three RBFs as a function of shape parameter for the 36-training point model. It is seen that, shape parameter played a role in minimizing the approximation error of the models. The MQ RBF-based models, which already provided the best performance among all five RBFs, did not show considerable change in error. In case of IMQ, the approximation error decreased a little bit as the value of the shape parameter was increased. The most dramatic change was observed for the GS RBF-based surrogate model, where the approximation error decreased almost exponentially from a value of $\sim$7.5\% to $\sim$2\%. For $c_j$ = 1.5 the error of the GS-based model was almost comparable to the other two RBFs.

This behavior is consistent with prior studies conducted to illustrate the applicability of RBFs as an interpolation tool~\cite{Larsson2003, Mongillo2011}. Higher values of the shape factor usually improve the overall accuracy of approximation; however, the \emph{optimal} number depends heavily on the nature of the problem, the RBF in use, and the number of training points. The shape parameter not only provides a way for removing ill-conditioning, it also serves as a measure of influence domain. If a particular unknown point is imagined in the center of the influence domain then only the data points inside or near the influence domain affect the quality of approximation of that point. As higher values of the shape parameter lead to a bigger influence domain, the approximation quality generally improves. However, larger shape parameter values also imply large condition number of a system that subsequently leads to larger error in the coefficients. 

\subsection{Effect of Random Noise} 
\begin{figure}
\begin{center}
\subfigure[36-training point model]{
\resizebox*{9cm}{!}{\includegraphics{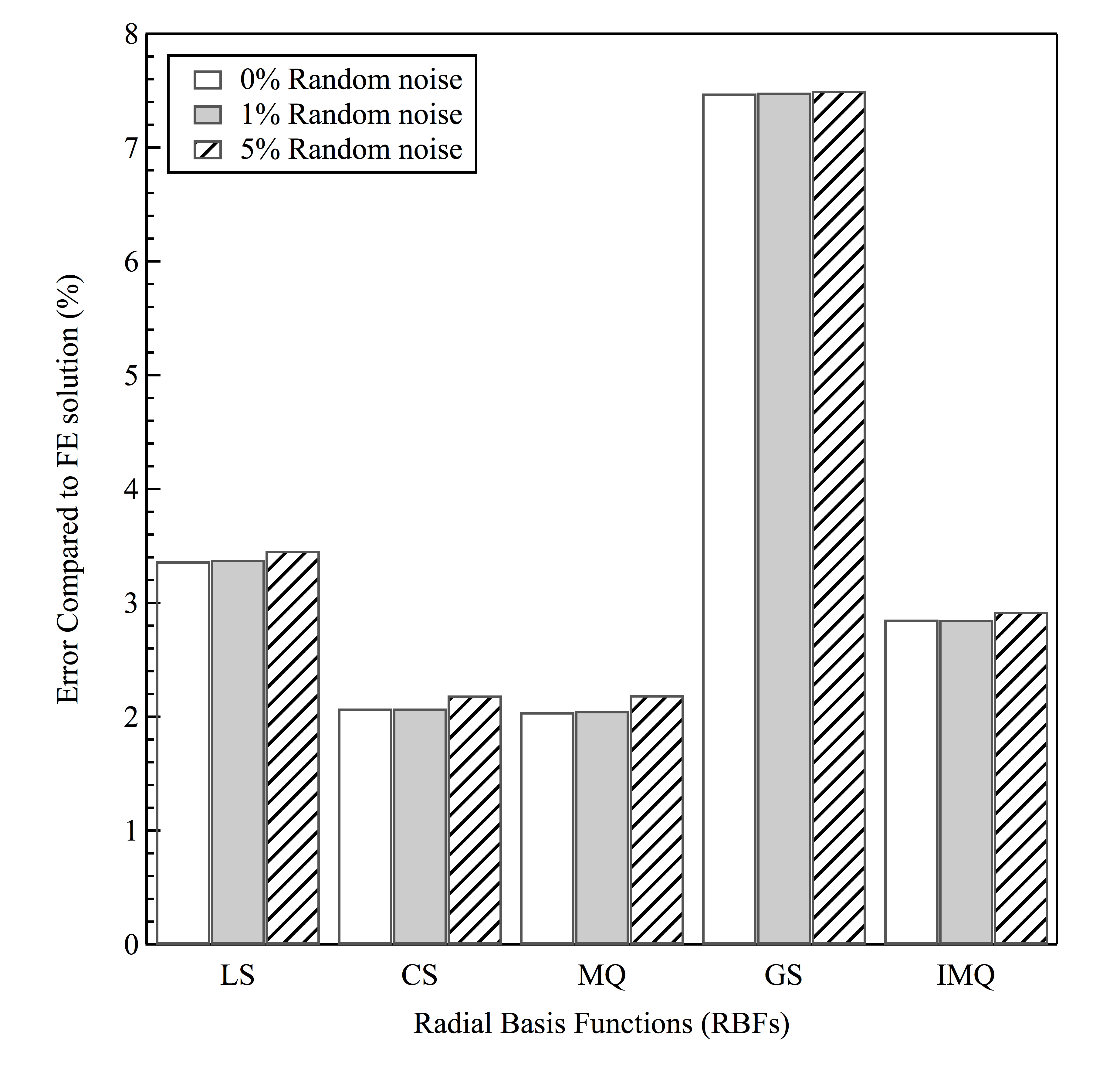}}}\hspace{5pt}
\subfigure[64-training point model]{
\resizebox*{9cm}{!}{\includegraphics{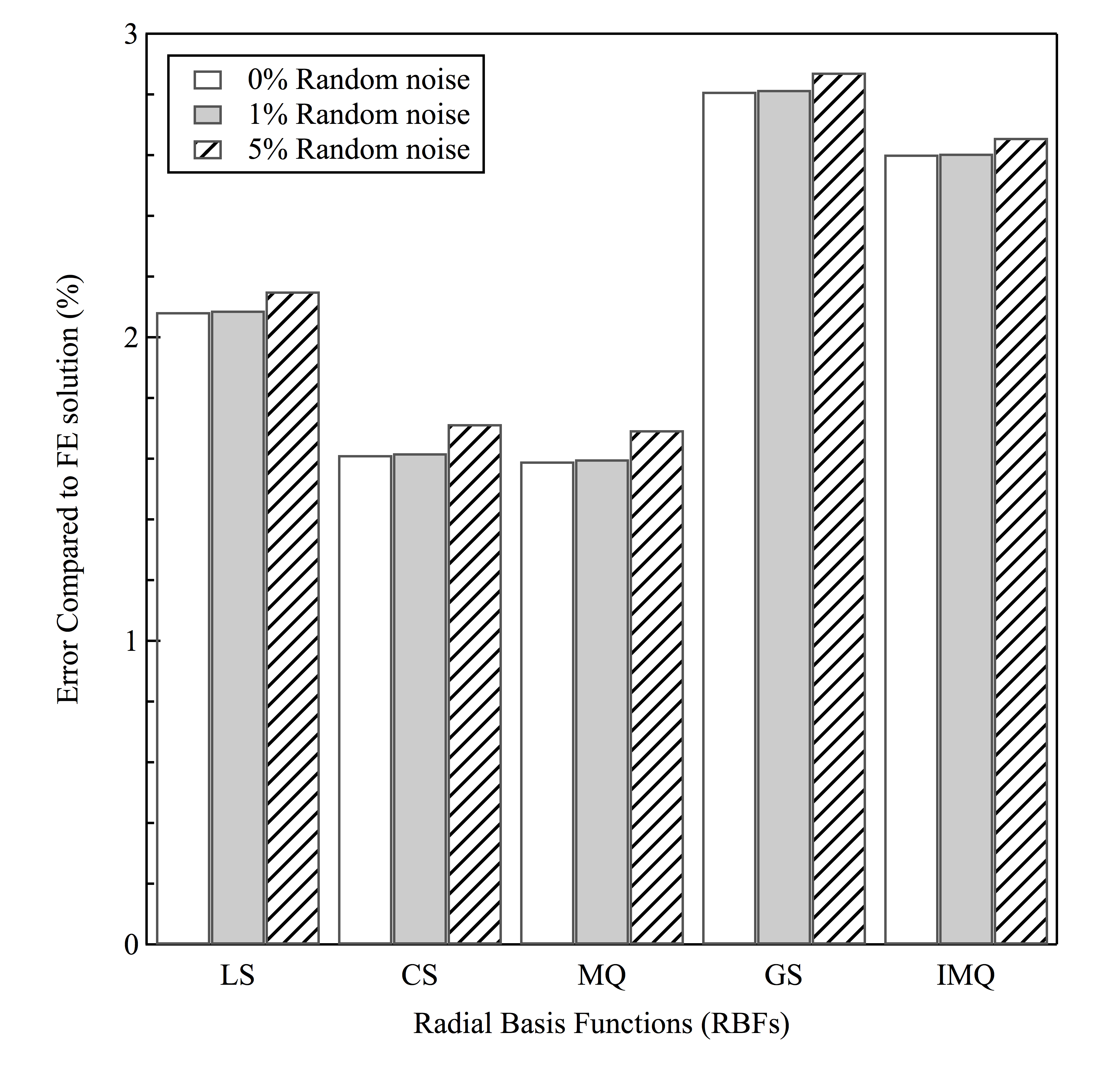}}}
\caption{Effect of random noise on the surrogate model performance calculated for unknown data (validation points) for various radial basis functions (RBF's), linear splines (LS), cubic splines (CS), multiquadratic (MQ), Gaussian (GS), and inverse multiquadratic (IMQ).}
\label{fig:noise}
\end{center}
\end{figure}

Random noise was introduced in to the training data to observe how well the POD--RBF based surrogate models perform when measurement errors are present in the training data. With regards to our specific study, it was particularly important to examine if approximation errors propagated for the surrogate model with reduced number of training data. Figure~\ref{fig:noise} shows the effects of 1\% and 5\% random noise on the performance of 36-training point and 64-training point based surrogate models. For both the surrogate models the introduction of  1\% or 5\% random noise resulted in only minimal increase in the approximation error. This demonstrated that the POD--RBF technique was very efficient in filtering out random measurement error in the training data. This was especially signification for the surrogate model trained with 36 points and illustrated that the POD--RBF technique was very effective in dealing with noisy data even when number of training points was low. 

Another important observation found from Fig.~\ref{fig:noise} was that RBFs played little to no role in changing the overall approximation error as a function of random noise being introduced in to the training data. For all the RBFs, both for 64-point and 36-point surrogate models, the error in approximation after training with noisy data was consistent with the approximation error trained with clean data. This finding suggested that although the choice of an RBF plays a crucial role for overall surrogate model performance, it does not provide any added benefit or disadvantage when dealing with noisy training data.

\section{Conclusion}

This study took a systematic approach towards understanding the role of training points quantity and the choice of RBF for surrogate model construction to solve a nanoindentation-based inverse problem. In particular, attention was concentrated to see if the information of system's sensitivity towards individual input parameters could be utilized to reduce the number of data points required to train the surrogate model without sacrificing considerable accuracy. 

A case study problem was formulated where an elastic--plastic model with three parameters was used to define the material behavior of single crystal aluminum. A Taguchi orthogonal array was used to design the experiments for the input parameters varying within a preselected range of values in few equidistant levels. By applying analysis of variance (ANOVA) on orthogonal array experiments, the sensitivities of nanoindentation output with respect to each model parameters were identified. This information was then used to reduce the number of levels for parameters that exhibited smaller effect on the nanoindentation output. Training data points were generated using finite element software ABAQUS by adopting a full factorial approach for input parameter sets. A systematic comparison was made between the performances of five different RBFs, namely,  linear splines (LS), cubic splines (CS), multiquadratic (MQ), Gaussian (GS), and inverse multi quadratic (IMQ).  This comparison allowed the investigation of the choice of RBF in terms of overall performance of the surrogate model. Finally, random noise was introduced in the training data in order to verify the stability of the surrogate model, especially as the number of training points was reduced. 

It was found that ANOVA analysis of a Taguchi orthogonal array-based experimental design could provide meaningful understanding about the sensitivity of each input parameter of the material model. This information could be utilized in two ways, a) to reduce the number of data points required for the less critical parameters, thus reducing the overall number of training points, and b) to combine this information with global optimization algorithms to reduce the computational effort for finding the global minima. In this study, it was successfully shown that an optimally trained surrogate model provided competitive quality of approximation when compared with a surrogate model trained higher number of training data. For multiquadratic (MQ) RBF-based surrogate model, approximately $\sim$0.5\% difference in accuracy was found for surrogate models trained with 64- and 36-points. 

Among the five RBFs that were compared in this study, multiquadratic (MQ) and Gaussian (GS) RBF provided best and worst performance, respectively, for both training schemes. While the difference between their performance was approximately $\sim$1\% for 64-points model, it increased dramatically to $\sim$5.5\% for 36-points model. Among the piece-wise continuous RBFs, cubic spline (CS) RBF's performance was comparable to MQ's performance, while linear spline (LS) RBF performed poorly. It was interesting to note that, while all the RBFs showed some increase in approximation error due to training point reduction, this effect was much more pronounced for LS and GS-based models. 

The poor performance of the LS RBF-based model could be attributed to LS's inability to replicate nonlinear input--output relationships of the nanoindentation experiment. To understand GS RBF's unexpectedly high approximation error, a parametric study was conducted to investigate the effect of shape parameter on the overall performance of the surrogates. As the value of the shape parameter increased the quality of approximation for GS-based surrogate improved dramatically due to the increase in radius of influence. Similar observation was made for IMQ-based surrogate, although the improvement was not as dramatics as GS. 

By introducing random noise in the training data, the stability of POD--RBF based surrogate models were investigated. It was found that POD--RBF was capable of providing good quality approximation even with noisy training data. Identical observations were made for both 36- and 64-training points model, where random noise did not significantly altered the approximation error.

This investigation demonstrated that through the use of sensitivity analysis it was possible to reduce the number of training points required for POD--RBF based surrogate model without sacrificing considerable accuracy. It was also found that due to the nonlinear behavior of input--output relationship of a nanoindentation experiment, a RBF which can conform into a nonlinear shape would perform better. The value of the shape parameter for continuously smooth functions was found to have effect on the overall quality of approximation. POD--RBF approach's power to effectively find the dominant nature of the data even from a smaller number of training points was observed through studying the effect of noisy training data. 
  
\chapter{ESTIMATION OF NONLINEAR VISCOELASTIC PARAMETERS}\label{chap:chap5}

\section{Study Details}

In order to demonstrate the applicability of the POD--RBF technique to determine the nonlinear viscoelastic Burgers model parameters, nanoindentation was carried out on epoxy. The finite element model was constructed using commercial finite element package ABAQUS (Dassault Syst\'emes, Providence, RI, USA). The nonlinear Burgers model was implemented in an user-defined subroutine (UMAT) via FORTRAN script. The information known from previous chapter about POD--RBF was combined to solve the problem of finding the model parameters.  

\subsection{Design of Experiments for Sensitivity Analysis}

Before generating finite element simulation data by varying the model parameters, a sensitivity study of the parameters was conducted. This information helps to reduce the number of finite element simulation used for training. This was demonstrated in \Cref{chap:chap4} for an elastic--plastic model. 

The nonlinear Burgers model that was chosen to represent the behavior of the epoxy has seven independent parameters. These parameters are E, \textnu, $C_s$, $m_s$, $C_t$, $m_t$, and $t_{\epsilon}$, as discussed earlier in \Cref{subchap:nonlin_model}. It is already known that a nanoindentation load--displacement response is not highly influenced by Poisson's ratio, \textnu~\cite{Magnenet2008, Ma2013, Clement2013}. Therefore, in order to keep the number of independent parameters to a minimum, \textnu~was given a constant value of 0.34, and was not included in the sensitivity analysis scheme. 

Sensitivity analysis was primarily carried out using Analysis of Variance (ANOVA) technique. The data required for ANOVA was generated using the Taguchi Design of Experiments (DOE) method. In this study, the six nonlinear model parameters were varied in three equidistant levels. A statistical software, Minitab (Minitab Inc., State College, PA, USA) was used to design the experiments. For six parameters, where each parameters were varied in three levels, Taguchi $L_{27}$ orthogonal array design was appropriate. \Cref{tab:burger_levels} shows the levels of the six individual parameters of the nonlinear Burgers model. The experimental design for this study according to the $L_{27}$ orthogonal array is listed in Table~\ref{tab:L_27}.

\begin{table}[htb]
\caption{Levels of nonlinear Burgers model parameters}
\begin{center}
\begin{tabular}[l]{@{}cccc}
\hline
  Parameters &  Level 1 & Level 2 & Level 3\\
\hline  
  E & 3 & 3.25 & 3.5 \\
  $C_s$ & 0.02 & 0.06 & 0.1 \\
  $m_s$ & 0.15 & 0.25 & 0.35 \\
  $C_t$ & 0.15 & 0.25 & 0.35 \\
  $m_t$ & 0.2 & 0.5 & 0.8 \\
  $t_{\epsilon}$ & 0.1 & 0.25 & 0.4 \\
\hline 
\end{tabular}
\end{center}
\label{tab:burger_levels}
\end{table}

\begin{table}
\caption{Experimental design based on the $L_{27}$ orthogonal array}
\begin{center}
\resizebox*{8.2cm}{!}{%
\begin{tabular}[l]{@{}ccccccc}
\hline
  Experiment & E & $C_s$ & $m_s$ &$C_t$ & $m_t$ & $t_{\epsilon}$\\
\hline  
  1   & 3      & 0.02 & 0.15 & 0.15 & 0.2 & 0.1 \\
  2   & 3      & 0.02 & 0.15 & 0.15 & 0.5 & 0.25 \\
  3   & 3      & 0.02 & 0.15 & 0.15 & 0.8 & 0.4 \\
  4   & 3      & 0.06 & 0.25 & 0.25 & 0.2 & 0.1 \\
  5   & 3      & 0.06 & 0.25 & 0.25 & 0.5 & 0.25 \\
  6   & 3      & 0.06 & 0.25 & 0.25 & 0.8 & 0.4 \\
  7   & 3      & 0.1   & 0.35 & 0.35 & 0.2 & 0.1 \\
  8   & 3      & 0.1   & 0.35 & 0.35 & 0.5 & 0.25 \\
  9   & 3      & 0.1   & 0.35 & 0.35 & 0.8 & 0.4 \\
  10 & 3.25 & 0.02 & 0.25 & 0.35 & 0.2 & 0.25 \\
  11 & 3.25 & 0.02 & 0.25 & 0.35 & 0.5 & 0.4 \\
  12 & 3.25 & 0.02 & 0.25 & 0.35 & 0.8 & 0.1 \\
  13 & 3.25 & 0.06 & 0.35 & 0.15 & 0.2 & 0.25 \\
  14 & 3.25 & 0.06 & 0.35 & 0.15 & 0.5 & 0.4 \\
  15 & 3.25 & 0.06 & 0.35 & 0.15 & 0.8 & 0.1 \\
  16 & 3.25 & 0.1   & 0.15 & 0.25 & 0.2 & 0.25 \\
  17 & 3.25 & 0.1   & 0.15 & 0.25 & 0.5 & 0.4 \\
  18 & 3.25 & 0.1   & 0.15 & 0.25 & 0.8 & 0.1 \\
  19 & 3.5   & 0.02 & 0.35 & 0.25 & 0.2 & 0.4 \\
  20 & 3.5   & 0.02 & 0.35 & 0.25 & 0.5 & 0.1 \\
  21 & 3.5   & 0.02 & 0.35 & 0.25 & 0.8 & 0.25 \\
  22 & 3.5   & 0.06 & 0.15 & 0.35 & 0.2 & 0.4 \\
  23 & 3.5   & 0.06 & 0.15 & 0.35 & 0.5 & 0.1 \\
  24 & 3.5   & 0.06 & 0.15 & 0.35 & 0.8 & 0.25 \\
  25 & 3.5   & 0.1   & 0.25 & 0.15 & 0.2 & 0.4 \\
  26 & 3.5   & 0.1   & 0.25 & 0.15 & 0.5 & 0.1 \\
  27 & 3.5   & 0.1   & 0.25 & 0.15 & 0.8 & 0.25 \\
\hline
\end{tabular}
}
\end{center}
\label{tab:L_27}
\end{table}

Each of these 27 computer simulations resulted in data in terms of indenter displacement. The resulting value of error function,~\textdelta~was calculated using the Eq.~\ref{eq:delta_2}. This was then utilized in ANOVA to determine the effect of each parameters on the error function.
\begin{equation}
\delta = \frac{1}{n} \sum \left[ (h^i_{exp}-h^i_{sim})^2 \right]
\label{eq:delta_2}
\end{equation}

In Eq.~\ref{eq:delta_2} \emph{i = 1, 2, 3, \dots, n}, and \emph{n} is the number of data points in a single nanoindentation simulation or experiment.

Sensitivity of the nanoindentation output was also determined in a different way, where the difference in output between lowest and highest limit of the individual parameter levels were determined. Unlike the Taguchi--ANOVA procedure explained above, here only the indenter depth at maximum load and the depth after unloading was studied. 

This type of parametric sensitivity analysis has been previously used in understanding nanoindentation experiments in general. In the current study, this sensitivity analysis was performed in order to complement the Taguchi--ANOVA procedure, and to get an objective understanding of how each parameters contribute to the variance of an indentation plot's two key features.

\subsection{Nanoindentation Experiment}

Nanoindentation experiments were conducted on an MTS Nanoindenter XP (Agilent Technologies, Santa Clara, CA, USA) using a load-controlled scheme with a Berkovich tip. The maximum load was set to be 0.5, 0.75, and 1.0 mN for the experiments. A triangular loading profile was chosen with 30, 45, 60, and 240 s durations. The durations were kept constant for both the loading and unloading segments. 

Before conducting the actual experiments the Berkovich tip was calibrated using a fused silica reference material. Also, the acceptable thermal drift rate was chosen to be 0.15 nm/s. After ensuring that the thermal drift rate was stable and below the target drift rate nanoindentation experiments were carried out.

\subsection{Material}

An epoxy polymer, named EPON 862, was selected for carrying out nanoindentation experiment. EPON 862 is a diglycidyl ether of bisphenol F (DGEBF). The curing agent used for this resin system was a moderately reactive, low viscosity aliphatic amine curing agent, called Epikure 3274. Both of these chemicals were supplied by Miller-Stephenson Chemical Company, Inc., Dunbury, Connecticut. 

Epoxy and hardener was mixed at 100:40 weight ratio and hand-mixed using a glass-rod for 5--10 minutes. The mixture was then degassed for around 10--20 minutes to remove any entrapped air bubbles. The mixture was then poured into an aluminum mold and cured at room temperature for 24 hours and subsequently post-cured at 121\textdegree C for 6 hours. The final sample was cut from the prepared epoxy plate using a bandsaw. Sample surface preparation was carried out by polishing using standard metallographic techniques.

\subsection{Genetic Algorithm}\label{subchap:ga}

A multi-objective genetic algorithm-based optimization procedure was used to identify the parameters of the nonlinear Burgers model. The procedure was implemented using MATLAB's (Mathworks Inc., Natick, MA, USA) global optimization toolbox. 

\emph{Double vector} was chosen as the population type. The initial population of 200 was randomly created with a uniform distribution. Scores of the first and all subsequent generations were determined by evaluating the fitness function that was submitted to the program via MATLAB script.

\emph{Selection} of the worthy candidates for being the next generation parent were carried out via tournament of size 2. Eighty percent of the next generation population was produced via \emph{crossover}, while the remainder of the  was created through \emph{mutation}. Gaussian mutation was selected, where a random number from a Gaussian distribution centered on zero was added to each vector entry of an individual. 

The standard deviation of the Gaussian distribution is controlled using two parameters, \emph{i.e.}~\emph{scale} and \emph{shrink}. Both of these two parameters were set to 1 for this study. The scale parameter defines the standard deviation of the Gaussian distribution for first generation, while the shrink parameter determines the amount of shrinking that will occur to the standard deviation by the time it reaches the last generation. 

In this study, the crossover function was chosen to be \emph{intermediate}. In case of intermediate crossover the creation of children from two parents is controlled by a single parameter ratio. The value of this parameter was selected to be 1 for this study. Next generation children were created through a random weighted average of the parents following Eq.~\ref{eq:cross}.
\begin{equation}
\text{child\textsuperscript{1} = parent\textsuperscript{1}+...rand*Ratio*(parent\textsuperscript{2} - parent\textsuperscript{1})}
\label{eq:cross}
\end{equation}

Every once in a while, the worst performing individuals of one subpopulation need to be replaced by the best performing individuals of a different subpopulation. This process is called \emph{migration}. Migration in a genetic algorithm-based optimization can be controlled using three parameters, \emph{i.e.} \emph{direction, fraction, and interval} of migration. 

The direction parameter specifies in which direction migration will take place. The fraction parameter controls the number of individuals that will be migrated from one subpopulation to another. The interval parameter dictates the number of generations that will be elapsed between each migration.

In this study, forward migration direction was chosen. This meant individuals from n\textsuperscript{th} subpopulation would replace individuals from (n+1)\textsuperscript{th} subpopulation and so on. The migration fraction and interval were chosen to be 0.2 and 20, respectively.

Total number of generations for the optimization algorithm was chosen to 100$\times$ number of parameters, \emph{i.e.} 100$\times$6 = 600 for this study. The fitness (error) function tolerance was chosen to be 1e\textsuperscript{-4}.

\subsection{Parametric Study: Friction Coefficient}

The performance of material model calibration using POD--RBF technique depends primarily on the quality of training data coming from finite element simulations. This means that the better the finite element model is in terms of replicating the real experiment scenario of a nanoindentation experiment the better the quality of training data, which in essence provides better quality approximation from the surrogate model.

One important factor to consider in nanoindentation experiment is the friction between the indenter tip and the sample or material surface. In real life situations it is possible that the two surfaces generate finite amount of friction. This can be taken into consideration by defining sliding contact with a finite friction coefficient between the surfaces. 

However, to simplify the finite element model most researchers have opted to assume frictionless contact between tip and sample surface~\cite{Wang2007, Warren2006}. Their assumption was based on the fact that nanoindentation load--displacement data was insensitive to friction. Nonetheless, a few researchers have shown that friction can influence the results in a simulation study~\cite{Taljat1998, Goh2004, Harsono2008}. This is because influence of friction in a simulated study depends on some other factors, such as the material model used and the geometry of the tip. If these factors change from one study to another, investigating the effect of friction becomes a necessity. 

Only few studies have previously used the nonlinear Burgers model in a nanoindentation-based finite element study. Therefore it is necessary to study the effect of friction coefficient on the load--displacement output. To facilitate the understanding this study performed a parametric study, where the friction coefficient value was varied in four steps ranging from 0---0.5. All the experimental load and strain rate levels were studied to improve the understanding.

\subsection{POD--RBF Surrogate Model}

As discussed in previous chapters the POD--RBF method requires creating snapshots (input--output relationships of the system) from which the surrogate model could be established. Each of the data that provides a one-to-one relationship between the input and the output is called a snapshot. As discussed in~\Cref{chap:chap4}, the more snapshots or training data points utilized to generate the surrogate model the better the approximation becomes.

However, the computational burden associated with generating large number of snapshots becomes the limiting factor in obtaining very high-fidelity predictions from the surrogate model. As shown in chapter~\ref{chap:chap4}, sensitivity analysis could be utilized to reduce the number of snapshots without sacrificing approximation error. Hence, in this study, a similar approach was adopted to reduce the computation burden of training the surrogate model for nonlinear Burgers model.

Once the appropriate number of levels for different parameters were selected using information from sensitivity analysis, a full factorial approach was taken to generate the input parameter sets. These parameter sets were combined to produce the input matrix, $\bm{P}$. Finite element simulation experiments were carried out for every individual parameter sets and their corresponding indenter displacement data was assimilated in the snapshot matrix, $\bm{U}$. 

In this study, four different experimental conditions were utilized for which the training data would be generated. In these experimental conditions, the maximum load was kept constant at 1mN, while the strain rate was varied from 1/30 s\textsuperscript{-1} to 1/240 s\textsuperscript{-1}. One surrogate model was created for each of those experimental conditions using finite element data. The approximations from each surrogate model was compared against their own experimental indenter displacement data to form the objective or error function.

In keeping with the findings of chapter~\ref{chap:chap4} the Multiquadratic RBF was chosen for this study. Since, the value of the shape parameter ($c_j$) does not influence the POD--MQ RBF surrogate model's performance significantly it was chosen 0.5 for this study~\cite{Hamim2016}. 

\section{Results and Discussion}

\subsection{Effect of Friction Coefficient}

\begin{figure}
\begin{center}
\subfigure[Effect of friction coefficient on maximum depth ]{
\resizebox*{10cm}{!}{\includegraphics{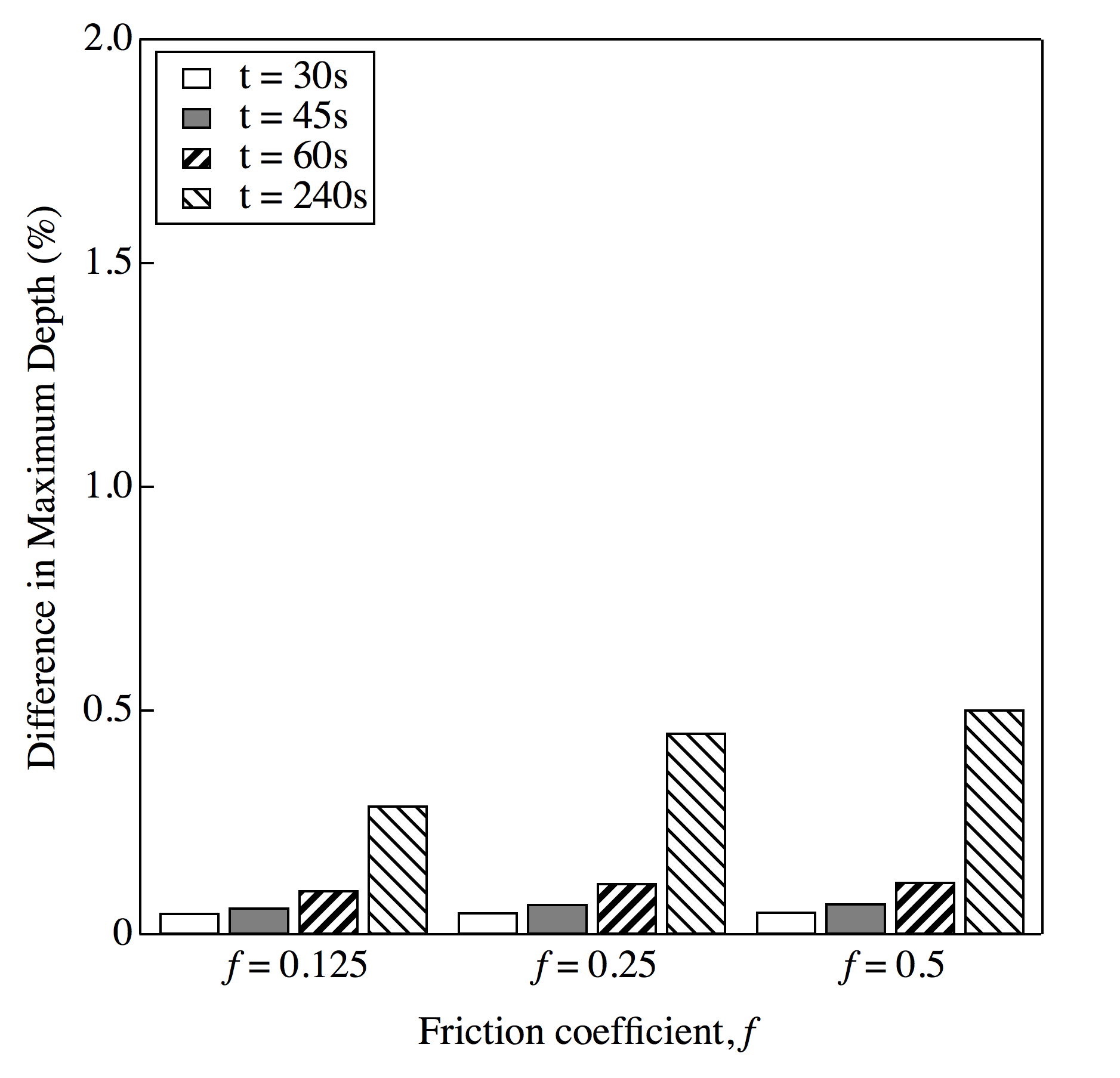}}}\hspace{5pt}
\subfigure[Effect of friction coefficient on residual depth]{
\resizebox*{10cm}{!}{\includegraphics{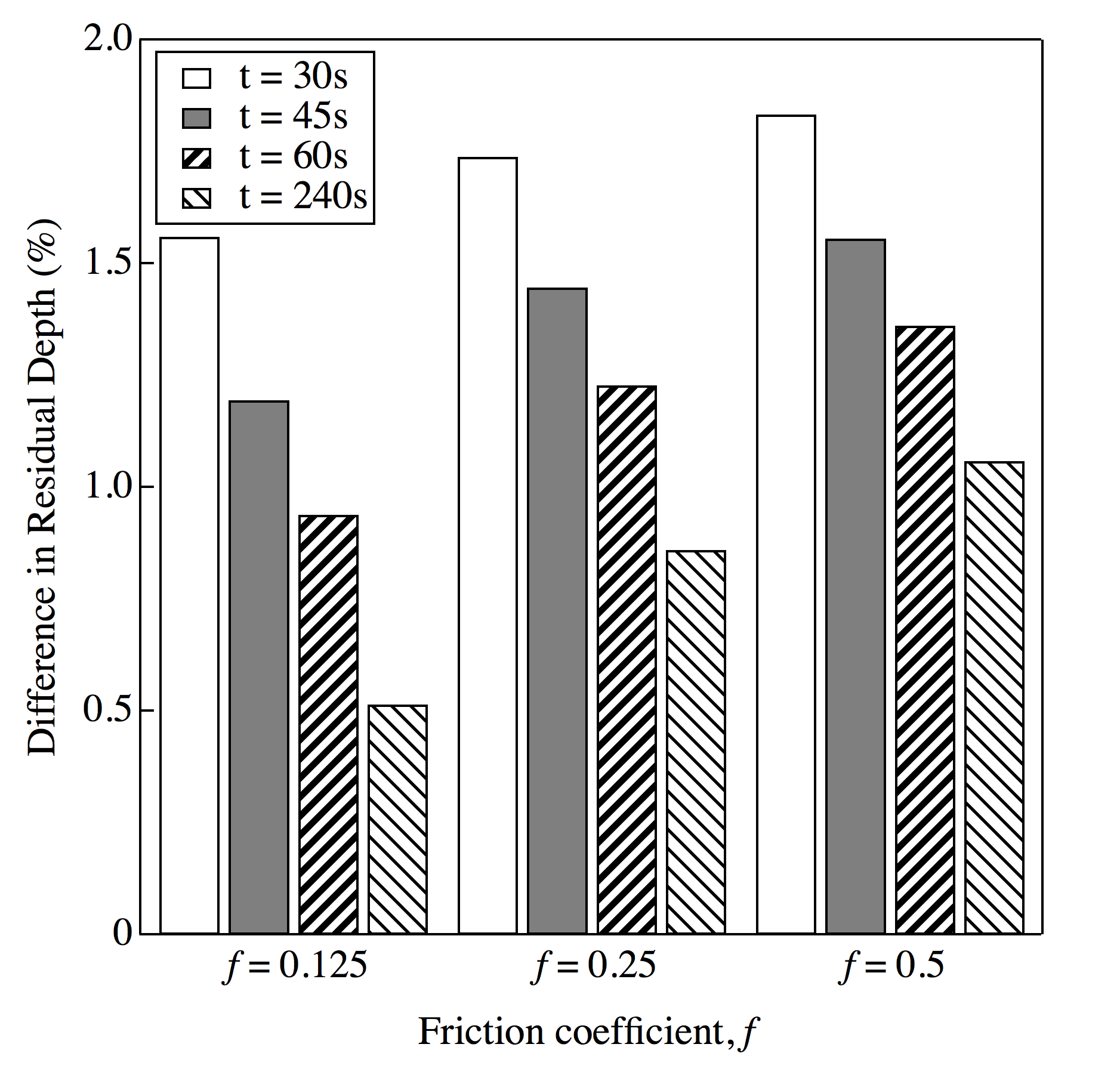}}}
\caption{Effect of friction coefficient, \emph{f} on nanoindentation data for different loading--unloading time and constant maximum load (difference = depth for frictionless -- depth for f = 0.125/0.25/0.5) }
\label{fig:fric_time}
\end{center}
\end{figure}

The effect of the friction coefficient on maximum and residual depths attained during nanoindentation was analyzed. \Cref{fig:fric_time} shows the effect for conditions represented by a fixed maximum load of 1.0 mN with different loading--unloading times. For any given value of the friction coefficient, the values of both maximum and residual depths decreased as compared to the corresponding frictionless case of indentation. The plots represent the reduction of depths between the simulations of a frictionless condition and a particular friction coefficient (e.g. \emph{f} = 0.125, 0.25, or 0.5). All other conditions, e.g. boundary conditions, maximum load, loading--unloading time, and material model parameters, were kept constant. Both Fig.~\ref{fig:fric_time}(a) and \ref{fig:fric_time}(b) are plotted at the same vertical scale for ease of comparison. 

\Cref{fig:fric_time} shows that for any value of the friction coefficient within the studied range, the value of maximum and residual depths were reduced in comparison to the frictionless condition. This behavior was found to be true for all studied conditions with varying loading--unloading times. When friction is considered in a nanoindentation study, part of the energy that could be utilized to displace the material gets dissipated as frictional energy. This loss of energy leads to a reduced displacement of the indenter. Similar behavior has been observed for simulation of elastic--plastic indentation. DiCarlo~\emph{et al.} observed that the introduction of friction in the model increased the calculated hardness by lowering the indenter displacement at maximum load~\cite{DiCarlo2003}. 

\Cref{fig:fric_time}(a) also illustrates that for a given friction coefficient, the reduction in maximum depth varied as a function of loading--unloading time. The greater the loading--unloading time the lower was the maximum depth observed in comparison to the frictionless condition. 

\Cref{fig:fric_time}(b) shows the reduction in residual depth values between a frictionless simulation and a finite friction coefficient simulation. Here, for any given value of friction coefficient, the difference diminished with the increase of loading--unloading time.  When lower loading--unloading time is used in an indentation experiment, the viscoelastic creep response is subdued. Hence, the elastic response has relatively higher dominance on the overall deformation behavior. The observed behavior may mean that friction has more effect on the residual depth when viscoelastic behavior has lower dominance over the nanoindentation data.

For both maximum and residual depths the reduction in depths was observed to be very small for all loading--unloading times. For instance in case of t = 240s, the condition which showed the highest deviation for maximum depth, the reduction was found to be $\approx$ 0.5\%. On the other hand, the highest reduction in residual depth was found to be $\approx$ 1.8\% 

\begin{figure}
\begin{center}
\subfigure[Effect of friction coefficient on maximum depth]{
\resizebox*{10cm}{!}{\includegraphics{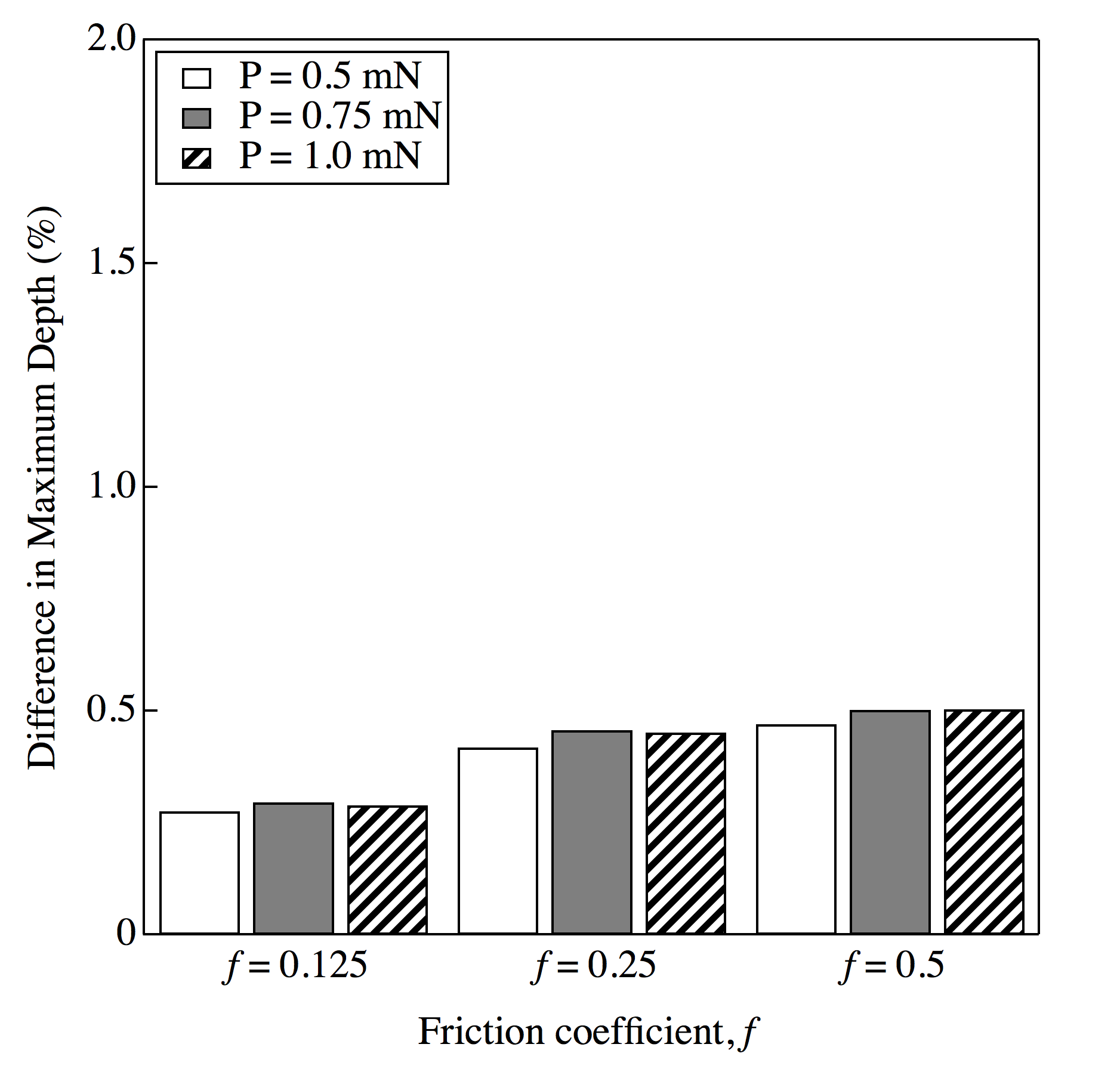}}}\hspace{5pt}
\subfigure[Effect of friction coefficient on residual depth]{
\resizebox*{10cm}{!}{\includegraphics{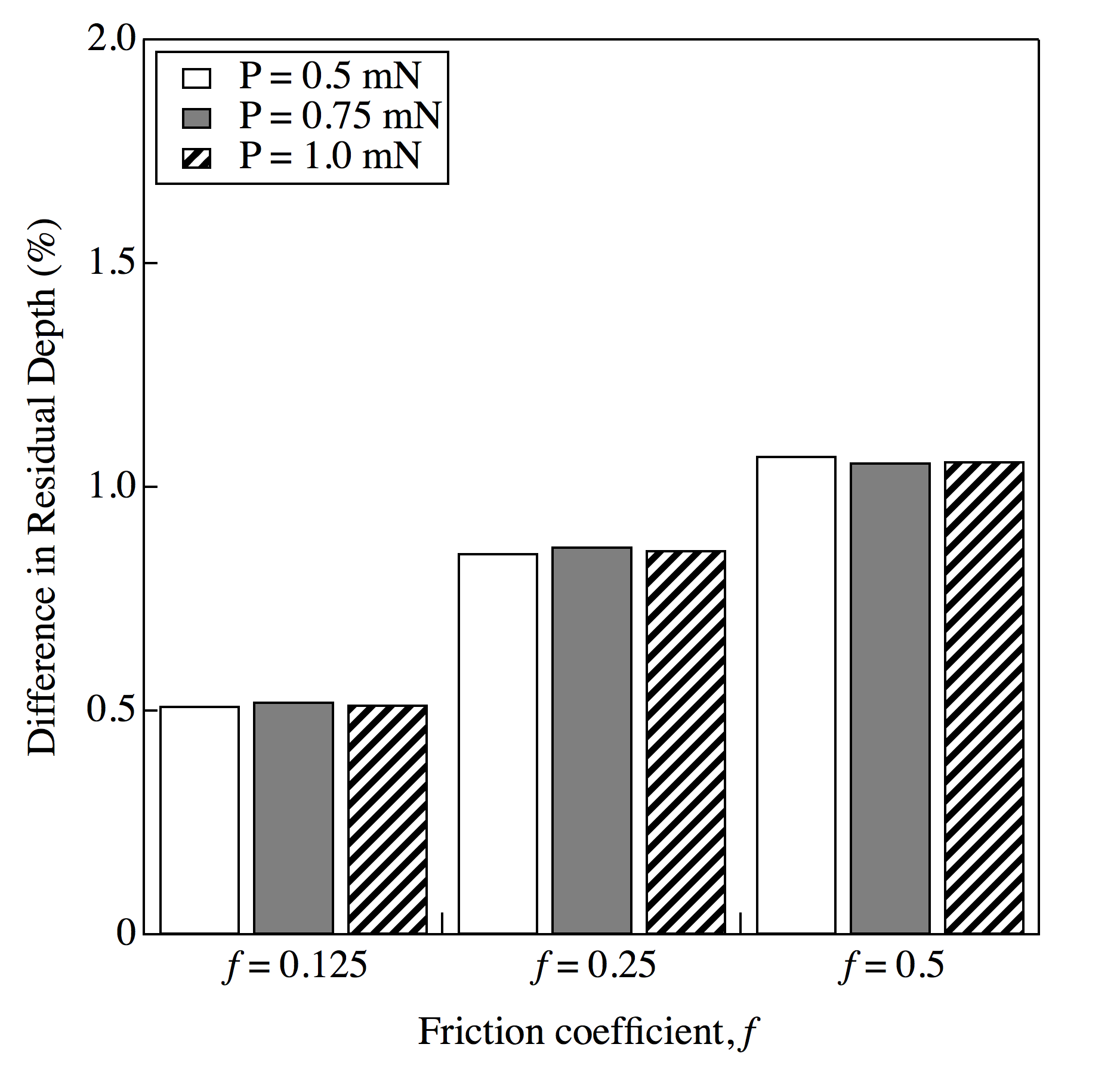}}}
\caption{Effect of friction coefficient, \emph{f} on nanoindentation data for different maximum load (constant loading--unloading time)}
\label{fig:fric_load}
\end{center}
\end{figure}

\Cref{fig:fric_load} shows the effect of the friction coefficient on maximum and residual depths when loading--unloading time was kept constant and the maximum load was varied from 0.5---1.0 mN. Similar to the investigation mentioned above, where loading--unloading time was varied within experimental range, values of maximum and residual depths were found to have decreased from the values obtained for the frictionless condition. 

\Cref{fig:fric_load}(a) shows that increase in the friction coefficient resulted in a higher reduction in maximum depth in comparison to the frictionless counterpart. This observation was common for all three different maximum load conditions. Similar behavior was observed for residual depth reductions as illustrated in Fig.~\ref{fig:fric_load}(b). Since higher friction coefficient would lead to greater frictionally dissipated energy, higher reduction compared to the frictionless conditions would therefore be expected.

Nevertheless, the overall differences were very small. As a matter of fact for \emph{f} = 0.5, which provided the maximum differences, reductions in maximum and residual depths were found to be $\approx$ 0.5 and 1\%, respectively. 

Another common observation between figure~\ref{fig:fric_load}(a) and~\ref{fig:fric_load}(b) was that between different maximum load conditions there were hardly any difference for a given coefficient of friction. It could mean that within the given range of loads (0.5mN---1.0mN), maximum load have no effect over the friction behavior of tip and sample surface. However, determining whether the maximum load insensitivity is an universal fact requires further investigation. 

This parametric study shows that the inclusion of friction in the finite element model leads to changes in the indentation load--displacement response. Nonetheless, the variations are small for the conditions of interest. Real nanoindentation experiment can never be entirely frictionless. Therefore, this study included the effect of friction in the model by using a coefficient \emph{f} = 0.25 for all sensitivity analysis and surrogate model development purposes. 

\subsection{Sensitivity Analysis}

\Cref{tab:taguchi_anova1} shows the result of sensitivity analysis carried out using Taguchi-based design of experiments. The data of 27 experiments carried out according to $L_{27}$ orthogonal array was used to get information about the sensitivity of output towards individual parameters. 

\begin{table}[htb]
\caption{Analysis of Variance for different parameters}
\begin{center}
\begin{tabular}{c c c c c c c}
\hline
Source & DF & Adj SS & Adj MS & F-Value & P-Value & \% Contribution\\
\hline
E & 2 & 5597885401 & 2798942700 & 3630.33 & 0.000 & 11.20\\
$C_s$ & 2 & 16004929654 & 8002464827 & 10379.48 & 0.000 & 32.01\\
$m_s$ & 2 & 22166899947 & 11083449973 & 14375.63 & 0.000 & 44.34\\ 
$C_t$ & 2 & 6207522908 & 3103761454 & 4025.69 & 0.000& 12.42\\ 
$m_t$ & 2 & 14961103 & 7480552 & 9.70 & 0.002& 0.03\\ 
$t_{\epsilon}$ & 2 & 92652 & 46326 & 0.06 & 0.942 & 0.00\\ 
Error & 14 & 10793843 & 770989 &  & \\
\hline
Total & 26 & 50003085508 &  &  & \\
\hline
\end{tabular}
\end{center}
\label{tab:taguchi_anova1}
\end{table}%

The `\% Contribution' data, which is a measure of variation contributed by individual parameters towards the output, shows that except for $t_{\epsilon}$ all other parameters contributed towards the overall variation of output. However, the contribution was significantly influenced by the `steady state' parameters ($C_s$ and $m_s$). 

\begin{figure}
\begin{center}
\subfigure[Effect on indentation maximum depth]{
\resizebox*{10cm}{!}{\includegraphics{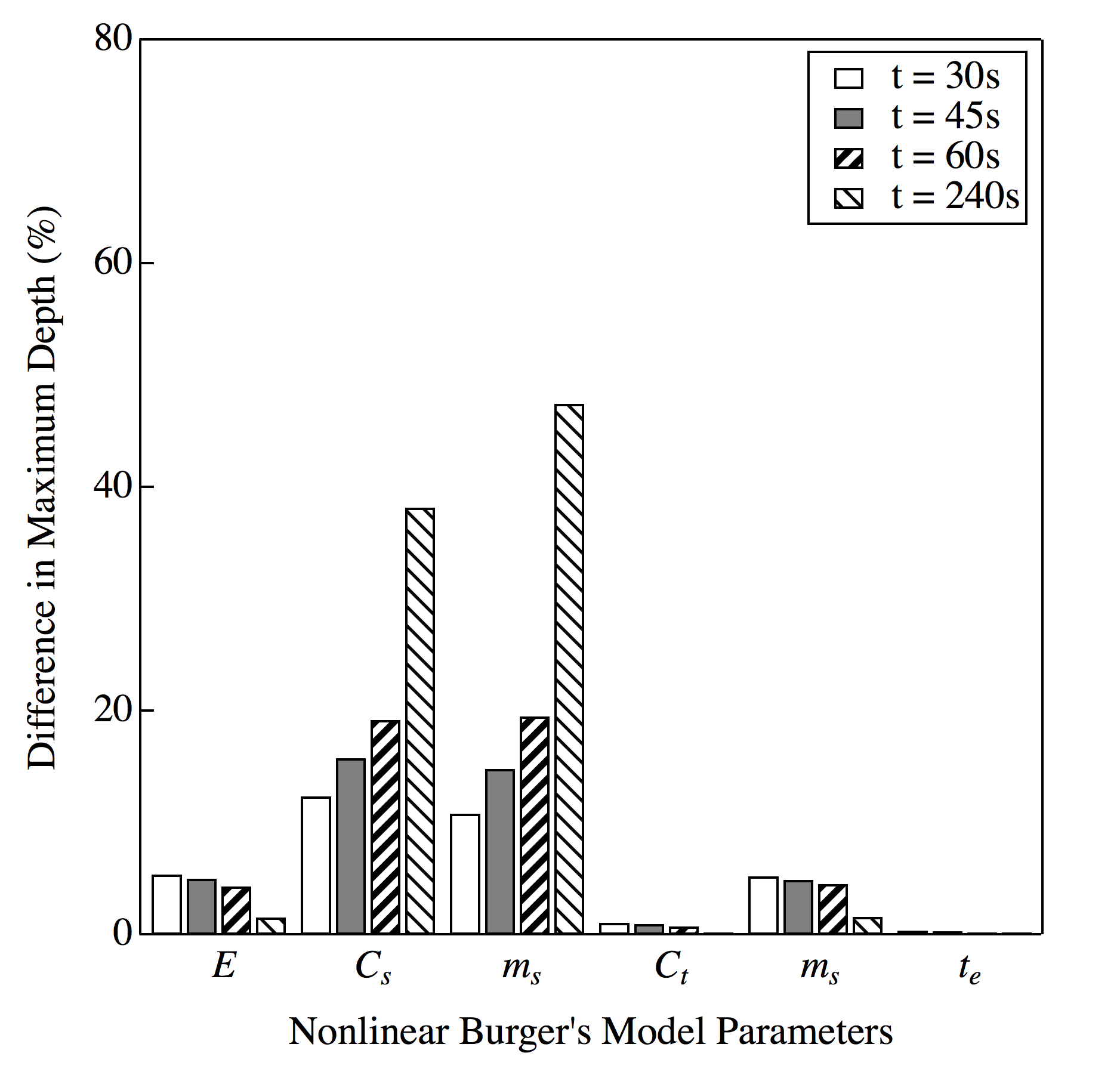}}}\hspace{5pt}
\subfigure[Effect on indentation residual depth]{
\resizebox*{10cm}{!}{\includegraphics{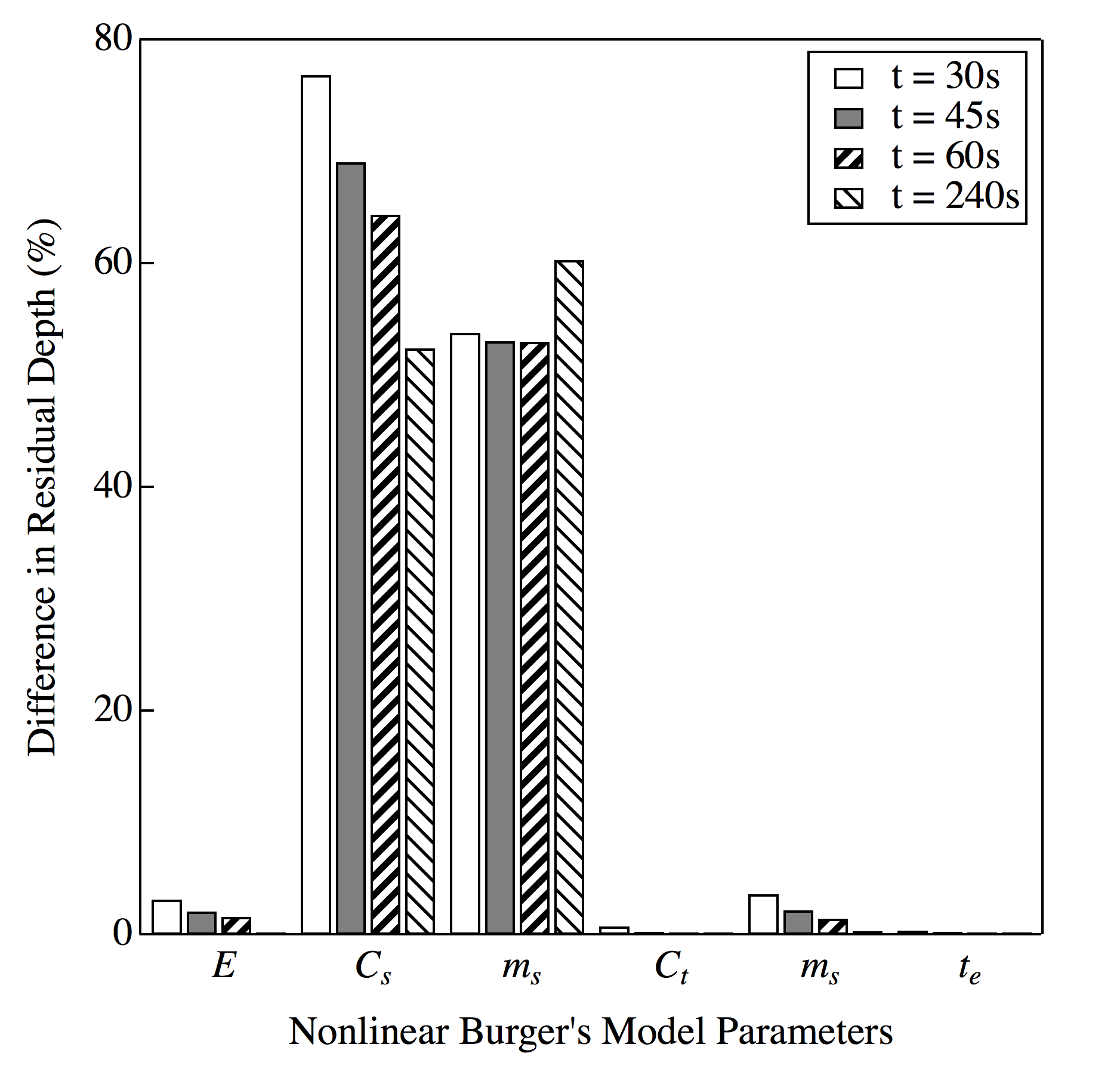}}}
\caption{Output sensitivity towards different nonlinear Burgers model parameters}
\label{fig:sensitive_param}
\end{center}
\end{figure}

\Cref{fig:sensitive_param} shows the sensitivity of indentation depth at maximum load and at the end of unloading, \emph{i.e.} maximum and residual depths. Similar to Taguchi--ANOVA based sensitivity results, it could be seen that $t_{\epsilon}$ has little to no impact on maximum or residual depths. $C_s$ and $m_s$ had the most significant impact on the output for both studied outputs. Furthermore, the level of significance is much more pronounced for residual depth compared to maximum depth.

Another observation that could be drawn from these results was elastic modulus has a positive correlation with the strain rate. In other words, both maximum and residual depth was comparatively more impacted by elastic modulus when the strain rate was higher. One explanation of this fact could be that when strain rate is higher viscoelastic response gets subdued due to inherent time lag between the elastic and viscoelastic response. As the strain rate gets slower and slower the viscoelastic or the creep response catches up with the elastic response. Hence, the elastic part of the displacement becomes less dominant in the overall displacement pattern of the material.

\Cref{fig:sensitive_param} shows another important observation, which is contrary to ANOVA results. The two parameters, $C_t$ and $m_t$ shows opposite trend in these two sensitivity tests. In ANOVA $C_t$ showed substantial influence towards the output, while it was fairly insignificant in \Cref{fig:sensitive_param}. Contrary to $C_t$, $m_t$ showed good sensitivity in fig.~\ref{fig:sensitive_param}, although very insignificant sensitivity in ANOVA results.

This is because fig.~\ref{fig:sensitive_param} represents the sensitivity of individual parameters towards two points in the nanoindentation load--displacement plot, namely maximum and residual depth. Although these two points are very important in understanding material's response, these cannot represent the entire plot. It is possible that two plots distinct in every other way can have the same maximum depth and residual depth pattern. That is why having multiple complementary means of determining sensitivity can provide a broader view of the problem.

\subsection{Surrogate Model Training and Inverse Analysis}

The findings from the sensitivity analysis was taken into account to revise the number of levels for each nonlinear Burgers model parameter. As discussed, $t_{\epsilon}$ showed no influence over the output of the nanoindentation simulations. This implies that either $t_{\epsilon}$ cannot be accurately determined from a Berkovich nanoindentation experiment or that it is a redundant parameter in describing the material response. Considering these facts, in order to reduce computational expense, $t_{\epsilon}$ was given a constant value.

The two parameters that were the most influential of the remaining five, $C_s$ and $m_s$, were varied at four levels. Meanwhile, moderately influential two parameters, E and $C_t$, were varied at three levels. According to ANOVA, $m_t$ was not significantly sensitive towards the output. However, the parametric study showed that $m_t$ had some influence over maximum indenter depth and residual indenter depth. For this reason, instead of assigning a constant value to $m_t$, it was varied in two levels.

\begin{table}[htb]
\caption{Parametric Space of Nonlinear Burgers Parameters for Surrogate Training}
\begin{center}
\begin{tabular}[l]{@{}ccc}
\hline
  Parameters &  No of Points in Space & Parametric Value Space\\
\hline  
  E & 3  & 3, 3.25, 3.5 \\
  $C_s$ & 4 &  0.02, 0.045, 0.07, 0.1 \\
  $m_s$ & 4  & 0.35 \\
  $C_t$ & 3  & 0.15, 0.25, 0.35 \\
  $m_t$ & 2  & 0.2, 0.8 \\
  $t_{\epsilon}$ & 1  & 0.25 \\
\hline 
\end{tabular}
\end{center}
\label{tab:surrogate_levels}
\end{table}

\Cref{tab:surrogate_levels} shows the corresponding levels for each parameters that were selected based on the sensitivity analysis. In a full factorial basis, a total of 3$\times$4$\times$4$\times$3$\times$2$\times$1 = 288 finite element simulations were carried out in order to generate the surrogate model for every single experimental conditions. In each of these simulations, 100 load--displacement data points were used to represent the nanoindentation plot. Since there were four individual experimental conditions to represent, a total of four surrogate models were developed. The snapshot matrix used to generate each of these surrogate model had dimensions of 100$\times$288.

After the POD model reduction process was carried out and the RBF coefficients were calculated, the POD--RBF surrogate model was ready to approximate nanoindentation data within the specified parametric space (see \Cref{tab:burger_levels}). An objective function was written in MATLAB where each surrogate model's output was compared against the corresponding experimental data. This objective function was used within the MATLAB Global Optimization Toolbox to run multi-objective genetic algorithm-based global optimization. The optimization algorithm was set to run in parallel mode until it met the stopping criteria described in \Cref{subchap:ga}. \Cref{tab:burger_optim} shows the result from the global optimization algorithm.

\begin{table}[htb]
\caption{Optimized Nonlinear Burgers Model Parameters}
\begin{center}
\begin{tabular}[l]{@{}cccccccc}
\hline
  Parameters &  E & \textnu & $C_s$ & $m_s$ & $C_t$ & $m_t$ & $t_{\epsilon}$\\
\hline  
  Optimized & 3.28 & 0.34 & 0.09& 0.20& 0.24& 0.47& 0.25\\
\hline 
\end{tabular}
\end{center}
\label{tab:burger_optim}
\end{table}

\begin{figure}
\begin{center}
\subfigure[Loading--unloading time = 30s]{
\resizebox*{7.35cm}{!}{\includegraphics{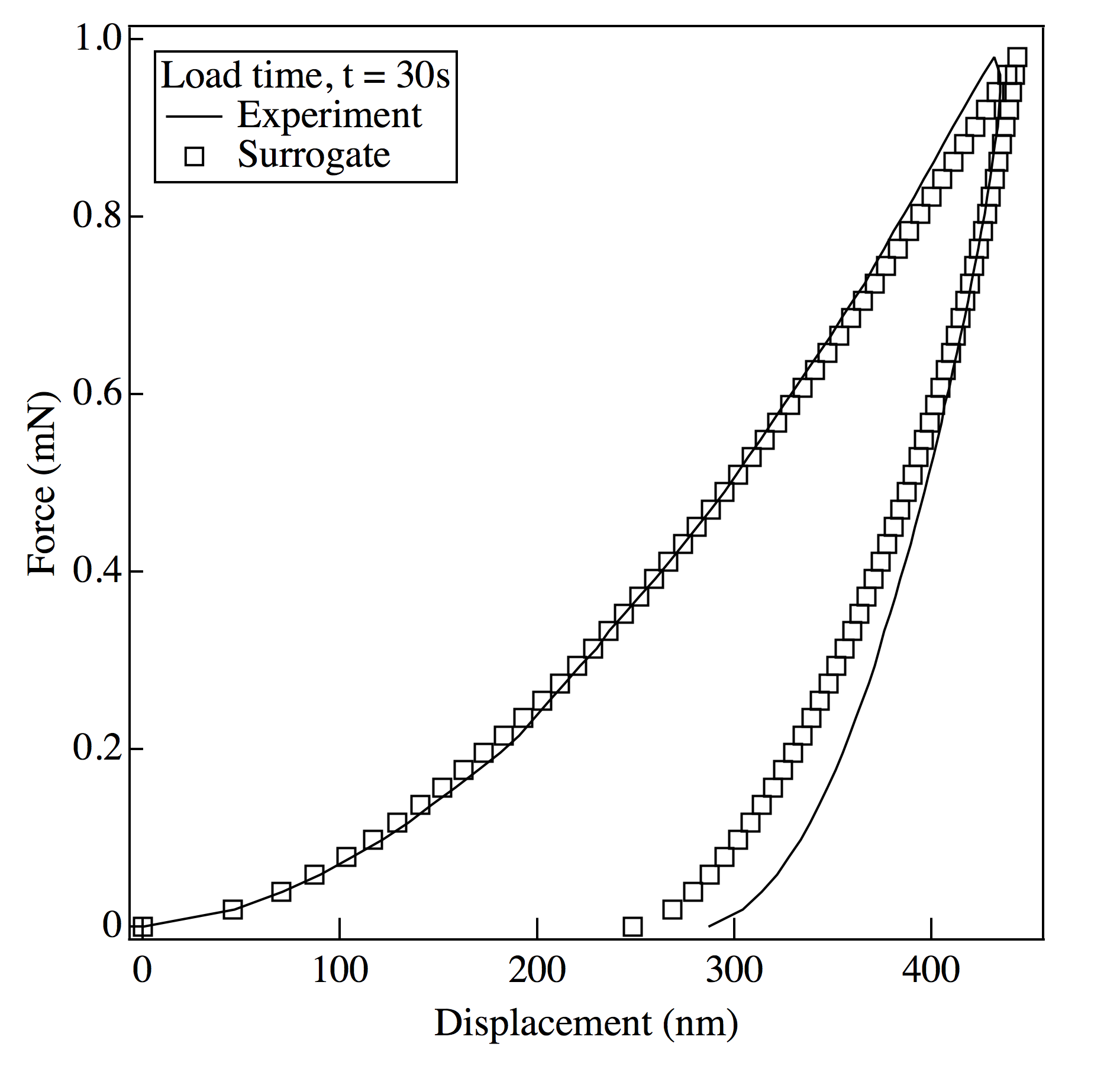}}}\hspace{2pt}
\subfigure[Loading--unloading time = 45s]{
\resizebox*{7.35cm}{!}{\includegraphics{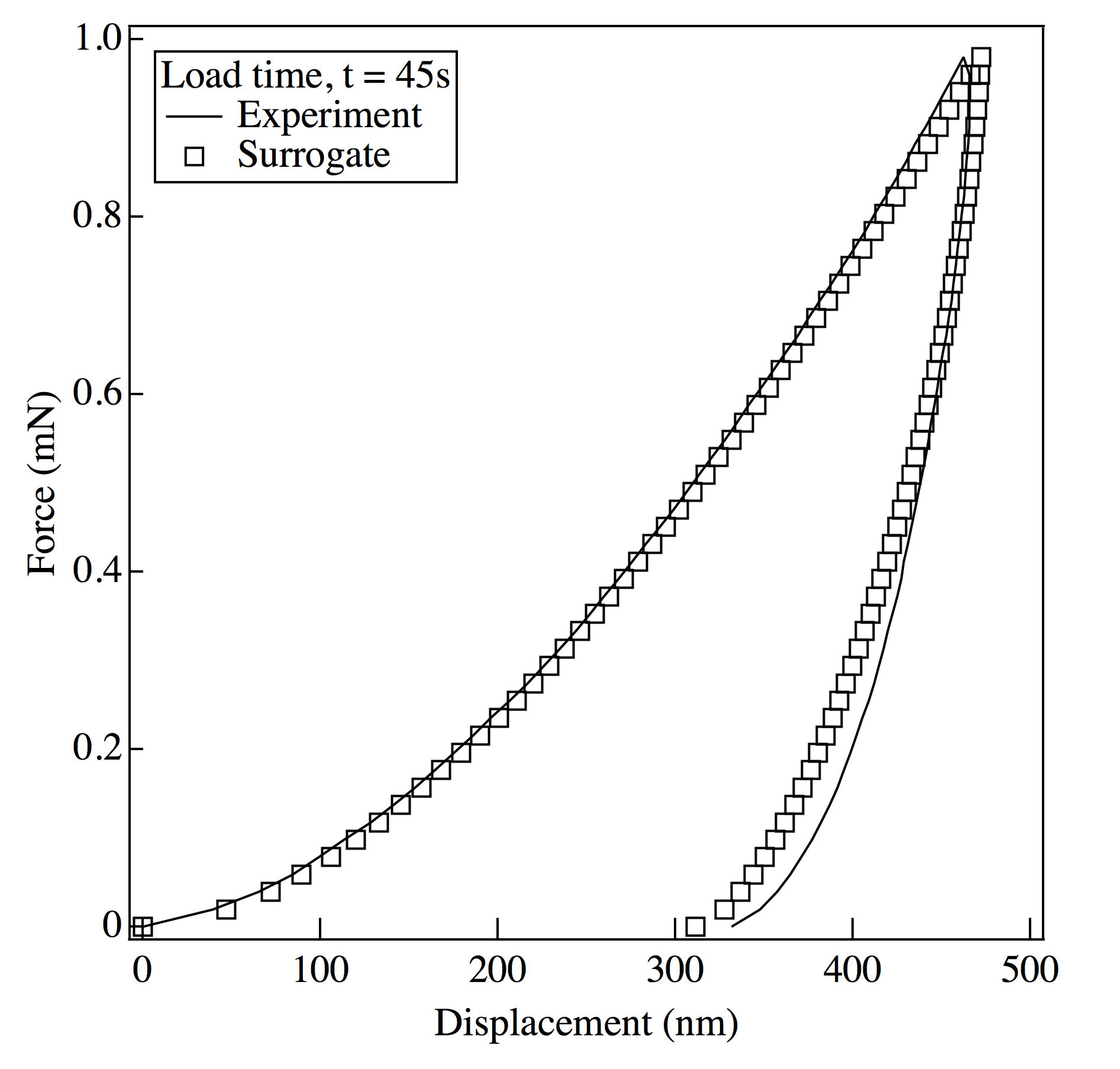}}}
\subfigure[Loading--unloading time = 60s]{
\resizebox*{7.35cm}{!}{\includegraphics{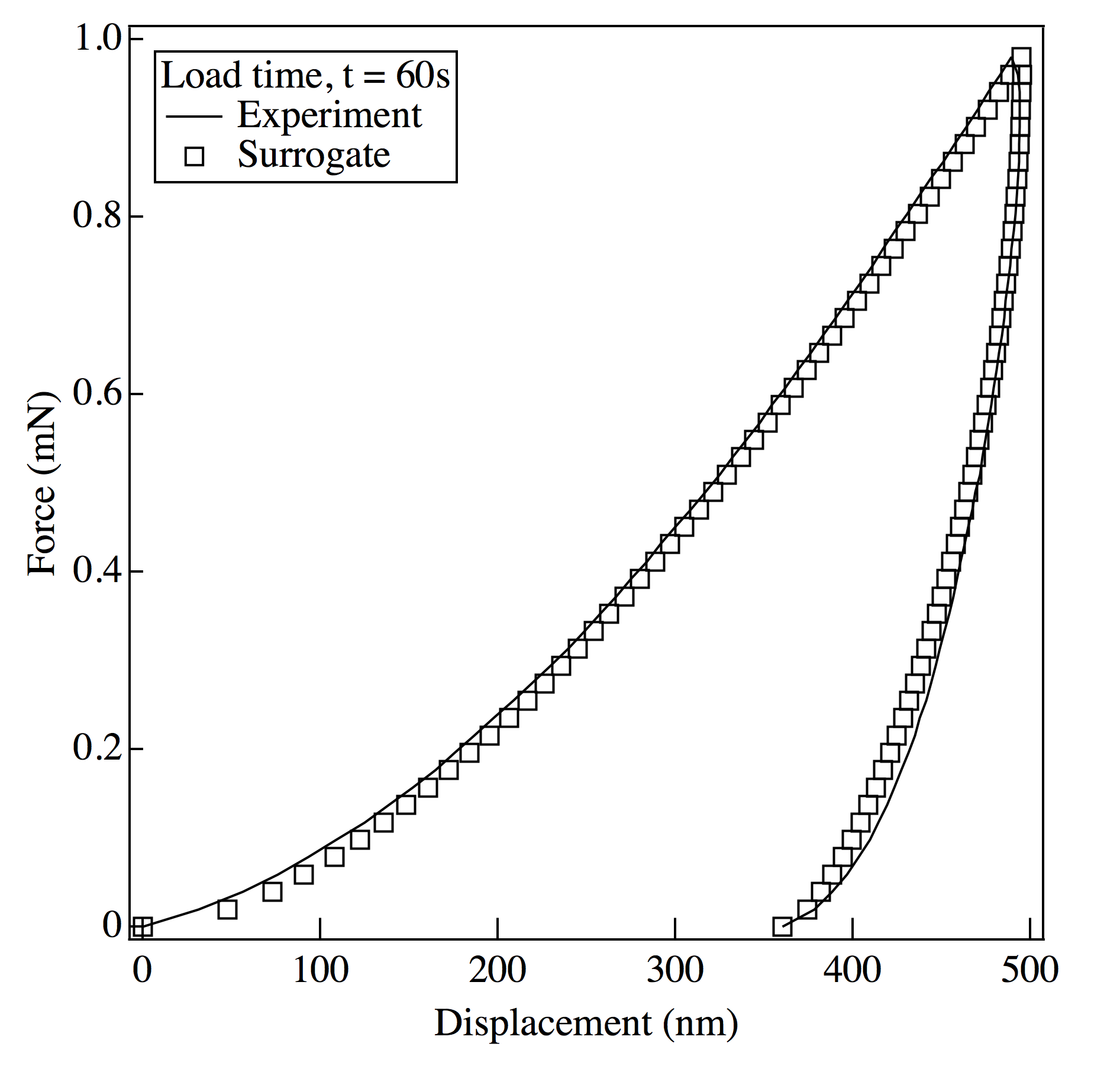}}}\hspace{2pt}
\subfigure[Loading--unloading time = 240s]{
\resizebox*{7.35cm}{!}{\includegraphics{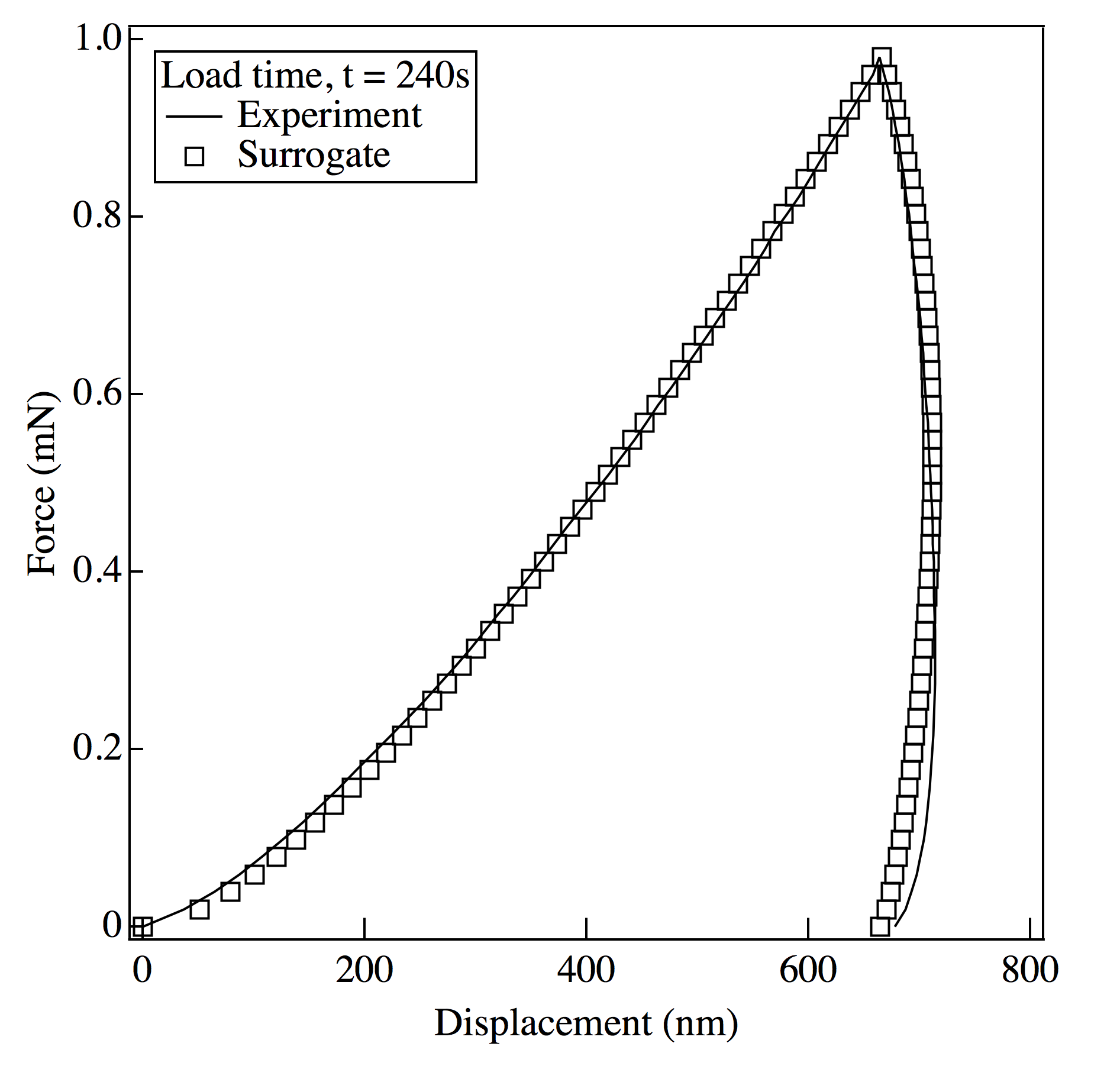}}}
\caption{Experiment vs.~surrogate model for calibrated nonlinear Burgers model parameters}
\label{fig:match_pod}
\end{center}
\end{figure}

The optimized set of parameters were the numerical best fits depending on the objective function that produces the numerical difference between the predicted and experimental data. \Cref{fig:match_pod} shows the comparison of predicted and experimental data for all four experimental cases. These were the experimental conditions that were closely followed in creating finite element models and were used to train the predictive or surrogate model. From \Cref{fig:match_pod} it can be seen that all four surrogate model outputs were very close to the corresponding experimental data. This demonstrated the fact that the multi-objective genetic algorithm-based optimization procedure was successful in finding a common minima taking the constraints in to consideration.

Although, the surrogate model prediction's were mostly close with the experimental data few inconsistencies were observed. For example, the final unloading portion data for the loading--unloading time t = 30s did not match very well. Similar behavior was observed for t = 45s, even though qualitatively the difference between prediction and experiment diminished. For higher loading--unloading time, e.g. t = 60s and 240s, the difference was noticeably very small. 

\begin{figure}
\begin{center}
\subfigure[Maximum load = 0.5 mN]{
\resizebox*{7.35cm}{!}{\includegraphics{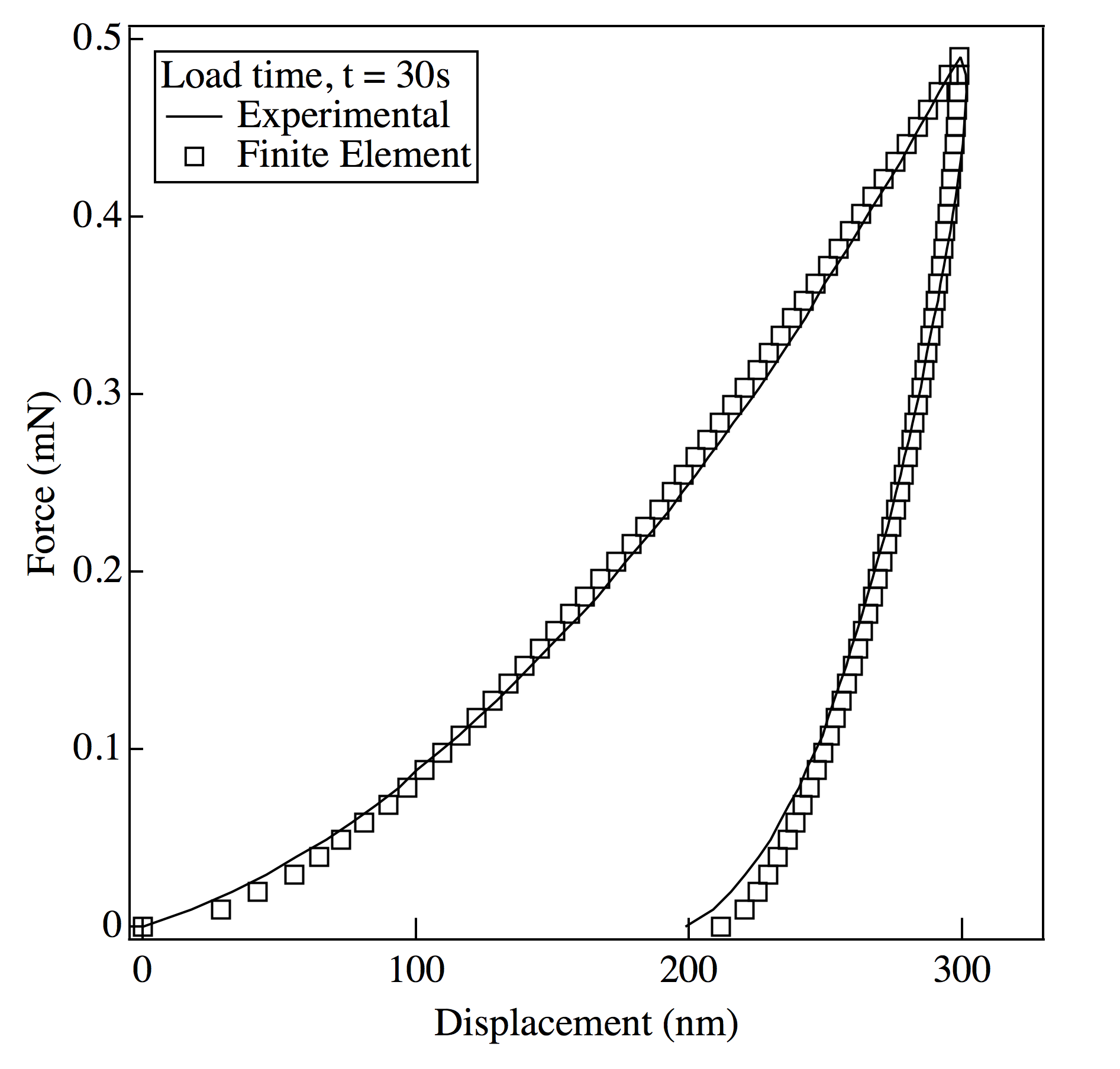}}}\hspace{2pt}
\subfigure[Maximum load = 0.5 mN]{
\resizebox*{7.35cm}{!}{\includegraphics{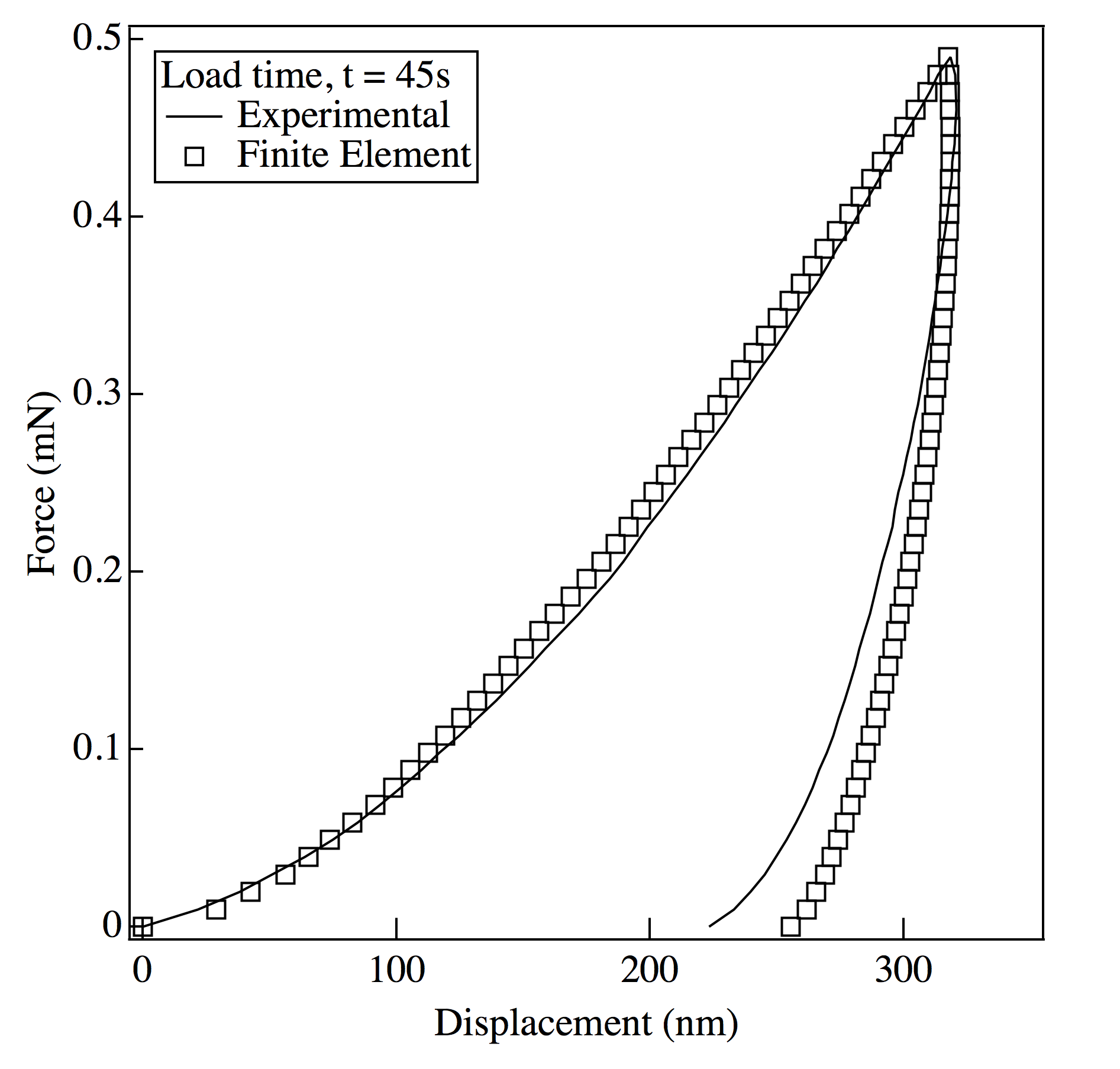}}}
\subfigure[Maximum load = 0.75 mN]{
\resizebox*{7.35cm}{!}{\includegraphics{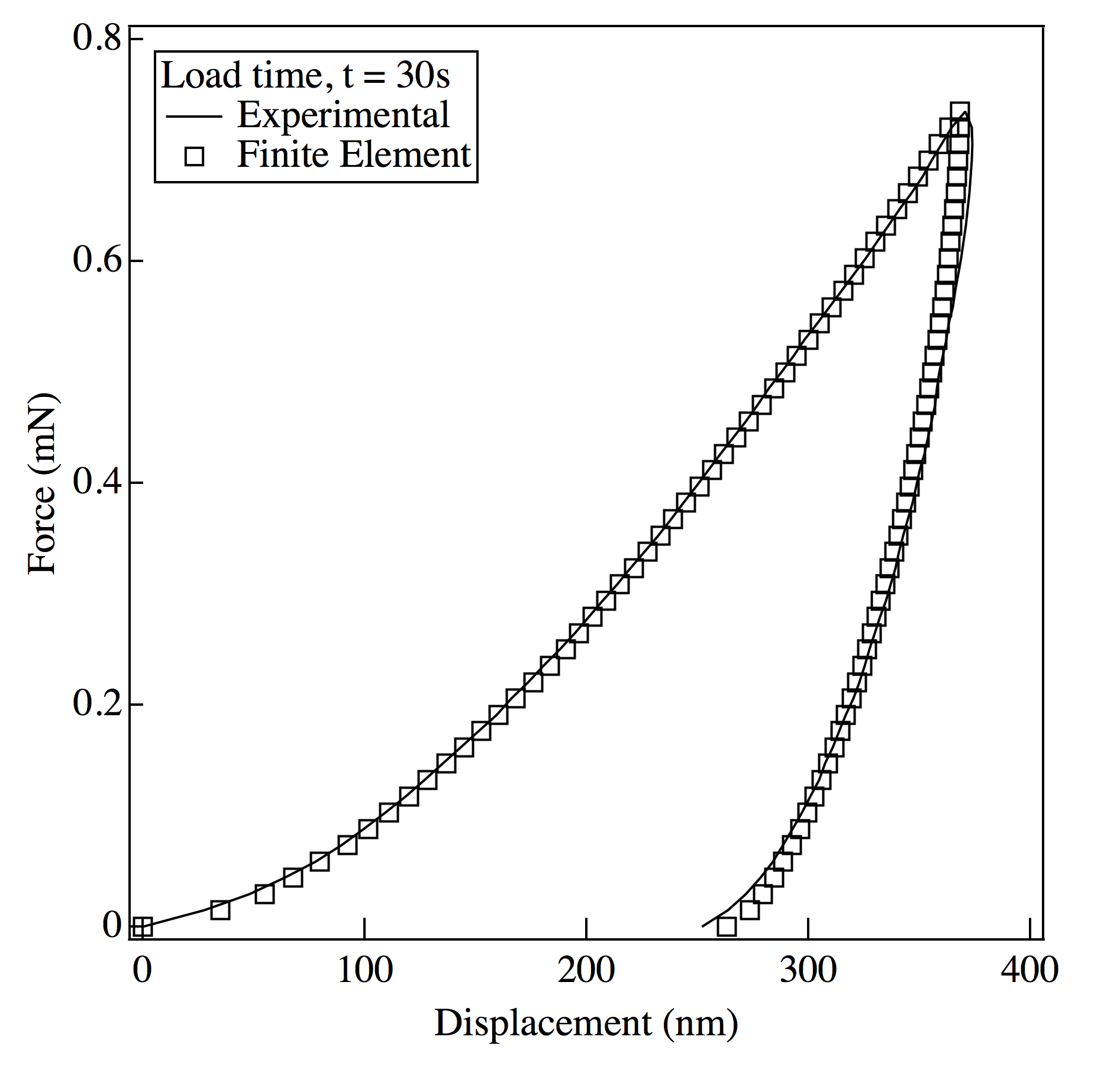}}}\hspace{2pt}
\subfigure[Maximum load = 0.75 mN]{
\resizebox*{7.35cm}{!}{\includegraphics{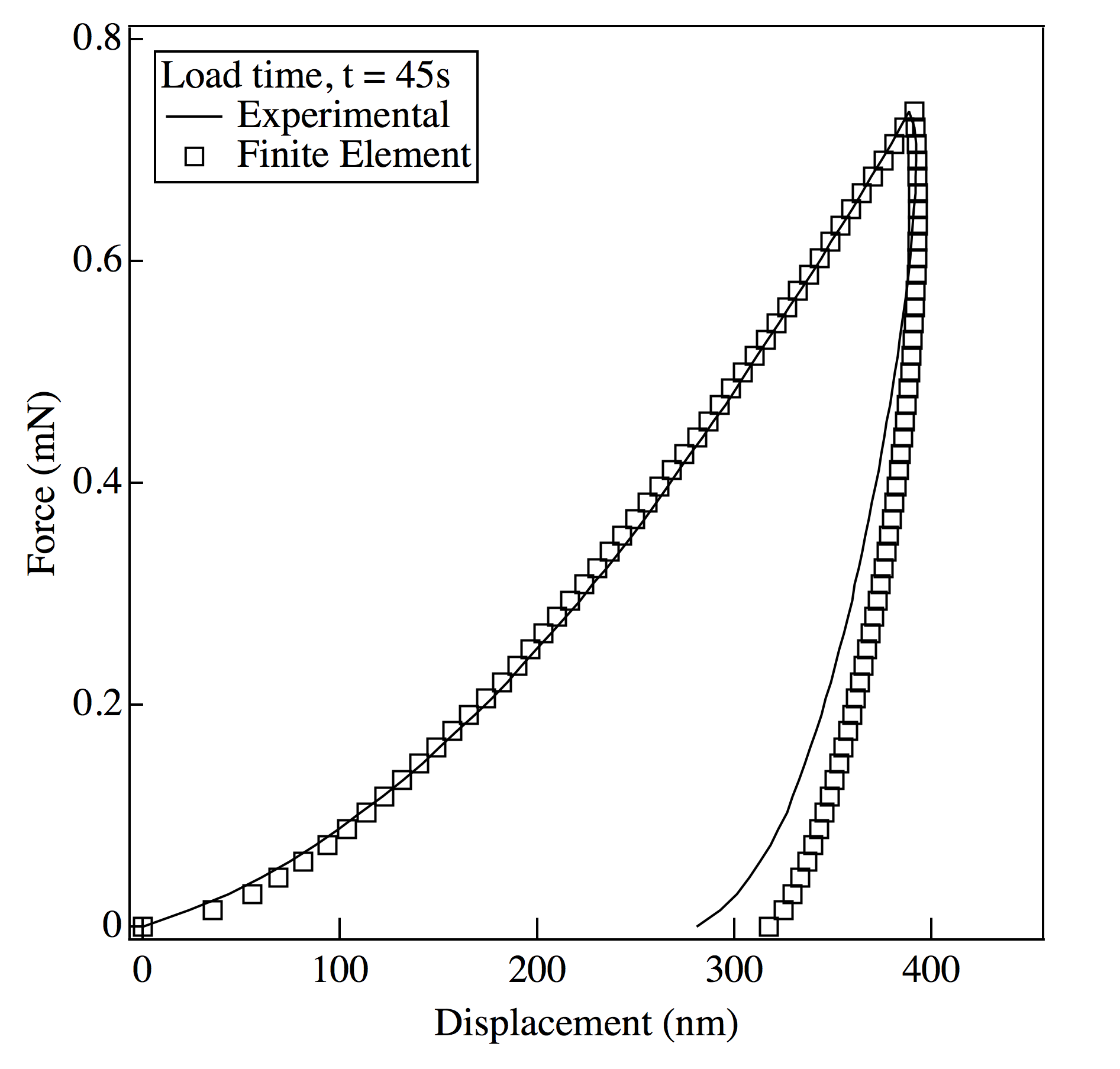}}}
\caption{Experiment vs.~finite element simulation for calibrated nonlinear Burgers model parameters}
\label{fig:match_fe}
\end{center}
\end{figure}

\Cref{fig:match_fe} shows the comparison of finite element model output for the identified Burgers model parameters vs. the corresponding nanoindentation experimental data. The surrogate model developed with finite element simulation data was not trained for these experimental conditions. These conditions were used to validate the optimized set of Burgers model parameters. 

In training the surrogate model for approximating nonlinear Burgers model output, experimental conditions with varying loading--unloading times were used. On the other hand, in these validation experiments maximum loads for which the model was not trained have been used. This decision was deliberately made in consideration of the fact that nonlinear viscoelastic behavior not only depends on strain rate but also on the strain levels associated with the experiment. For nanoindentation experiments it could be safely assumed that changing the load levels would result in change of strain levels.

From the validation plots for untrained experimental conditions, it could be observed that the finite element simulation output closely matched with the experimental data. In two of the cases (see \Cref{fig:match_fe}b \& \ref{fig:match_fe}d) a portion of the unloading curve showed some discrepancies in a qualitative sense. 

\Cref{tab:error} shows quantitative variation between the various plots in Fig.~\ref{fig:match_pod} and \ref{fig:match_fe}. Variations between the plots has been represented in terms of RMSE, R\textsuperscript{2}, Avg.~Error, and \% Error. As it can be seen, various quantitative discrepancies for \Cref{fig:match_fe}b \& \ref{fig:match_fe}d were found to be comparable with the other plots that showed little discrepancies in qualitative sense. 

Another important observation that could be made was that \Cref{fig:match_pod}b and \ref{fig:match_pod}d both showed almost same quantitative variation. Although, \Cref{fig:match_pod}d's match looked slightly better than \Cref{fig:match_pod}b if perceived visually.  

\begin{table}[htb]
\begin{center}
\caption{Variation Between Different Plots} 
($P_{max}$ = maximum load, t = loading--unloading time)
\par\smallskip
\begin{tabular}[l]{@{}ll|cccc}
\hline
  \multicolumn{2}{c|}{Conditions} &  RMSE & R\textsuperscript{2} & Avg.~Err.~(nm) & \% Error\\
\hline  
  \multirow{4}{*}{$P_{max}$ = 1.0mN} & t = 30s & 13.23  & 0.9821 & 9.11 & 2.81\\%
  & t = 45s & 9.17  & 0.9867 & 7.11 & 2.34\\%
  & t = 60s & 6.72  & 0.9891 & 5.70 & 2.81\\%
  & t = 240s & 9.01  & 0.9893 & 7.07 & 2.33\\%
  \cline{1-2}
  \multirow{2}{*}{$P_{max}$ = 0.5mN} & t = 30s  & 4.19  & 0.9884 & 3.29 & 2.73\\
  & t = 45s & 10.06  & 0.9827 & 7.89 & 3.84\\%
  \cline{1-2}
 \multirow{2}{*}{$P_{max}$ = 0.75mN} & t = 30s & 3.48  & 0.9895 & 2.78 & 1.41\\%
  & t = 45s & 10.69  & 0.9858 & 7.59 & 3.38\\%
\hline 
\end{tabular}
\end{center}
\label{tab:error}
\end{table}

The variation between different plots for optimized set of model parameters could have stemmed from different factors. One such factor could be the friction coefficient used in the finite element model. As it can be seen the effect of friction coefficient varied depending on the experimental conditions that were being replicated. As a result the error associated with using a particular friction coefficient also varied from one experimental condition to the other. Since the whole process of inverse analysis depended on numerical manipulations, different error in the FEA data could have skewed the parameter optimization in one way or the other.

Ascertaining that the material model parameter set that has been extracted from the inverse analysis procedure is indeed the global parameter set that would satisfy all possible material response is a challenge. In order to deal with this challenge, material responses from other experiments, such as compression, tension, or flexural tests could be included in the process. For some materials carrying out the aforementioned tests may not be feasible, e.g. thin films, coatings, biological cells. In those cases improving the confidence in the optimized parameter set could be established by obtaining material response data from multiple nanoindentation experiments, such as changing the cone angle for a pyramidal indenter tip, or using spherical tips with different radii. 

Another way of finding additional constraints for the numerical analysis would be use additional experimental data from the same nanoindentation experiment. For example, if imprint geometry or residual depth profile data could be harnessed from a nanoindentation experiment and used in the objective function, the probability of finding the unique model parameter set increases.

 \chapter{CONCLUSIONS}\label{chap:conclusions}

The focus of this study was to identify the nonlinear viscoelastic model parameters for a soft material. Mechanical characterization of soft materials, such as polymers and biomaterials is often challenging due to various size and shape restrictions of the bulk testing methods. Nanoindentation, or Depth Sensing Indentation (DSI) is particularly useful in characterizing material behavior since sample preparation is very straightforward. 

While application of the nanoindentation technique for identifying elastic--plastic material model parameters has been extensively studied, identification of viscoelastic behavior still required further investigation. This is because viscoelastic behavior is a much more complex material behavior to analyze due to the time dependence of material response. 

Earlier studies that investigated viscoelastic behavior using nanoindentation utilized correspondence principal-based analytical solutions to define material response. This is a rather simplistic way of defining material behavior because it assumes viscoelastic response to be linear. Furthermore, the analytical solutions are often valid until the load is monotonically increasing, i.e. loading portion of the nanoindentation curve. Since soft materials are nonlinearly viscoelastic and material response information from only the loading curve is incomplete, analytical solutions are unable to capture the full spectrum of material response.

To circumvent the problem associated with analytical solutions, a hybrid approach named inverse analysis can be used to model complex material behaviors. Various studies have shown that by reducing the numerical difference between simulated and experimental data, material model parameters can be identified. Even though comparatively simple elastic--plastic nanoindentation has been widely studied through inverse analysis, complex time-dependent material response specially nonlinear viscoelastic behavior has not been investigated at depth.

In this study, load--displacement material response of a nanoindentation experiment conducted on soft epoxy material has been modeled using nonlinear Burger's model. The model parameters of Burger's model was identified using a global optimization algorithm that reduced the differences between the simulation and the experimental data. The traditional method of inverse analysis-based parameter identification requires finite element simulation to run inside the optimization algorithm. The computational expense required to identify parameters thus becomes very large. In order to solve the computation expense problem a predictive or surrogate model was trained using finite element simulation data. A surrogate model, once trained, is a few order of magnitude faster than actual finite element simulation. Hence, instead of using the finite element analysis within the optimization algorithm, the surrogate model was used to approximate the simulation data. 

In this study, a POD--RBF based surrogate model was used. The performance or quality of the POD--RBF surrogate model is dependent on a few parameters. These parameters, such as number of training data and choice of basis functions, were studied at depth before using surrogate model to calibrate material model. From this investigation it was found that the information from a sensitivity analysis of the model parameters could be utilized to reduce the number of sampling points without conceding quality of approximation. 

This study utilized Taguchi--ANOVA based sensitivity analysis to identify the key parameters that has the most influence over the nanoindentation output. A parametric sensitivity analysis was also performed in order to understand the effect of each parameter over the maximum and residual depths from load--displacement data. The model parameters representing steady state viscoelasticity ($C_s$ and $m_s$) was contributing majority of the variation towards the output. Depending on the results of the sensitivity analysis the parametric space was defined within which the snapshots for the POD--RBF method were determined.

Another key aspect of finite element material modeling, i.e. effect of friction has been investigated in this study. Studies conducted in the area of nanoindentation-based material model calibration suggested that nanoindentation load--displacement data can be affected by the friction between tip and sample surface. This phenomenon is primarily dependent upon the material model and the tip geometry under investigation. In this investigation it was found that friction has a small influence over the nanoindentation data for the studied material model and the Berkovich tip.

The snapshots for the POD--RBF method were generated via finite element simulations with varying parameter sets within the parametric space. To include friction effect within the model a finite valued friction coefficient was used. The POD--RBF surrogate model was trained through numerical manipulations as described in earlier chapters. The objective or cost function was defined as the mean squared errors between the experimental and numerical (surrogate approximation) data. A genetic algorithm based optimization method was used to reduce the objective function to determine the model parameter set that satisfies the given constraints. 

It was observed that the differences between the experimental and surrogate model predicted data for the optimized parameter set was small. This meant that for the trained conditions the optimization process coupled with surrogate model was able to provide a satisfactory parameter set. In order to check the validity of the calibrated model parameters, another set of comparison was drawn between nanoindentation data and finite element simulation data. These nanoindentation experiments were carried out in different conditions for which the predictive model was not trained. Simulation data for the optimized parameter set matched well with these experimental data. This demonstrated that the optimized parameter set was able to capture material behavior for various experimental scenarios. 
 
In this study, it has been shown that the importance of developing an analysis technique to characterize materials exhibiting nonlinear viscoelastic behavior is enormous. Along with improving the fundamental knowledge about polymers and biomaterials, this would also help in understanding disease progression, designing better artificial organs, providing service life prediction of composites, etc. This study presented a robust approach to determine nonlinear viscoelastic model parameters that could be utilized to predict soft material's behavior under different kinds of loading. By strengthening the existing weaknesses in the literature this study opened up possibilities of characterizing soft materials using an alternative technique that is unobtrusive, non-cumbersome, and virtually nondestructive.    

%
%\nocite{*} % Use to exclude specific citations from *.bib file
\bibliography{thesis_main}
\bibliographystyle{ieeetr}%{alpha}%{ieeetr}%%
%\appendix
%\input{appendix}
%\input{vita}
%       \pagestyle{empty}

\end{document}